\documentclass[twocolumn,showpacs,aps,prc,superscriptaddress]{revtex4}

\usepackage[dvips]{graphicx}
\usepackage{color}
\usepackage{latexsym}

\newcommand{\thgg}{\theta_{\gamma\gamma}^\ast}
\newcommand{\cthgg}{\cos \theta_{\gamma\gamma}^\ast}
\newcommand{\cthcm}{\cos \theta_{\mathrm{c.m.}}}
\newcommand{\thcm}{\theta_{\mathrm{c.m.}}}
\newcommand{\phigg}{\phi}
\begin{document}

\title{Virtual Compton Scattering and Neutral Pion  Electroproduction 
       in the Resonance Region up to the Deep
       Inelastic Region at Backward Angles}

\author{G.~Laveissi\`{e}re}
\affiliation{LPC-Clermont,  Universit\'{e} Blaise Pascal, CNRS/IN2P3, F-63177 Aubi\`{e}re Cedex, France}
\author{N.~Degrande}
\affiliation{University of Gent, B-9000 Gent, Belgium}
\author{S.~Jaminion}
\affiliation{LPC-Clermont,  Universit\'{e} Blaise Pascal, CNRS/IN2P3, F-63177 Aubi\`{e}re Cedex, France}
\author{C.~Jutier}
\affiliation{LPC-Clermont,  Universit\'{e} Blaise Pascal, CNRS/IN2P3, F-63177 Aubi\`{e}re Cedex, France}
\affiliation{Old Dominion University, Norfolk, VA 23529}
\author{L.~Todor}
\affiliation{Old Dominion University, Norfolk, VA 23529}
\author{R.~Di Salvo}
\affiliation{LPC-Clermont,  Universit\'{e} Blaise Pascal, CNRS/IN2P3, F-63177 Aubi\`{e}re Cedex, France}
\author{L.~Van Hoorebeke}
\affiliation{University of Gent, B-9000 Gent, Belgium}
\author{L.C.~Alexa}
\affiliation{University of Regina, Regina, SK S4S OA2, Canada}
\author{B.D.~Anderson}
\affiliation{Kent State University, Kent OH 44242}
\author{K.A.~Aniol}
\affiliation{California State University, Los Angeles, Los Angeles, CA 90032}
% ...
\collaboration{The Jefferson Lab Hall A Collaboration}
\noaffiliation

\makeatletter
\global\@specialpagefalse
\def\@oddhead{\hfill {G. Laveissi\`ere {\it et al.,}} {Virtual Compton Scattering and Neutral Pion Electroproduction ... }}
\let\@evenhead\@oddhead
\def\@oddfoot{\reset@font\rm\hfill \thepage\hfill
} \let\@evenfoot\@oddfoot
\makeatother

\begin{abstract}
%
%
We have made the first measurements of the virtual Compton scattering (VCS) process via the H$(e,e'p)\gamma$ exclusive reaction in the nucleon resonance region, at backward angles.  Results are presented for the $W$-dependence at fixed $Q^2=1$ GeV$^2$, and for the $Q^2$-dependence at fixed $W$ near 1.5 GeV. The VCS data show resonant structures in the first and second resonance regions. The observed $Q^2$-dependence is smooth. The measured ratio of H$(e,e'p)\gamma$ to H$(e,e'p)\pi^0$ cross sections emphasizes the different sensitivity of these two reactions to the various nucleon resonances. Finally, when compared to Real Compton Scattering (RCS) at high energy and large angles, our VCS data at the highest  $W$ (1.8-1.9 GeV) show a striking $Q^2$-independence, which may suggest a transition to a perturbative scattering mechanism at the quark level.
%
%
\end{abstract}


\maketitle


\section{Introduction}
\label{sec:Introduction}

Understanding nucleon structure in terms of the non-perturbative dynamics
of quarks and gluons requires new and diverse experimental data to guide
theoretical approaches and to constrain models. Purely electro-weak processes
are privileged tools since they can be interpreted directly in terms of the
current carried by the quarks.
This paper presents a study of the virtual Compton scattering (VCS) process~:
$\gamma^\star p \rightarrow \gamma p$, in the nucleon resonance region via the
photon electroproduction reaction~: H$(e,e'p)\gamma$, together with results in the 
neutral pion electroproduction channel H$(e,e'p)\pi^0$. 
This study is based on part of the data of the E93-050 experiment~\cite{Laveissiere:2004nf,Laveissiere:2003jf} performed at the Thomas Jefferson National Accelerator Facility (JLab).
Its motivations were twofold: 1) investigate the very low-energy region, below the pion production threshold, to determine the Generalized Polarizabilities of the proton~\cite{Laveissiere:2004nf}; 2) make an exploratory study of the VCS process in the region of the nucleon resonances, which is the subject of the present paper. A first set of E93-050 results in the H$(e,e'p) \pi^0$ channel were published in~\cite{Laveissiere:2003jf}. This experiment was part of the Hall A commissioning phase, and was therefore conducted  prior to the Real Compton Scattering (RCS) and Deep VCS (DVCS) program at JLab. Lastly, in this  experiment the photon electroproduction process was for the first time cleanly separated from the dominant H$(e,e'p)\pi^0$ reaction above pion threshold.

The Constituent Quark Model of Isgur and Karl~\cite{Isgur:1979wd,Koniuk:1980vw}
reproduces many features of the nucleon spectrum. However, the structure
of the nucleon resonances, particularly the electro-weak transition form
factors, remain incompletely understood.
The simultaneous study of both $(N\pi)$ and $(N\gamma)$ final states of the
electroproduction process on the nucleon offers probes with very different
sensitivities to the resonance structures.
Another motivation for the present study is to explore the exclusive 
H$(e,e'p)\gamma$ reaction at high $W$, where 
perturbative current quark
degrees of freedom may become as important as those of constituent quarks
and resonances.
Quark-hadron duality implies that even at modest
$Q^2$, inelastic electron scattering in the resonance region can be analyzed
in terms of quark rather than nucleon resonance degrees of 
freedom~\cite{Niculescu:2000tk}.

\begin{figure}[ht]
\includegraphics[width=8.6cm]{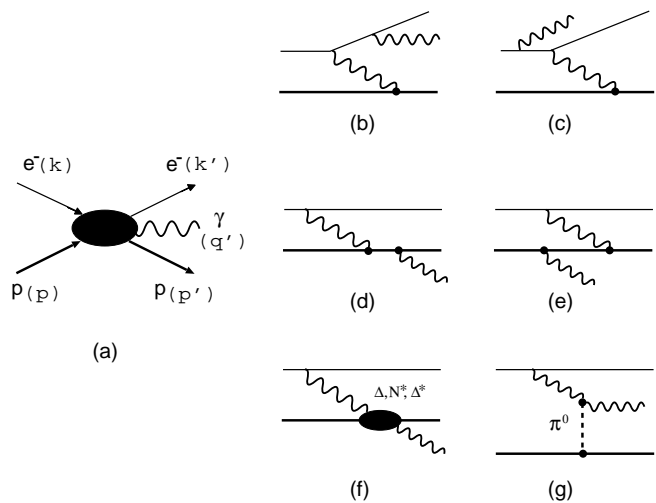}
\caption{\label{fig:feynman}
Kinematics for photon electroproduction on the proton (a) and lowest order amplitudes for  Bethe-Heitler ($\rm{b}$, $\rm{c}$),  VCS Born ($\rm{d}$, $\rm{e}$), VCS Non-Born (f), and $t$-channel  $\pi^0$-exchange (g) processes. The particle  four-momenta are indicated in parenthesis in a).
}
\end{figure}

\subsection{Kinematics}
\label{sec:Kinematics}

The kinematics of the H$(e,e'p)\gamma$ reaction are represented 
in  Fig.~\ref{fig:feynman}a. A common set of invariant kinematic variables is defined as $-Q^2=(k-k')^2=q^2$, $s=W^2 = (q+p)^2$, $t=(p-p')^2$,
and $u=(p-q')^2$.
The $\vec{q}$-direction defines the polar axis of the coordinate system~:
$\thgg$ and $\phigg$
are the polar and azimuthal angles in the $\gamma^\star p \rightarrow \gamma p$
subprocess c.m. frame. The scattered electron direction defines
$\phigg =0$.
The H$(e,e'p)\gamma$ reaction was measured  below the pion threshold in several experiments~\cite{Roche:2000ng,Laveissiere:2004nf,Bourgeois:2006js,Janssens:2008qe} and in the region of the $\Delta(1232)$ resonance~\cite{Bensafa:2006wr,Sparveris:2008jx}. 

We present in this paper the first measurements of the H$(e,e'p)\gamma$ cross section that were made through the entire nucleon resonance region.  We measured the photon electroproduction  cross section in two scans:
\begin{itemize}
\item The nucleon excitation function from threshold to $W=1.9$~GeV at $Q^2=1$~GeV$^2$;
\item The $Q^2$-dependence near $W=1.5$ GeV. 
\end{itemize}

The cross section for the H$(e,e'p)\pi^0$ process was determined simultaneously in the experiment, at the same kinematics. All these measurements were performed in backward kinematics, i.e. within a cone ($\cthgg<-0.5$) centered on the backward axis  ($\vec{q}^{\, \, \prime}$ opposite to $\vec{q}$). This angular domain, traditionally dominated by $u$-channel exchanges, is opposite to the DVCS kinematics which are at forward $\thgg$.

\subsection{Interference of Bethe-Heitler and VCS Amplitudes}
\label{sec:BHVCS}

In the one-photon exchange approximation, the photon electroproduction
amplitude (Fig.~\ref{fig:feynman}a) includes the coherent superposition of 
the VCS Born (Fig.~\ref{fig:feynman}d
and \ref{fig:feynman}e) and Non-Born (Fig.~\ref{fig:feynman}f) amplitudes,
and the Bethe-Heitler (BH) one (Fig.~\ref{fig:feynman}b and
\ref{fig:feynman}c)~\cite{Bethe:1934za}.
Note that in the BH amplitude, the mass-squared of the virtual photon (elastically
absorbed by the proton) is $t$. In the VCS amplitude, the mass-squared of the
virtual photon (inelastically absorbed) is $-Q^2$.
The BH amplitude dominates over VCS when the photon is emitted in either
the direction of the incident or scattered electron. It also breaks the symmetry of the electroproduction amplitude around the virtual photon direction.
Thus, in the data analysis we have not expanded the $\phigg$-dependence of the H$(e,e'p) \gamma$  cross section in terms of the usual electroproduction $LT$ and $TT$ interference terms. This would be possible for $W$ well above the $\Delta(1232)$-resonance, where the BH amplitude becomes negligible. However, in this region ($W \ge 1.4$ GeV) our data are mostly $\phigg$-independent  within statistics.

\section{ Experiment and Analysis} 
\label{sec:Experiment}

\begin{figure}[t]
\includegraphics[width=8.6cm]{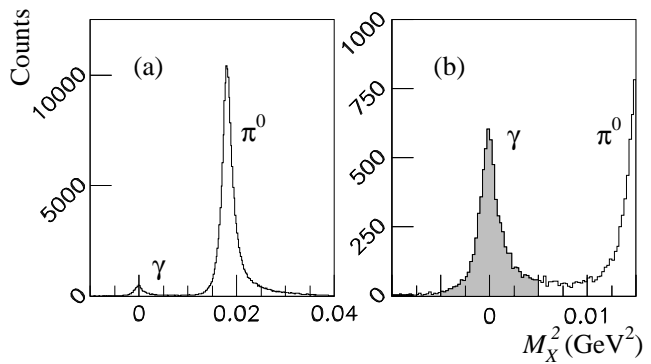}
\caption{\label{fig:missing-mass}Squared missing mass $M_X^2$ for an experimental setting at $W=1.2$~GeV (a) and zoom around the $\gamma$ peak (b). The shaded window [--0.005, 0.005] GeV$^2$ is used to select the $\gamma$ events. The FWHM of the peak increases from 0.0022 to 0.0050 GeV$^2$ when $W$ goes from 1.1 to 1.9 GeV.
}
\end{figure}

We performed the experiment at JLab in Hall~A. The continuous electron beam of energy $4.032$~GeV with an intensity of 60-120~$\mu$A bombarded a 15~cm liquid hydrogen target. The scattered electron and recoil proton were detected in coincidence in two high-resolution spectrometers. The emitted photon or $\pi^0$ was identified by reconstruction of the mass of the missing particle.   A spectrum of the squared missing mass $M_X^2=(k+p-k'-p')^2$ is displayed in Fig.~\ref{fig:missing-mass} and shows the good resolution achieved in the separation of the two electroproduction channels. The apparatus is described in detail in~\cite{Alcorn:2004}, and the detector acceptance and spectrometer settings in~\cite{Laveissiere:2003jf}.

We extract the five-fold differential cross section
$d^5\sigma (ep \to ep \gamma) \ = \ d^5\sigma / (dk'_{lab} \, d[\Omega_e]_{lab} \, d[\Omega_p]_{\mathrm{c.m.}})$
using the method described in~\cite{Laveissiere:2003jf};
$dk'_{lab}$ and $d[\Omega_e]_{lab}$ are the scattered electron differential
momentum and solid angle in the lab frame, and $d[\Omega_p]_{\mathrm{c.m.}}$ is the proton c.m. differential solid angle.
The calculations of the solid angle and radiative corrections
are based on a simulation~\cite{Janssens:2006} including the coherent sum
of the BH and VCS-Born amplitudes (Fig.~\ref{fig:feynman}b,c,d and e)
only. The inclusion of the BH-amplitude ensures that our simulation reproduces the strong $\phigg$-dependence near pion threshold. Corrections were applied for acceptance, trigger efficiency, acquisition and electronic dead times, tracking efficiency, target boiling, target impurity and proton absorption~\cite{Laveissiere:2003jf}.
In addition, a correction ($-0.1$ to $-1.7\%$) for the exclusive $\pi^0$
background in the $M_X^2$ window $[-0.005, 0.005]$~GeV$^2$ was made using
our simulation, based on the results of \cite{Laveissiere:2003jf}.

The data are binned in the variables $\cthgg, \phigg$ and $W$. In each bin the cross section is determined at a fixed point, using the model dependence of the BH+Born calculation.  This fixed point is at the center of the bins in  $\cthgg, \phigg$ and $W$. We define three bins in $\cthgg$ :  [-1.0, -0.95],\ [-0.95,-0.80], and [-0.80,-0.50], therefore the cross section is determined at  $\cthgg$= -0.975, -0.875 and -0.650. The phase space in $\phigg$ is divided in six bins from 0$^{\circ}$ to 180$^{\circ}$, of 30$^{\circ}$ width. The statistics from $\phigg=-180^{\circ}$ to $\phigg=0^{\circ}$ are added using the symmetry property of the unpolarized cross section w.r.t. to the lepton plane,  $d \sigma (\phigg)= d \sigma (2 \pi - \phigg)$. The elementary bin size in $W$ is 20 MeV. \\
The two other variables needed to complete the kinematics are the photon virtuality $Q^2$ (constant in the first scan and variable in the second scan) and the beam energy in the lab, which is always kept fixed: $k_{lab}=4.032$ GeV. As a consequence, the virtual photon polarization 
$\epsilon=  \left[    1+2 ({\vec q}^{ \, \, 2} / Q^2) \tan^2 (\theta_e / 2 ) \right]^{-1} $
is not constant but decreases monotonically from 0.95 at $W$=1 GeV to 0.75 at $W$=1.9 GeV.  \\
Full results, including statistical and instrumental uncertainties are presented in the Tables of the Appendix. The cross-section values are statistically independent, bin-to-bin. Systematic errors on the cross section are studied in~\cite{Laveissiere:2001th}. They mainly originate from uncertainties in the absolute normalization (integrated beam charge), the radiative corrections, and the knowledge of the experimental acceptance. They are mostly correlated  bin-to-bin. Another source of systematic error is due to the physical background subtraction. It is mostly independent  bin-to-bin in $W$,  and it affects the $\gamma$ channel more than the $\pi^0$ channel (due to lower VCS statistics). As a result the total systematic error is larger in the $\gamma$ channel than in the  $\pi^0$ channel (cf. the Tables of the Appendix).

The most detailed cross section is five-fold differential. However, for relevant studies we will use a two-fold cross section. Throughout this analysis the parametrization of ref.~\cite{Brash:2001qq} is used for the proton electromagnetic form factors, namely to compute the BH+Born cross section. The next sections present our results.


\section{Results}
\label{sec:Results-Res}

\subsection{VCS Resonance Data, Scan in $W$ at $Q^2=1$ GeV$^2$}
\label{sec:result-scan1}

\begin{figure}[htb]
\includegraphics[width=8.6cm]{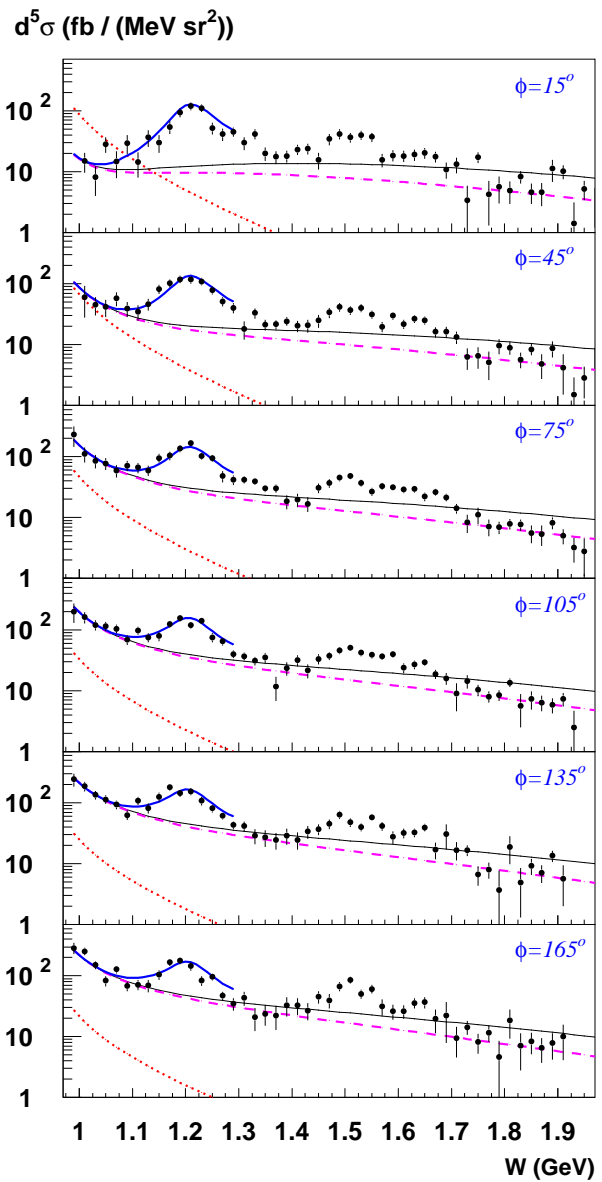}
\caption{\label{fig:excitation-curve}
 (Color online) Excitation curves for the H$(e,e'p)\gamma$ reaction at $Q^2=1$~GeV$^2$, $\cthgg =-0.975$, $k_{lab}=4.032$ GeV  and six values of $\phigg$, as marked. The thick solid curve  up through the $\Delta$-resonance is our DR fit of the Generalized Polarizabilities~\cite{Laveissiere:2004nf}.  The thin solid curve is the  BH+Born cross section, and the dashed curve is the BH+Born+$\pi^0$-exchange  cross section~\cite{Vanderhaeghen:1996iz}.  The dotted curve is the pure Bethe-Heitler cross section.}
\end{figure}

This first scan provides an overall picture of the nucleon excitation spectrum induced by the electromagnetic probe, conditioned by the $(\gamma p)$  specific de-excitation channel. Tables~\ref{tablehf:q2-1.0-ctcm1} to \ref{tablehf:q2-1.0-ctcm3}  contain the numerical values of the five-fold differential cross section $d^5\sigma (ep \to ep \gamma) $ at $Q^2=1$~GeV$^2$, for the six bins in $\phigg$ as a function of $W$,  and  $\cthgg = -0.975$, $-0.875$, and $-0.650$. In Fig.~\ref{fig:excitation-curve} we present this cross section in the most backward bin, at $\cthgg = -0.975$.

 
The strong rise that the data show towards very low $W$ and $\phigg \sim 180^{\circ}$, is due to the BH tail of elastic electron scattering. In this region there is obviously a strong interference between the BH and the VCS amplitudes, evolving from destructive at $\phigg=15^{\circ}$ to constructive at larger $\phigg$. The cross section calculated from the coherent sum of the BH and Nucleon-Born amplitudes (thin solid curve) is in excellent agreement with the data. \\

 
In the first resonance region, the thick solid curve shows the calculation based on Dispersion Relations (DR) by B.Pasquini {\it et al.\/}~\cite{Pasquini:2001yy}. In this theoretical framework, our data were previously analyzed in terms of Generalized Polarizabilities for $W<1.28$ GeV~\cite{Laveissiere:2004nf}. The DR model is able to predict the 12 independent VCS scattering amplitudes, in terms of the $\gamma^*N\rightarrow N\pi$ multipoles, $t$-channel $\pi^0$ exchange, and two phenomenological functions: $\Delta\alpha(Q^2)$ and $\Delta\beta(Q^2)$.  These two functions  parameterize the contributions to the electric and magnetic polarizabilities from high-energy virtual channels. In particular, the term $\Delta\beta(Q^2)$ is modeled by $t$-channel $\sigma$-meson exchange. In~\cite{Hyde-Wright:2004gh}, it was suggested that the combination $[\Delta\alpha+\Delta\beta](Q^2)]$ is likely dominated by the  $N\pi\pi$ and $N\eta$ multipoles, which are not presently included in the DR formalism. When a comprehensive partial wave analysis of the $\gamma^* N\rightarrow N\pi\pi$ multipoles becomes available, the DR formalism could be extended to the second resonance region ($W\approx 1.5$ GeV). A comparison with the present data would improve our understanding of the spatial distribution of the polarization response of the proton, by identifying more explicitly which channels and excitations contribute to the Generalized Polarizabilities. 
 

In  Fig.~\ref{fig:excitation-curve} we also see strong resonance phenomena in the second resonance region. The higher resonances are not distinguishable, due to a combination of the limited statistical precision and the interference of many open channels in the intermediate state of the VCS amplitude. For $W>$ 1.3 GeV, there is no complete model calculation of VCS incorporating resonances. Nevertheless, the data follow the general trend of the  BH+Born calculation at high $W$, when we include the destructive $t$-channel $\pi^0$ exchange graph~\cite{Vanderhaeghen:1996iz} of Fig.~\ref{fig:feynman}g. This is somewhat surprising, given the spectrum of baryon resonances. In the resonance model of Capstick and Keister~\cite{Capstick:1993qi} for RCS, the positive-parity intermediate states contribute constructively and the negative-parity states contribute destructively to the backward-angle cross section. Although diffractive minima can cause some amplitudes to change sign with $Q^2$,  this basic effect will remain in VCS. Thus the high-level density of resonances at large $W$ does not necessarily enhance the backward cross section, and leads to a smooth behavior. In section~\ref{sec:compar-rcs-vcs}, we explore the question of which degrees of freedom are essential for the high-energy backward Compton amplitude.

\subsection{$Q^2$-Dependence in the Region of $W=1.5$ GeV }
\label{sec:result-scan2}

A second set of data was taken in order to study the $Q^2$-dependence of the cross section
at a fixed c.m. energy. Ideally, such data provide information on the transition form factors of the  nucleon resonances. Here we have performed an exploratory scan in the second resonance region, around $W=1.53$ GeV, where the strongest excitations are  the $D_{13}(1520)$ and $S_{11}(1535)$ resonances.

This study was performed for both channels H$(e,e'p)\gamma$ and H$(e,e'p)\pi^0$. Measuring the two processes at the same kinematics allows to compare the sensitivity to the various resonances in two different exit channels.

The detailed $Q^2$-dependence of our experimental data is obtained by subdividing the spectrometer acceptance of three separate kinematic settings centered at $Q^2=0.6$, 1.0, and 2.0 GeV$^2$. Tables \ref{tab-gam-q2dep-1} to \ref{tab-piz-q2dep-2} contain the differential cross section in each elementary bin in $(Q^2,W,\phi)$. For the figures we define a two-fold cross section. To this aim we first divide $d^5 \sigma$ by the virtual photon flux factor: 
\begin{eqnarray}
{ d^2 \sigma \over d [ \Omega_p ] _{\mathrm{c.m.}} } &=&
{ d^5 \sigma \over dk'_{lab} \, d [ \Omega_e ] _{lab} \, d [ \Omega_p ] _{\mathrm{c.m.}}}
\ \times \ { \displaystyle 1 \over  \displaystyle  \Gamma} \ . 
\label{eq01}
\end{eqnarray}
The flux factor (Hand convention~\cite{Hand:1963bb}) is defined by:
\begin{eqnarray}
\begin{array}{lll}
 \Gamma  &=& 
\displaystyle
{\alpha \over 2 \pi^2 } \cdot { k'_{lab} \over k_{lab} }  \cdot 
{W^2-M_p^2 \over 2 M_p Q^2 } \cdot  {1 \over 1-\epsilon} \\
\label{eq02}
\end{array}
\end{eqnarray}
where $\alpha$ is the fine structure constant  and $M_p$ the proton mass. We then extract the $\phigg$-independent term of $d^2 \sigma / d [ \Omega_p ] _{\mathrm{c.m.}}$, which will be called reduced cross section and noted $ \langle d^2 \sigma \rangle$. Since in each small $Q^2$-bin the coverage in $\phigg$ is often not complete, this extraction is performed by fitting the experimental data to the  $\phigg$-dependence of a model. The chosen model is (BH+Born) for photon electroproduction and MAID2000 for pion electroproduction. We just fit a global scale parameter from model to experiment; then from this parameter and the model it is straightforward to determine $\langle d^2 \sigma \rangle $ in the bin. The data represented in  Figs.~\ref{fig:VCS-Q2} and \ref{fig:pi0-Q2} are given in Table~\ref{tab-q2dep-figs}.


\subsubsection{The H$(e,e'p)\gamma$ Process }
\label{sec:result-scan2gam}

\begin{figure}
\includegraphics[width=8.6cm]{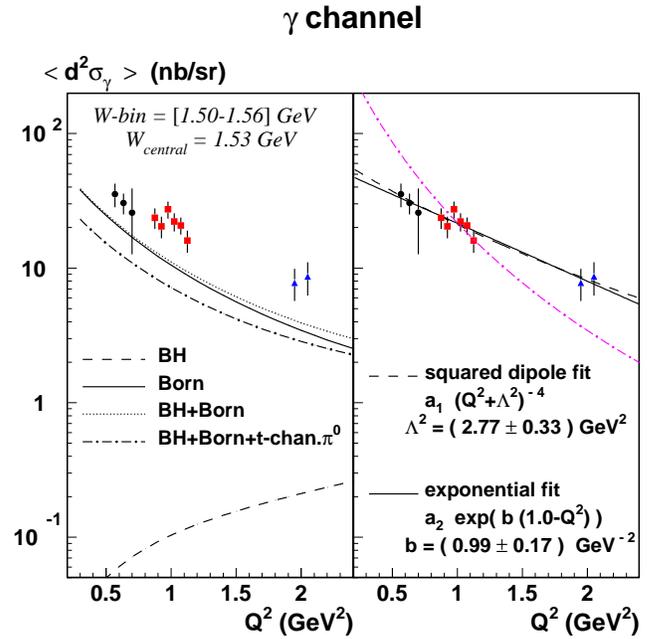}
\caption{\label{fig:VCS-Q2} 
 (Color online) The $Q^2$-dependence of the reduced cross section $ \langle d^2 \sigma_{\gamma} \rangle $ in photon electroproduction (see text), at fixed $\cthgg = -0.975$, $k_{lab}=4.032$ GeV and $W=1.53$ GeV (statistical error only). 
The three different datasets are labelled by circles, squares and triangles. Left panel: comparison with theoretical calculations at $W=1.53$ GeV. Right panel: the same experimental points with two different types of fits, having ($a_1, \Lambda^2$) or ($a_2, b$) as free parameters. The dot-dashed curve is the standard nucleon dipole squared,  $G_D^2=(1+Q^2(\mbox{GeV}^2)/0.71)^{-4}$, normalized arbitrarily to the data point at $Q^2=1$ GeV$^2$.
}
\end{figure}


If the BH process was fully negligible, the obtained cross section $\langle  d^2 \sigma_{\gamma} \rangle $  would represent the usual term \ $d^2 \sigma_T + \epsilon d^2 \sigma_L$ of the VCS subprocess $(\gamma^* p \to \gamma p )$. However, this is only approximately true. In the kinematics considered here, the modulus of the BH amplitude  still represents 6-15\%  of the modulus of the BH+Born amplitude.

In  Fig.~\ref{fig:VCS-Q2} we plot the $Q^2$-dependence of the reduced VCS cross section $ \langle  d^2 \sigma_{\gamma} \rangle $  at $W=1.53$ GeV and $\cthgg =-0.975$. A large bin width  (60 MeV) is chosen in $W$ in order to gain statistical accuracy.  The measured values are a factor 2-3 above the BH+Born calculation, which may not be surprising since the model does not include any resonance structure. The $Q^2$-dependence of the data is rather smooth. It is well reproduced in relative by the (BH+Born) or the (BH+Born + $t$-channel $\pi^0$-exchange) calculation.

The data are well fitted by a dipole or an exponential behavior, as illustrated in the right panel of Fig.~\ref{fig:VCS-Q2}. We note that the dipole mass parameter $\Lambda^2$ is much larger than for the standard nucleon dipole form factor $G_D$. Without doing a complete analysis in terms of helicity amplitudes of the resonances as in refs.~\cite{Stoler:1993yk} or~\cite{Tiator:2003uu}, it is clear from our data that the involved transition form factors have a much slower decrease with $Q^2$ than $G_D$, in the explored $Q^2$-range.  Interpretation of these data will require a systematic treatment of both the on-shell and off-shell intermediate states, entering the imaginary and real parts of the VCS amplitude, respectively. Strong contributions to the real part of the VCS amplitude are expected from resonances distant in $W$.



\subsubsection{The H$(e,e'p)\pi^0$ Process}
\label{sec:result-scan2piz}

\begin{figure}
\includegraphics[width=8.6cm]{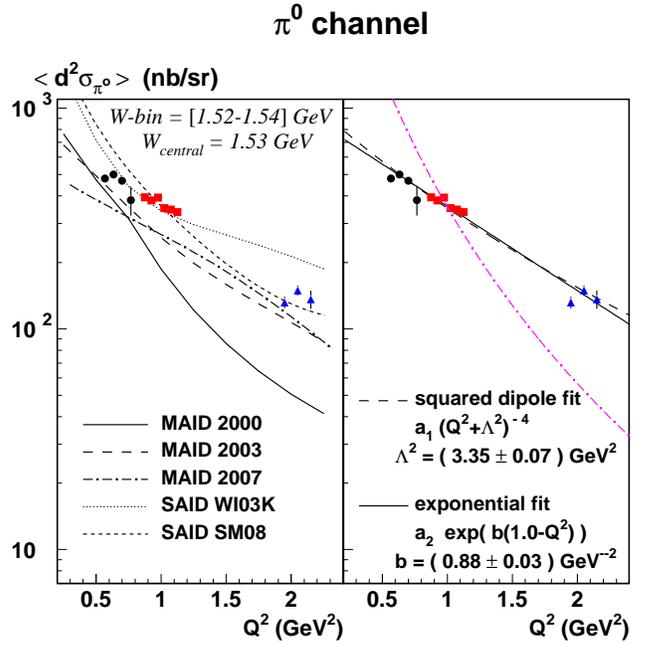}
\caption{\label{fig:pi0-Q2}  
 (Color online) The $Q^2$-dependence of the reduced cross section $ \langle d^2 \sigma_{\pi^0} \rangle $ in $\pi^0$ electroproduction, at fixed $\cthgg = -0.975$, $k_{lab}=4.032$ GeV and $W=1.53$ GeV (statistical error only). Left panel: comparison with theoretical calculations at $W=1.53$ GeV. Right panel: the same experimental points with two different types of fits. Same comments as in the previous figure.
}
\end{figure}

In the $\pi^0$ channel, the reduced cross section $\langle d^2 \sigma_{\pi^0} \rangle $ strictly corresponds to the $\phi$-independent term $d^2 \sigma_T + \epsilon d^2 \sigma _L$ of pion electroproduction. The data at the $Q^2=1$ GeV$^2$ setting were previously published in~\cite{Laveissiere:2003jf} (but without subdividing into small $Q^2$-bins).

Figure~\ref{fig:pi0-Q2} shows the $Q^2$-dependence of the reduced cross section $ \langle d^2 \sigma_{\pi^0} \rangle $ at the same kinematics as Fig.~\ref{fig:VCS-Q2}. The enhanced statistics in the $\pi^0$ channel allow us to choose a smaller bin width in $W$, of 20 MeV. The observed $Q^2$-dependence is again rather smooth. Among the various versions of the MAID unitary isobar model~\cite{Drechsel:1998hk}, the most recent ones (2003 or 2007) better reproduce the $Q^2$-dependence of the data, however they still underestimate the cross section in absolute by $\sim$20-30\%.
The SAID WI03K~\cite{Arndt:2003zk} curve is a global fit including our $Q^2=1$ GeV$^2$ data~\cite{Laveissiere:2003jf}, i.e. the points labelled by a square in Fig.~\ref{fig:pi0-Q2}. Therefore this model works well around $Q^2=1$ GeV$^2$, but gives poorer agreement with the data around  $Q^2=2$ GeV$^2$. The most recent SAID calculation SM08~\cite{Arndt:2006ym} is in  good agreement with the data for $Q^2 = $ 1 and 2 GeV$^2$.


\subsubsection{$W$-Dependence of the $Q^2$-Dependence}
\label{sec:result-scan2-wdep}

\begin{figure}
\includegraphics[width=8.6cm]{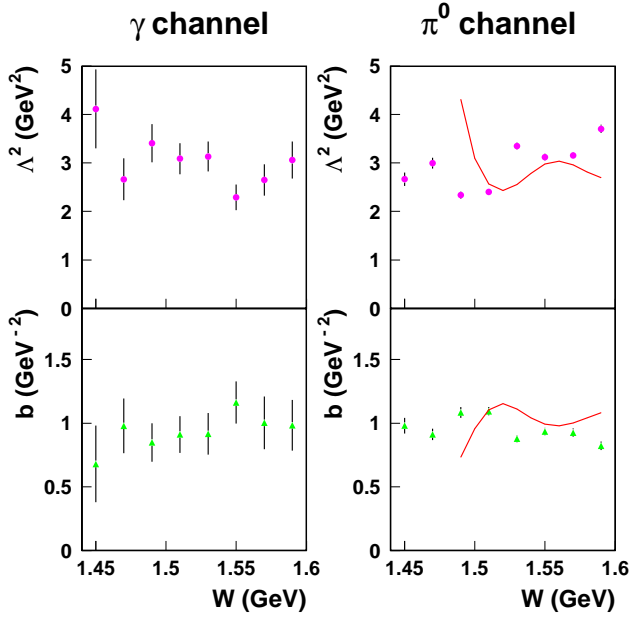}
\caption{\label{fig:bparam}
 (Color online) The slope parameters  $\Lambda^2$ and  $b$ introduced in the two previous figures, determined here in each elementary $W$-bin of width 20 MeV (statistical error only). Left and right panels are for photon and $\pi^0$ electroproduction, respectively. 
The solid curve ($\pi^0$ channel) gives the result of the MAID2003 calculation, limited to $W \ge $ 1.49 GeV. This is the only region in $W$ for which the MAID calculation is well described by a simple dipole or exponential fit in the $Q^2$-range [0.5, 2.0] GeV$^2$.
}
\end{figure}

From both inclusive and exclusive data, it is known~\cite{Stoler:1993yk} that in the second resonance region the virtual photo-absorption cross section is dominated by the $D_{13}(1520)$ resonance at low $Q^2$ ($<$1 GeV$^2$), while for $Q^2>2$  GeV$^2$ it is dominated by the $S_{11}(1535)$ resonance. Furthermore, some of the transition multipoles of these two resonances do not have simple dipole shapes in $Q^2$ \cite{Tiator:2003uu}. This should result in a complicated $Q^2$-dependence of electroproduction cross sections. However, surprisingly, the behavior observed in Figs. \ref{fig:VCS-Q2} and \ref{fig:pi0-Q2} at $W=$ 1.53 GeV can be described by a single dipole fall-off. \\
To further explore the $Q^2$-behavior of our reduced cross-sections, we have performed the same fits as in Figs.~\ref{fig:VCS-Q2} and \ref{fig:pi0-Q2}, i.e. dipolar or exponential, in each elementary $W$-bin of 20 MeV width, in the $W$-range [1.45, 1.59] GeV. The result is presented in Fig.~\ref{fig:bparam}. One first observes that the fitted parameters take globally the same value for the H$(e,e'p)\gamma$ and H$(e,e'p)\pi^0$ processes, i.e.  $\Lambda^2 \simeq 3$ GeV$^2$, or $b \simeq 0.9$ GeV$^{-2}$ everywhere. This slope value for $b$ is intermediate between the values found for the $S_{11}(1535)$ and the $D_{13}(1520)$ resonances ($b = $ 0.38 and 1.60 GeV$^{-2}$, respectively~\cite{Brasse:1984vm} (see also~\cite{Armstrong:1998wg,Denizli:2007tq})). Assuming that the $Q^2$-dependence of the  virtual photo-production of a resonance is given only by the coupling $(\gamma^* p \to $ resonance) and does not depend on the exit channel, then our $\sim$ constant $b$ suggests that approximately the ``same mixing'' of resonances is seen in the two exit channels ($\gamma p$ or $\pi^0 p$), in the explored $W$-range. 

However, at a finer scale the data of Fig.~\ref{fig:bparam} do show some variations with $W$, which appear to be non-trivial, and of opposite sign in the two exit channels $\gamma p$ and $\pi^0 p$. Such variations are also present in model calculations, e.g. MAID2003 in the figure ($\pi^0$ channel). One concludes that  the competition from multiple resonance channels results in a complicated $W$-dependence of the $Q^2$-dependence of electroproduction cross sections.

Note that the $b$ parameter of the exponential fit was determined  previously in ref.~\cite{Laveissiere:2003jf} for the $\pi^0$ channel \footnote{In ref.~\cite{Laveissiere:2003jf}, eq.(17) has a misprint. The exponential fit should read $e^{+b_{exp}(1\mbox{\tiny GeV}^2-Q^2)}$.}. This fit used our data in the limited  $Q^2$-range of [0.85, 1.15] GeV$^2$ instead of the present range [0.4, 2.2] GeV$^2$, and it turned out that the obtained $b$ values were usually smaller than the present ones. In the bin $W \in$ [1.5, 1.6] GeV, this limited fit yielded $b=0.6 \pm 0.1$ GeV$^{-2}$. The present global $Q^2$ fit in the same $W$-bin yields $b=0.93 \pm 0.02$ GeV$^{-2}$. This latter value better represents the average $Q^2$-evolution of the cross section, in the full $Q^2$-range [0.4, 2.2] GeV$^2$.

\begin{figure}[t]
\includegraphics[width=8.6cm]{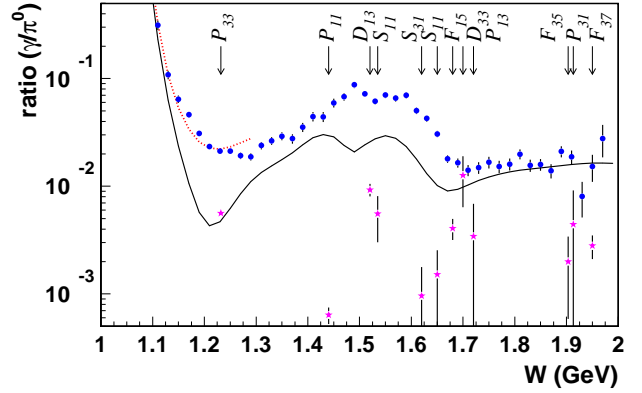}
\caption{\label{fig:ratio-gam-pi0}
 (Color online) Ratio of the reduced cross sections $ \langle d^2 \sigma  \rangle $ of the processes H$(e,e'p)\gamma$ and H$(e,e'p)\pi^0$, at $Q^2=1$~GeV$^2$,  $\cthcm = -0.975$ and $k_{lab}=4.032$ GeV (full circles; statistical error only). The full curve and the dotted curve for $W<1.3$ GeV are this ratio, at the same kinematics, calculated using different theoretical models (see text section~\ref{sec:ratio-gamma-pizero}). The stars represent the ratio $r_{N^*}$ (see text section~\ref{sec:ratio-gamma-pizero}) for the listed individual resonances, as obtained from~\cite{Yao:2006px}.
}
\end{figure}

\subsection{VCS to $\pi^0$ Ratio}
\label{sec:ratio-gamma-pizero}

From the present results and those published in~\cite{Laveissiere:2003jf}, we have determined the experimental ratio between the H$(e,e'p)\gamma$ and H$(e,e'p)\pi^0$ cross sections at $\cthgg = -0.975$ and $Q^2=1$~GeV$^2$ for the entire resonance region (see Table~\ref{tab-ratio}). In Fig.~\ref{fig:ratio-gam-pi0} we show the value of the ratio of the $\phigg$-independent cross sections,  $r = \langle d^2 \sigma_{\gamma} \rangle / \langle d^2 \sigma_{\pi^0} \rangle $. 
Two theoretical calculations of this observable are also displayed: the full curve is obtained with BH+Born+$\pi^0$-exchange for the H$(e,e'p)\gamma$ reaction (numerator) and MAID2003~\cite{Drechsel:1998hk} for the H$(e,e'p)\pi^0$ reaction (denominator); the dotted curve is obtained by changing the numerator to the DR model for VCS~\cite{Pasquini:2001yy}. This latter calculation agrees well with our data in the $\Delta$(1232) resonance region. As a reference, we have also indicated (in star symbols)  the value of the simple ratio of the branching ratios of the individual resonances~\cite{Yao:2006px} :  $r_{N^*} = BR(N^*\to p \gamma) / BR (N^* \to N \pi)$. This ratio $r_{N^*}$ is integrated over 4$\pi$ in the final state, therefore it has different dynamical sensitivity than our backward data and should not be directly compared to them. Furthermore, there are important interference effects between individual resonances, at the amplitude level, which are not considered in  $r_{N^*}$, while they are -at least partially- taken into account in the theoretical curves of  Fig.~\ref{fig:ratio-gam-pi0}.

Below the $\Delta$(1232) resonance, the large enhancement of the ratio is due to the rising (BH+B) cross section in the VCS channel. We note the large enhancement of the measured ratio also in the second resonance region ($P_{11}$, $D_{13}$, $S_{11}$). These particular resonances have large couplings to the $N\pi\pi$ and $N\eta$ channels, which contribute as virtual channels to the VCS process.  We have already noted the likely significance of these resonances for the Generalized Polarizabilities. The observed variation of the ratio $r$ with $W$ illustrates our initial motivation:  the VCS and $\gamma^* p\rightarrow \pi^0 p$ channels have very different sensitivities to the resonances.


\subsection{VCS-RCS Comparison} 
\label{sec:compar-rcs-vcs}


The RCS reaction $\gamma p \to \gamma p$  has been intensively investigated in the $\Delta(1232)$-resonance~\cite{Wissmann:1999vi} and in the high-energy diffractive region~\cite{Bauer:1978iq}. It was also studied above the $\Delta(1232)$ at Bonn~\cite{Jung:1981wm}, Saskatoon~\cite{Hallin:1993ft}, and Tokyo~\cite{Wada:1984sh,Ishii:1985ei}. The Cornell experiment~\cite{Shupe:1979vg} measured the RCS process at photon energies $E_\gamma$ in the range 2--6 GeV and angles from $45^\circ$ to $128^\circ$ in the c.m. frame.  
There are no high-energy fully backward RCS data. The recent JLab experiment E99-114 \cite{Hamilton:2004fq, Danagoulian:2007gs} measured the RCS process at $E_\gamma$ in the range 3--6 GeV.  The large-angle data at fixed $W$ are roughly independent of $\thcm$ (within a factor of two) in the range $[90^\circ,120^\circ]$ \cite{Danagoulian:2007gs}.

\begin{figure}[t]
\includegraphics[width=8.6cm]{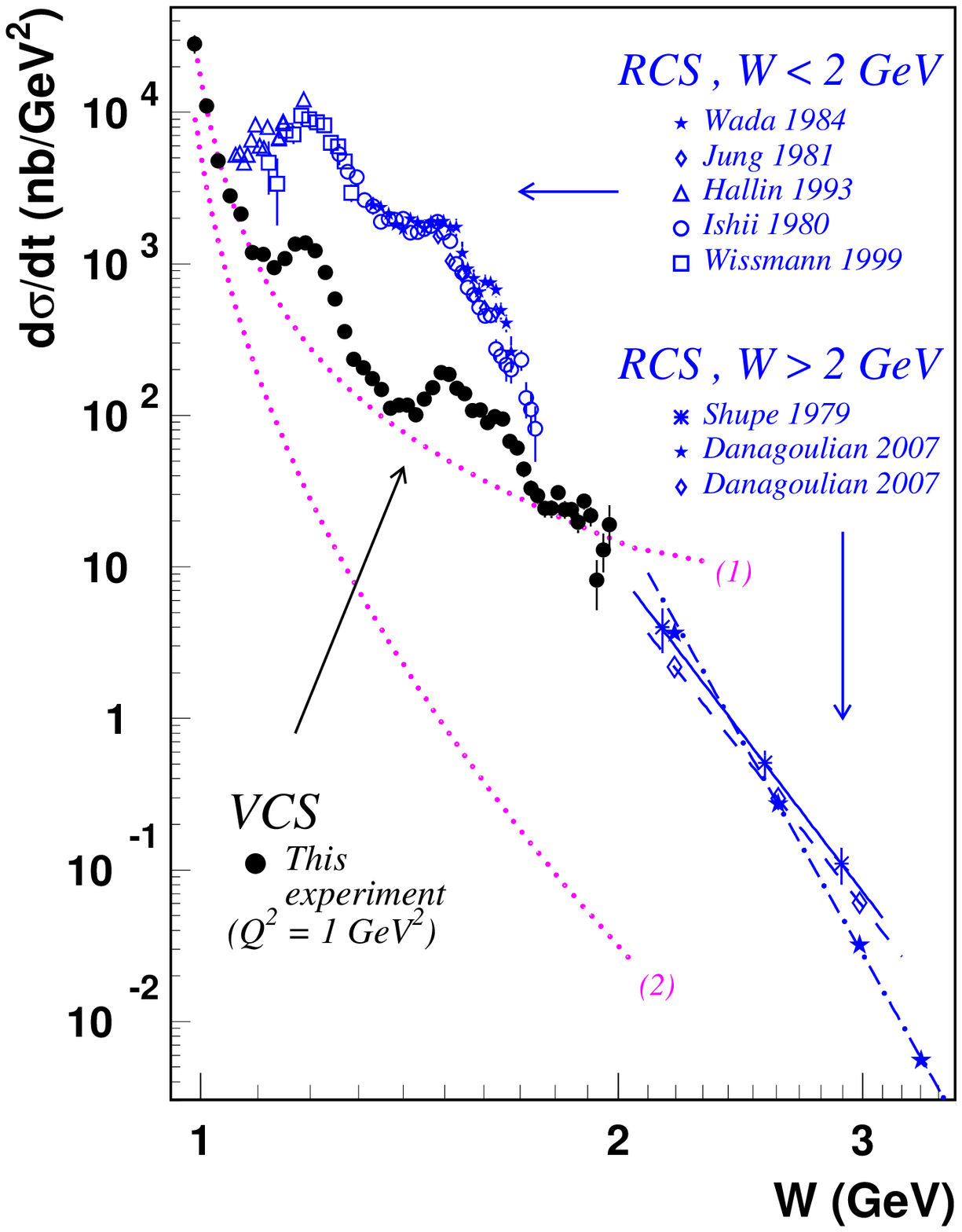}
\caption{\label{fig:rcs-vcs-dt}
(Color online) Comparison of VCS data from this experiment 
($\bullet$)
         at ($Q^2=1$ GeV$^2$, $\thgg = 167.2^\circ$) and RCS data,
    for $W < $ 2 GeV at
         $\thcm = 159-162^\circ$ ($\star$)~\cite{Wada:1984sh},
         $\thcm = 128-132^\circ$ ($\diamondsuit$)~\cite{Jung:1981wm},
         $\thcm = 141^\circ$     ($\triangle$)~\cite{Hallin:1993ft},
         $\thcm = 130-133^\circ$ ($\circ$)~\cite{Ishii:1979bq} and
         $\thcm = 131^\circ$     ($\Box$)~\cite{Wissmann:1999vi},
    and for $W > $ 2 GeV at
         $\thcm = 105-128^\circ$ ($\ast$)~\cite{Shupe:1979vg},
         $\thcm = 113^\circ$ ($\star$) and 
                 $128^\circ$ ($\diamondsuit$)~\cite{Danagoulian:2007gs}.
The dotted curves labelled (1) and (2) are the BH+Born+$\pi^0$-exchange cross 
section (see text) and the BH one, respectively. For  $W > $ 2 GeV,
         the solid curve is an $s^{-6}$ power law normalized to the
         $W=2.55$~GeV Cornell point of ref.~\cite{Shupe:1979vg},
         the dot-dashed and dashed curves are  $s^{-8.1}$ and  $s^{-5.3}$
         power laws fitted to the JLab data~\cite{Danagoulian:2007gs}
         at $\thcm = 113^\circ$ and $128^\circ$, respectively.
        }
\end{figure}

In Fig.~\ref{fig:rcs-vcs-dt} we compare our VCS data at backward angle with
existing large-angle RCS data.  For this purpose we have used the VCS reduced
cross section defined in section~\ref{sec:result-scan2},  
determined in the experimental scan at fixed $Q^2=1$ GeV$^2$ and 
$\cthgg = -0.975$, and the data have been converted in terms of $d \sigma /dt$ (see Table~\ref{tab-ratio}). In this figure, at low $W$, we see again the rapid rise in the VCS
cross section due to the coherent sum of the BH and Born amplitudes.
As illustrated previously in Fig.~\ref{fig:excitation-curve}, the VCS excitation in the
$\Delta(1232)$ region is accurately fitted by the Dispersion Relations,
including both the onshell $N\rightarrow\Delta$ transition form factors
and the Generalized Polarizabilities \cite{Laveissiere:2004nf}.  Above the
$\Delta$-resonance  we do not have an explicit model
of the VCS process.  Through the second resonance region ($W\approx 1600$ MeV)
the RCS and VCS data show on-shell $s$-channel resonances.  The VCS/RCS
comparison in this region shows a strong decrease of the cross section from 
$Q^2=0$ to $Q^2=1$  GeV$^2$,
as expected from $s$-channel resonance form factors.

The VCS/RCS comparison for $W \ge 1.8$ GeV is in marked contrast with the
behavior at lower $W$.  At high $W$ the VCS cross section intercepts the trend 
of the largest-angle RCS cross sections ($\thcm \approx 130^\circ$), around
$W=2$ GeV. Also, for $W > 2$ GeV the $W^{-2n}$-scaling of the RCS data 
has a completely different trend than the (BH+Born+$\pi^0$-exchange) VCS curve, 
which seems to form a baseline for the VCS data at lower $W$.

We briefly review the high-energy behavior of the Compton
amplitude in three kinematic domains: $-t\ll W^2$ (forward Compton scattering);
$-t\approx W^2/2$ (wide-angle Compton scattering), and the present domain
of $-t\approx W^2$ (backward Compton scattering).  
\begin{itemize}
\item
Forward  Compton scattering
at high energy and at $Q^2 =0$ (RCS)
can be described by $t$-channel Regge exchange
processes \cite{Bauer:1978iq}.  As the photon 
virtuality increases, the Regge exchange  amplitudes
are suppressed by factors of $m_V^2/(m_V^2+Q^2$) from the vector meson
poles (of mass $m_V$) in the entrance channel.  
At high $Q^2$ (but empirically
only several GeV$^2$) the forward Compton amplitude is dominated
by the perturbative, leading-twist "handbag" amplitude of
Deep Inelastic Scattering (DIS) \cite{Yao:2006px}.
Similarly, a recent QCD factorization theorem \cite{Ji:1998xh,Collins:1998be}
predicts that in the off-forward deeply virtual limit ($-t/Q^2\ll 1$),
the  DVCS $\gamma^* p\rightarrow p\gamma$ amplitude factorizes the perturbative 
$\gamma^* q \rightarrow q \gamma$ amplitude on an elementary quark in the
target, convoluted with twist-2 quark (or gluon, at low Bjorken scaling variable $x_B$) 
matrix elements called Generalized Parton Distributions (GPDs). 
Recent DVCS experiments have found evidence for the perturbative mechanism at $Q^2$-scales
of several GeV$^2$~\cite{MunozCamacho:2006hx,:2007jq,Airapetian:2001yk,:2007cz,Chekanov:2003ya};
\item
Wide-angle Compton scattering at sufficiently high energies
will be dominated by the perturbative two-gluon exchange kernel \cite{Brodsky:1975vy}. 
The presently available RCS data with $(W^2, -t, M^2-u)$ all being large 
\cite{Shupe:1979vg,Hamilton:2004fq, Danagoulian:2007gs}
are consistent with a sub-asymptotic model
based on the elementary Klein-Nishina 
(or ``$\gamma q\rightarrow q\gamma$ handbag'')
 process on a single quark,
convoluted with high-momentum configurations in the proton
\cite{Radyushkin:1998rt,Huang:2001ej,Miller:2004rc};
\item
We expect that the high-energy RCS amplitude  at $\thcm \approx \pi$ 
is dominated by $u$-channel Regge exchange. 
 This $u$-channel Regge behavior is seen, for example in the $\gamma p\rightarrow n \pi^+$ reaction in the backward direction \cite{Guidal:1997by}.
On the other hand, as $Q^2$ increases it is likely that the Regge exchange mechanism is strongly suppressed in the backward direction (just as it is for forward Compton scattering), and thus we do not expect it to be dominant in our VCS kinematics.
In RCS, only at high transverse momentum $p_T$, corresponding to both $-t$ and $M^2-u$ being large, is the perturbative mechanism expected to be dominant.  
 Inspired by the $Q^2$-scaling behavior in DIS and DVCS, we may suggest that, as $W^2$ and $Q^2$ increase, in the backward VCS cross section there is a transition to a  hard-scattering process at the quark level. Of course more VCS data at high $W$ ($\ge 2$ GeV) and backward angles would be needed in order to explore this conjecture.
\end{itemize}
A QCD description of VCS in the backward region has been proposed
\cite{Pire:2004ie}.  This is a different kind of factorization relative to 
to DVCS in the forward region.  In the forward kinematics,  
the hard subprocess is the
exchange of two partons (the handbag diagram) and the GPDs encode the hadronic
part of the amplitude.  In the backward kinematics, the hard subprocess
is the exchange of three quarks, and $N\rightarrow qqq\gamma$
transition distribution
amplitudes (TDAs) replace the GPDs.  This picture should be valid at large
enough values of $Q^2$ and $W^2$, and must be tested independently
in each channel 
({\em e.g.}
backward VCS, $p\overline{p}\rightarrow \gamma^*\gamma
\ldots$).
In particular, the matrix element of
the $\gamma^* p\rightarrow \gamma p$ scattering amplitude
is predicted to have the following asymptotics in the backward direction at
fixed $x_B = Q^2/(W^2-M^2+Q^2)$ ({\bf not} fixed $W^2$)
\cite{Lansberg:2006uh,Lansberg:2007ec}:
\begin{eqnarray}
{\mathcal M} &\sim \displaystyle \frac{\alpha_S^2(Q^2)}{Q^3} 
\end{eqnarray}
where $\alpha_S$ is the strong coupling constant.
Neglecting terms of order $M^2/W^2$, this scaling law is obtained for $Q^2/W^2$ fixed.
Following the Hand convention~\cite{Hand:1963bb} utilized in \cite{Lansberg:2007ec}, the $\gamma^* p \rightarrow \gamma p$ differential cross section
will have the following scaling:
\begin{eqnarray}
\frac{d\sigma}{dt} &= \displaystyle \frac{1}{16\pi (W^2-M^2)^2}   
\left|{\mathcal M}\right|^2
\label{eq:ScalingHand} \\
\smallskip
&\sim \displaystyle \frac{\alpha_S^4(Q^2)}{(W^2)^5}  \ \times \   
f\left(\frac{Q^2}{W^2},u\right) \ .
\label{eq:ScalingPire}
\end{eqnarray}
This asymptotic scaling law of $W^{-10}$ predicted for  backward VCS 
at fixed $(Q^2/W^2,u)$ is different from the $W^{-12}$ scaling predicted for
wide-angle RCS at fixed $-t/W^2$ \cite{Brodsky:1975vy}.
A second scaling law applies to backward electroproduction, 
that the ratio of pion electroproduction to photon electroproduction
should be $Q^2$-independent at large $W^2$ and fixed $x_B$
\cite{Lansberg:2006uh}:
\begin{eqnarray}
\frac{d\sigma(\gamma^* p \to \gamma p)}{d\sigma(\gamma^*p \to \pi^0 p)}
&\sim \left(Q^2\right)^0
\qquad {\rm for}\quad \thcm \approx \pi \ .
\end{eqnarray}

With the advent of the 12 GeV upgrade to JLab, it will be feasible to extend both RCS, VCS, and pion electroproduction  measurements to higher $W^2$ and higher $Q^2$. These data can establish empirically the scaling laws of the Compton amplitude in these new kinematic domains.


\section{Conclusion}

In summary, the JLab experiment E93-050 studied for the first time the  H$(e,e'p)\gamma$ process  in the nucleon resonance region. This experiment provides a dataset of cross sections that is unique at backward angles. 
For $W \ge 1.4$ GeV, the BH contribution to photon electroproduction is small, and the reaction is dominated by the VCS process. The $W$-dependence of the VCS cross section shows resonance phenomena, as observed in RCS.  Our data allow to compare the sensitivity to the various nucleon resonances in two different exit channels, $\gamma p$ and $\pi^0 p$, namely by studying the $Q^2$-dependence of the cross section  for the two reactions H$(e,e'p)\gamma$ and H$(e,e'p)\pi^0$. The  $\gamma$-to-$\pi^0$ ratio shows strong variations with $W$ across the resonance region. At our highest $W$ (1.8-1.9 GeV) the comparison with wide-angle RCS may suggest that the VCS process undergoes a transition to a hard-scattering mechanism at the quark level. Therefore the data presented in this paper emphasize the interest of exploring a new kinematic domain of exclusive electroproduction reactions (high $W$, high $Q^2$, backward angles) in which new conjectures involving the fundamental degrees of freedom of QCD could be tested. \\

\begin{acknowledgments}
We would like to thank B.Pire for discussions, and B.Pasquini and M.Vanderhaeghen for providing their codes (regarding Dispersion Relations and radiative corrections, respectively).

We wish to acknowledge essential work of the JLab accelerator staff
and Hall A technical staff.
This work was supported by DOE contract DE-AC05-84ER40150 under
which the Southeastern Universities Research Association (SURA)
operated the Thomas Jefferson National Accelerator Facility. We acknowledge additional grants from the US DOE and NSF, the French
Centre National de la Recherche Scientifique and Commissariat \`a
l'Energie Atomique, the Conseil R\'egional d'Auvergne, the
FWO-Flanders (Belgium) and the BOF-Gent University.
\end{acknowledgments}

\bibliography{common}

\appendix
\section{Cross Section Tables} \label{app-1}

This Appendix lists in detail the experimental cross section  corresponding to the different studies presented in the paper. All cross sections are determined at a fixed incoming electron energy of 4.032 GeV. Ascii files of the tables are available at URL: \newline
http://clrwww.in2p3.fr/sondem/E93050-tables-RES/ \newline
or upon request to the authors.
Due to the choice of method~\cite{Janssens:2006}, the cross section is determined at well-defined points in phase space. These kinematic values have no error, and the uncertainty is entirely reported on the cross section. Error bars are given as r.m.s.

Tables \ref{tablehf:q2-1.0-ctcm1},  \ref{tablehf:q2-1.0-ctcm2},  \ref{tablehf:q2-1.0-ctcm3}  correspond to the study of section~\ref{sec:result-scan1}.  They contain the H$(e,e'p)\gamma$ five-fold differential cross section \ $ d^5\sigma / ( dk'_{lab} \, d[\Omega_e]_{lab} \, d[\Omega_p]_{\mathrm{c.m.}})$  at fixed $Q^2=1$ GeV$^2$, for $\cthgg= -0.975, -0.875$ and $-0.650$ respectively, and six values of $\phigg$. A bin is empty if the number of events is smaller than 5,  or the systematic error very large ($ >  1.5 \times d^5 \sigma$).

Tables \ref{tab-gam-q2dep-1} to \ref{tab-piz-q2dep-2}  correspond to the study of section~\ref{sec:result-scan2}.  They give the measured cross section in the most elementary bins, covering the $Q^2$-range [0.43, 2.15] GeV$^2$, the $W$-range [1.45,  1.59] GeV, at fixed $\cthgg$ or $\cthcm= -0.975$ and for six bins in $\phigg$.
Tables \ref{tab-gam-q2dep-1} and \ref{tab-gam-q2dep-2} give the  five-fold cross section \ $ d^5\sigma / ( dk'_{lab} \, d[\Omega_e]_{lab} \, d[\Omega_p]_{\mathrm{c.m.}})$ \  for the H$(e,e'p)\gamma$ process, while tables  \ref{tab-piz-q2dep-1} and \ref{tab-piz-q2dep-2} give the two-fold cross section \ $ d^2 \sigma / d [ \Omega_p ] _{\mathrm{c.m.}} $ \  of Eq.~\ref{eq01}  for the H$(e,e'p)\pi^0$  process. A bin is empty if the number of events is smaller than 2.

Table~\ref{tab-q2dep-figs} contains the data  depicted in Figs.~\ref{fig:VCS-Q2} and \ref{fig:pi0-Q2}. As explained in the text, for the  H$(e,e'p)\gamma$ process this cross section $ \langle d^2 \sigma_{\gamma} \rangle $ is obtained from the raw data of Tables \ref{tab-gam-q2dep-1} and \ref{tab-gam-q2dep-2}, by dividing by the virtual photon flux, performing a (model-based) $\phi$-analysis, keeping only the $\phi$-independent term and then grouping three elementary bins in $W$ (at 1.51, 1.53 and 1.55 GeV). For the  H$(e,e'p)\pi^0$ process this cross section $ \langle d^2 \sigma_{\pi^0} \rangle $ is obtained from the raw data of Table \ref{tab-piz-q2dep-2} at $W=1.53$ GeV,  by performing only the $\phi$-analysis step.

Table~\ref{tab-ratio} gives the ratio \ $r = \langle d^2 \sigma_{\gamma} \rangle / \langle d^2 \sigma_{\pi^0} \rangle $ \ depicted in Fig.~\ref{fig:ratio-gam-pi0}. This table also provides the values of the reduced cross section in the photon electroproduction channel corresponding to Fig.~\ref{fig:rcs-vcs-dt}. Note that in VCS, the conversion from \ $d^2 \sigma / d [ \Omega_p ] _{\mathrm{c.m.}}$ \ to \ $d \sigma /dt$  \  is the following:
\begin{eqnarray*}
\begin{array}{lll} 
 \displaystyle
{  d^2 \sigma \over d [ \Omega_p ] _{\mathrm{c.m.}} } \ = \ {  d \sigma \over dt}  \cdot J 
\ \ \ \ \ \ \  , \ \ \ \ \mbox{with:} \\
 \displaystyle J = 
{ 1 \over 2 \pi} \cdot
{ (s-M_p^2) \cdot {\big [} 4 Q^2 s + (s-M_p^2-Q^2)^2 {\big ]} ^{{1 \over 2}}
\over 2 s} \\
\end{array}
\end{eqnarray*}
where $s$ and $t$ are the Mandelstam variables defined in
section~\ref{sec:Kinematics}.




\begingroup\squeezetable
\begin{table*}\begin{ruledtabular}
\caption{\label{tablehf:q2-1.0-ctcm1}
H$(e,e'p)\gamma$ cross section $ d^5\sigma / ( dk'_{lab} \, d[\Omega_e]_{lab} \, d[\Omega_p]_{\mathrm{c.m.}})$ ($\pm$ statistical $\pm$ systematic error) at 
$Q^2$=1.0~GeV$^2$ and $\cthgg = -0.975$, in fb/(MeV $\cdot$ sr$^2$). 
}
\begin{tabular}{|c|rrrrr|rrrrr|rrrrr|rrrrr|rrrrr|rrrrr|}
$W$ & \multicolumn{5}{c|}{$\phi=15^\circ$}
&\multicolumn{5}{c|}{$\phi=45^\circ$}
&\multicolumn{5}{c|}{$\phi=75^\circ$}
&\multicolumn{5}{c|}{$\phi=105^\circ$}
&\multicolumn{5}{c|}{$\phi=135^\circ$}
&\multicolumn{5}{c|}{$\phi=165^\circ$} \\  (GeV) & &  & &  & &  & &  & &  & &  & & & & & & & & & & & & & & & &
 & & \\ \hline
 0.99   & \ & \ & \  & \  & \    & \ & \ & \  & \  & \   &     231 $\:\:$&$\pm$&      84 $\:\:$&$\pm$&     101 $\:\:$   &     200 $\:\:$&$\pm$&      70 $\:\:$&$\pm$&      77 $\:\:$   &     243 $\:\:$&$\pm$&      59 $\:\:$&$\pm$&     108 $\:\:$   &     288 $\:\:$&$\pm$&      59 $\:\:$&$\pm$&     104 $\:\:$   \\ 
 1.01  &   14.9 &$\pm$&    5.4 &$\pm$&    5.6  &      60 $\:\:$&$\pm$&      32 $\:\:$&$\pm$&      36 $\:\:$   &     111 $\:\:$&$\pm$&      30 $\:\:$&$\pm$&      20 $\:\:$   &     161 $\:\:$&$\pm$&      33 $\:\:$&$\pm$&      30 $\:\:$   &     188 $\:\:$&$\pm$&      34 $\:\:$&$\pm$&      64 $\:\:$   &     254 $\:\:$&$\pm$&      36 $\:\:$&$\pm$&      45 $\:\:$   \\ 
 1.03  &    8.1 &$\pm$&    4.1 &$\pm$&    4.3  &      45 $\:\:$&$\pm$&      14 $\:\:$&$\pm$&       6 $\:\:$   &      85 $\:\:$&$\pm$&      21 $\:\:$&$\pm$&      14 $\:\:$   &     119 $\:\:$&$\pm$&      22 $\:\:$&$\pm$&      24 $\:\:$   &     134 $\:\:$&$\pm$&      22 $\:\:$&$\pm$&      20 $\:\:$   &     152 $\:\:$&$\pm$&      22 $\:\:$&$\pm$&      33 $\:\:$   \\ 
 1.05  &   28.4 &$\pm$&    8.3 &$\pm$&    4.7  &      41 $\:\:$&$\pm$&      13 $\:\:$&$\pm$&      15 $\:\:$   &      77 $\:\:$&$\pm$&      17 $\:\:$&$\pm$&      23 $\:\:$   &     114 $\:\:$&$\pm$&      19 $\:\:$&$\pm$&      23 $\:\:$   &     114 $\:\:$&$\pm$&      19 $\:\:$&$\pm$&      15 $\:\:$   &      83 $\:\:$&$\pm$&      15 $\:\:$&$\pm$&      62 $\:\:$   \\ 
 1.07  &   14.6 &$\pm$&    6.8 &$\pm$&    4.0  &      57 $\:\:$&$\pm$&      14 $\:\:$&$\pm$&       9 $\:\:$   &      59 $\:\:$&$\pm$&      13 $\:\:$&$\pm$&      13 $\:\:$   &     104 $\:\:$&$\pm$&      17 $\:\:$&$\pm$&      18 $\:\:$   &      94 $\:\:$&$\pm$&      15 $\:\:$&$\pm$&      25 $\:\:$   &     128 $\:\:$&$\pm$&      18 $\:\:$&$\pm$&      24 $\:\:$   \\ 
 1.09  &      29 $\:\:$&$\pm$&      10 $\:\:$&$\pm$&       4 $\:\:$   &      39 $\:\:$&$\pm$&      11 $\:\:$&$\pm$&      10 $\:\:$   &      71 $\:\:$&$\pm$&      14 $\:\:$&$\pm$&       9 $\:\:$   &      68 $\:\:$&$\pm$&      12 $\:\:$&$\pm$&      10 $\:\:$   &      63 $\:\:$&$\pm$&      11 $\:\:$&$\pm$&      12 $\:\:$   &      67 $\:\:$&$\pm$&      12 $\:\:$&$\pm$&      24 $\:\:$   \\ 
 1.11  &   14.4 &$\pm$&    6.4 &$\pm$&    5.0  &      34 $\:\:$&$\pm$&      10 $\:\:$&$\pm$&       2 $\:\:$   &      66 $\:\:$&$\pm$&      12 $\:\:$&$\pm$&       9 $\:\:$   &      97 $\:\:$&$\pm$&      14 $\:\:$&$\pm$&      11 $\:\:$   &     109 $\:\:$&$\pm$&      16 $\:\:$&$\pm$&       9 $\:\:$   &      71 $\:\:$&$\pm$&      13 $\:\:$&$\pm$&      18 $\:\:$   \\ 
 1.13  &      36 $\:\:$&$\pm$&      11 $\:\:$&$\pm$&       7 $\:\:$   &      46 $\:\:$&$\pm$&      10 $\:\:$&$\pm$&       7 $\:\:$   &      59 $\:\:$&$\pm$&      11 $\:\:$&$\pm$&      15 $\:\:$   &      75 $\:\:$&$\pm$&      13 $\:\:$&$\pm$&       8 $\:\:$   &      82 $\:\:$&$\pm$&      15 $\:\:$&$\pm$&      17 $\:\:$   &      69 $\:\:$&$\pm$&      15 $\:\:$&$\pm$&       3 $\:\:$   \\ 
 1.15  &      30 $\:\:$&$\pm$&       9 $\:\:$&$\pm$&      15 $\:\:$   &      82 $\:\:$&$\pm$&      14 $\:\:$&$\pm$&      15 $\:\:$   &      95 $\:\:$&$\pm$&      15 $\:\:$&$\pm$&       9 $\:\:$   &      79 $\:\:$&$\pm$&      14 $\:\:$&$\pm$&      14 $\:\:$   &     125 $\:\:$&$\pm$&      20 $\:\:$&$\pm$&      43 $\:\:$   &     105 $\:\:$&$\pm$&      18 $\:\:$&$\pm$&      27 $\:\:$   \\ 
 1.17  &      54 $\:\:$&$\pm$&      12 $\:\:$&$\pm$&       9 $\:\:$   &     103 $\:\:$&$\pm$&      16 $\:\:$&$\pm$&       8 $\:\:$   &     104 $\:\:$&$\pm$&      17 $\:\:$&$\pm$&      12 $\:\:$   &     124 $\:\:$&$\pm$&      17 $\:\:$&$\pm$&       9 $\:\:$   &     180 $\:\:$&$\pm$&      22 $\:\:$&$\pm$&      24 $\:\:$   &     168 $\:\:$&$\pm$&      22 $\:\:$&$\pm$&      10 $\:\:$   \\ 
 1.19  &      94 $\:\:$&$\pm$&      16 $\:\:$&$\pm$&       7 $\:\:$   &     118 $\:\:$&$\pm$&      17 $\:\:$&$\pm$&      13 $\:\:$   &     135 $\:\:$&$\pm$&      17 $\:\:$&$\pm$&      13 $\:\:$   &     154 $\:\:$&$\pm$&      18 $\:\:$&$\pm$&      23 $\:\:$   &     145 $\:\:$&$\pm$&      19 $\:\:$&$\pm$&      14 $\:\:$   &     178 $\:\:$&$\pm$&      24 $\:\:$&$\pm$&      15 $\:\:$   \\ 
 1.21  &     119 $\:\:$&$\pm$&      18 $\:\:$&$\pm$&      17 $\:\:$   &     117 $\:\:$&$\pm$&      17 $\:\:$&$\pm$&      20 $\:\:$   &     167 $\:\:$&$\pm$&      18 $\:\:$&$\pm$&       9 $\:\:$   &     119 $\:\:$&$\pm$&      15 $\:\:$&$\pm$&      10 $\:\:$   &     153 $\:\:$&$\pm$&      21 $\:\:$&$\pm$&       6 $\:\:$   &     144 $\:\:$&$\pm$&      24 $\:\:$&$\pm$&      20 $\:\:$   \\ 
 1.23  &     111 $\:\:$&$\pm$&      16 $\:\:$&$\pm$&      13 $\:\:$   &     108 $\:\:$&$\pm$&      15 $\:\:$&$\pm$&      18 $\:\:$   &     102 $\:\:$&$\pm$&      13 $\:\:$&$\pm$&      11 $\:\:$   &     141 $\:\:$&$\pm$&      16 $\:\:$&$\pm$&       8 $\:\:$   &     107 $\:\:$&$\pm$&      17 $\:\:$&$\pm$&      18 $\:\:$   &      83 $\:\:$&$\pm$&      15 $\:\:$&$\pm$&      14 $\:\:$   \\ 
 1.25  &      51 $\:\:$&$\pm$&      11 $\:\:$&$\pm$&      12 $\:\:$   &      78 $\:\:$&$\pm$&      12 $\:\:$&$\pm$&      11 $\:\:$   &      94 $\:\:$&$\pm$&      12 $\:\:$&$\pm$&      11 $\:\:$   &      74 $\:\:$&$\pm$&      13 $\:\:$&$\pm$&      10 $\:\:$   &      81 $\:\:$&$\pm$&      11 $\:\:$&$\pm$&       6 $\:\:$   &      96 $\:\:$&$\pm$&      12 $\:\:$&$\pm$&      10 $\:\:$   \\ 
 1.27  &   41.3 &$\pm$&    9.2 &$\pm$&    7.1  &   51.5 &$\pm$&    9.3 &$\pm$&    6.3  &   48.1 &$\pm$&    9.8 &$\pm$&    6.9  &   64.4 &$\pm$&    9.5 &$\pm$&    4.1  &   61.4 &$\pm$&    8.3 &$\pm$&    4.8  &   47.1 &$\pm$&    8.0 &$\pm$&    5.9  \\ 
 1.29  &   45.2 &$\pm$&    7.9 &$\pm$&    4.5  &   40.3 &$\pm$&    7.6 &$\pm$&    7.1  &   42.1 &$\pm$&    7.6 &$\pm$&    3.6  &   40.1 &$\pm$&    5.5 &$\pm$&    3.9  &   43.0 &$\pm$&    6.4 &$\pm$&    3.6  &   35.0 &$\pm$&    8.2 &$\pm$&    5.0  \\ 
 1.31  &      29 $\:\:$&$\pm$&       7 $\:\:$&$\pm$&      15 $\:\:$   &   18.0 &$\pm$&    5.9 &$\pm$&    5.4  &   42.3 &$\pm$&    6.0 &$\pm$&    2.4  &   37.0 &$\pm$&    5.0 &$\pm$&    3.1  &   41.4 &$\pm$&    7.7 &$\pm$&    3.5  &      43 $\:\:$&$\pm$&      11 $\:\:$&$\pm$&       7 $\:\:$   \\ 
 1.33  &      41 $\:\:$&$\pm$&       8 $\:\:$&$\pm$&      10 $\:\:$   &   33.0 &$\pm$&    6.1 &$\pm$&    3.1  &   39.7 &$\pm$&    4.8 &$\pm$&    3.2  &   31.1 &$\pm$&    5.1 &$\pm$&    3.4  &   28.8 &$\pm$&    8.0 &$\pm$&    6.5  &   20.9 &$\pm$&    8.6 &$\pm$&    1.4  \\ 
 1.35  &   19.9 &$\pm$&    4.8 &$\pm$&    4.4  &   21.2 &$\pm$&    4.2 &$\pm$&    4.4  &   30.2 &$\pm$&    4.0 &$\pm$&    2.1  &   35.1 &$\pm$&    6.6 &$\pm$&    6.7  &   27.1 &$\pm$&    8.3 &$\pm$&    8.0  &   23.9 &$\pm$&    8.8 &$\pm$&    5.4  \\ 
 1.37  &   17.5 &$\pm$&    4.1 &$\pm$&    4.3  &   21.6 &$\pm$&    3.7 &$\pm$&    2.1  &   30.5 &$\pm$&    4.6 &$\pm$&    2.4  &   11.8 &$\pm$&    5.0 &$\pm$&    9.5  &   24.4 &$\pm$&    7.3 &$\pm$&    2.1  &   22.5 &$\pm$&    9.6 &$\pm$&    8.2  \\ 
 1.39  &   18.0 &$\pm$&    3.8 &$\pm$&    3.1  &   24.2 &$\pm$&    3.9 &$\pm$&    4.0  &   18.5 &$\pm$&    4.7 &$\pm$&    4.4  &   23.5 &$\pm$&    5.8 &$\pm$&    4.8  &   29.1 &$\pm$&    8.5 &$\pm$&    6.8  &      33 $\:\:$&$\pm$&      11 $\:\:$&$\pm$&       3 $\:\:$   \\ 
 1.41  &   23.0 &$\pm$&    4.3 &$\pm$&    1.3  &   20.3 &$\pm$&    4.2 &$\pm$&    2.0  &   19.8 &$\pm$&    5.1 &$\pm$&    4.8  &   31.8 &$\pm$&    6.2 &$\pm$&    5.1  &   24.7 &$\pm$&    7.8 &$\pm$&    3.6  &      32 $\:\:$&$\pm$&      10 $\:\:$&$\pm$&       4 $\:\:$   \\ 
 1.43  &   23.9 &$\pm$&    4.8 &$\pm$&    2.3  &   20.8 &$\pm$&    5.0 &$\pm$&    5.3  &   16.8 &$\pm$&    4.5 &$\pm$&    3.0  &   21.6 &$\pm$&    5.6 &$\pm$&    2.4  &   33.7 &$\pm$&    8.0 &$\pm$&    4.7  &   27.0 &$\pm$&    8.5 &$\pm$&    5.5  \\ 
 1.45  &   15.6 &$\pm$&    4.9 &$\pm$&    4.1  &   24.8 &$\pm$&    6.1 &$\pm$&    3.2  &   30.8 &$\pm$&    5.8 &$\pm$&    2.2  &   33.4 &$\pm$&    6.6 &$\pm$&    4.5  &   36.7 &$\pm$&    7.3 &$\pm$&    5.1  &      45 $\:\:$&$\pm$&      10 $\:\:$&$\pm$&       6 $\:\:$   \\ 
 1.47  &   34.3 &$\pm$&    7.4 &$\pm$&    2.9  &   33.8 &$\pm$&    6.5 &$\pm$&    3.6  &   37.0 &$\pm$&    6.2 &$\pm$&    2.3  &   37.5 &$\pm$&    6.0 &$\pm$&    4.9  &   44.8 &$\pm$&    8.0 &$\pm$&    6.0  &      39 $\:\:$&$\pm$&      10 $\:\:$&$\pm$&       4 $\:\:$   \\ 
 1.49  &   41.5 &$\pm$&    7.8 &$\pm$&    4.0  &   41.5 &$\pm$&    6.8 &$\pm$&    4.2  &   45.5 &$\pm$&    6.3 &$\pm$&    8.8  &   45.4 &$\pm$&    5.9 &$\pm$&    7.0  &      64 $\:\:$&$\pm$&       9 $\:\:$&$\pm$&      10 $\:\:$   &      67 $\:\:$&$\pm$&      11 $\:\:$&$\pm$&      10 $\:\:$   \\ 
 1.51  &   36.5 &$\pm$&    6.7 &$\pm$&    4.6  &   36.6 &$\pm$&    6.1 &$\pm$&    4.0  &   48.6 &$\pm$&    5.6 &$\pm$&    5.9  &   50.8 &$\pm$&    6.1 &$\pm$&    5.8  &   47.6 &$\pm$&    7.4 &$\pm$&    5.2  &      85 $\:\:$&$\pm$&      11 $\:\:$&$\pm$&      11 $\:\:$   \\ 
 1.53  &   39.8 &$\pm$&    6.7 &$\pm$&    4.7  &   39.9 &$\pm$&    6.0 &$\pm$&    5.4  &   37.0 &$\pm$&    4.9 &$\pm$&    4.6  &   42.4 &$\pm$&    5.6 &$\pm$&    4.5  &   39.9 &$\pm$&    6.2 &$\pm$&    3.7  &   50.4 &$\pm$&    8.7 &$\pm$&    6.3  \\ 
 1.55  &   37.1 &$\pm$&    6.3 &$\pm$&    2.5  &   31.4 &$\pm$&    4.8 &$\pm$&    3.9  &   26.8 &$\pm$&    4.0 &$\pm$&    4.6  &   39.0 &$\pm$&    4.9 &$\pm$&    3.6  &   57.8 &$\pm$&    7.0 &$\pm$&    5.2  &      61 $\:\:$&$\pm$&      10 $\:\:$&$\pm$&       7 $\:\:$   \\ 
 1.57  &   15.6 &$\pm$&    3.9 &$\pm$&    3.8  &   19.7 &$\pm$&    3.8 &$\pm$&    3.5  &   32.7 &$\pm$&    4.1 &$\pm$&    2.8  &   36.8 &$\pm$&    4.3 &$\pm$&    3.2  &   41.3 &$\pm$&    6.9 &$\pm$&    7.3  &      31 $\:\:$&$\pm$&       9 $\:\:$&$\pm$&      10 $\:\:$   \\ 
 1.59  &   18.5 &$\pm$&    3.9 &$\pm$&    1.3  &   30.0 &$\pm$&    4.2 &$\pm$&    3.1  &   31.5 &$\pm$&    3.8 &$\pm$&    3.5  &   40.1 &$\pm$&    4.7 &$\pm$&    4.7  &   27.8 &$\pm$&    6.4 &$\pm$&    5.0  &   26.0 &$\pm$&    7.2 &$\pm$&    8.9  \\ 
 1.61  &   18.1 &$\pm$&    3.7 &$\pm$&    3.4  &   21.7 &$\pm$&    3.7 &$\pm$&    3.5  &   28.9 &$\pm$&    3.5 &$\pm$&    3.7  &   23.8 &$\pm$&    4.5 &$\pm$&    3.5  &   31.8 &$\pm$&    5.9 &$\pm$&    2.4  &   26.2 &$\pm$&    6.4 &$\pm$&    4.0  \\ 
 1.63  &   19.0 &$\pm$&    4.0 &$\pm$&    2.2  &   26.4 &$\pm$&    3.8 &$\pm$&    1.8  &   29.4 &$\pm$&    3.8 &$\pm$&    2.6  &   27.0 &$\pm$&    4.4 &$\pm$&    4.3  &   32.6 &$\pm$&    5.2 &$\pm$&    4.2  &   35.8 &$\pm$&    6.7 &$\pm$&    2.7  \\ 
 1.65  &   20.4 &$\pm$&    4.1 &$\pm$&    3.3  &   25.1 &$\pm$&    3.7 &$\pm$&    2.2  &   22.3 &$\pm$&    3.6 &$\pm$&    3.8  &   29.2 &$\pm$&    3.8 &$\pm$&    2.4  &   39.5 &$\pm$&    5.6 &$\pm$&    5.0  &   36.8 &$\pm$&    7.5 &$\pm$&    6.8  \\ 
 1.67  &   17.7 &$\pm$&    3.6 &$\pm$&    3.4  &   16.2 &$\pm$&    3.2 &$\pm$&    1.6  &   26.1 &$\pm$&    3.5 &$\pm$&    2.2  &   18.8 &$\pm$&    3.1 &$\pm$&    2.4  &   17.0 &$\pm$&    5.3 &$\pm$&    4.7  &   19.7 &$\pm$&    8.3 &$\pm$&    3.2  \\ 
 1.69  &   10.8 &$\pm$&    3.2 &$\pm$&    2.6  &   16.2 &$\pm$&    3.4 &$\pm$&    3.6  &   21.5 &$\pm$&    3.0 &$\pm$&    4.5  &   16.0 &$\pm$&    3.6 &$\pm$&    2.9  &      31 $\:\:$&$\pm$&      13 $\:\:$&$\pm$&      26 $\:\:$   &      22 $\:\:$&$\pm$&      14 $\:\:$&$\pm$&       9 $\:\:$   \\ 
 1.71  &   13.1 &$\pm$&    3.7 &$\pm$&    3.8  &   13.3 &$\pm$&    3.1 &$\pm$&    1.5  &   14.1 &$\pm$&    2.6 &$\pm$&    3.5  &    8.9 &$\pm$&    4.3 &$\pm$&    1.9  &   16.8 &$\pm$&    5.6 &$\pm$&    2.3  &    9.5 &$\pm$&    5.0 &$\pm$&    2.5  \\ 
 1.73  &    3.4 &$\pm$&    2.4 &$\pm$&    3.0  &    6.2 &$\pm$&    2.4 &$\pm$&    4.5  &    8.4 &$\pm$&    2.8 &$\pm$&    2.0  &   14.4 &$\pm$&    3.6 &$\pm$&    2.0  &   16.7 &$\pm$&    3.4 &$\pm$&    3.4  &   14.3 &$\pm$&    3.7 &$\pm$&    2.3  \\ 
 1.75  &   17.2 &$\pm$&    3.5 &$\pm$&    2.0  &    6.5 &$\pm$&    2.5 &$\pm$&    1.9  &   11.1 &$\pm$&    3.3 &$\pm$&    1.3  &   10.3 &$\pm$&    2.2 &$\pm$&    2.6  &    6.7 &$\pm$&    2.4 &$\pm$&    1.3  &    8.2 &$\pm$&    3.0 &$\pm$&    1.4  \\ 
 1.77  &    4.2 &$\pm$&    2.9 &$\pm$&    5.4  &    5.1 &$\pm$&    2.5 &$\pm$&    0.9  &    7.1 &$\pm$&    2.2 &$\pm$&    1.7  &    8.0 &$\pm$&    1.7 &$\pm$&    1.0  &    8.0 &$\pm$&    2.4 &$\pm$&    2.3  &   11.6 &$\pm$&    3.4 &$\pm$&    1.6  \\ 
 1.79  &    5.6 &$\pm$&    2.6 &$\pm$&    3.3  &    9.6 &$\pm$&    2.6 &$\pm$&    1.8  &    7.0 &$\pm$&    1.6 &$\pm$&    1.8  &    8.4 &$\pm$&    1.7 &$\pm$&    0.8  &    3.7 &$\pm$&    4.0 &$\pm$&    3.5  &    4.6 &$\pm$&    4.0 &$\pm$&    3.5  \\ 
 1.81  &    4.9 &$\pm$&    2.0 &$\pm$&    2.0  &    8.8 &$\pm$&    2.0 &$\pm$&    0.9  &    7.9 &$\pm$&    1.5 &$\pm$&    0.8  &   13.6 &$\pm$&    2.2 &$\pm$&    3.1  &      19 $\:\:$&$\pm$&       9 $\:\:$&$\pm$&      10 $\:\:$   &   18.6 &$\pm$&    9.3 &$\pm$&    9.8  \\ 
 1.83  &    8.2 &$\pm$&    1.9 &$\pm$&    1.1  &    5.7 &$\pm$&    1.6 &$\pm$&    1.1  &    7.7 &$\pm$&    1.6 &$\pm$&    0.9  &    5.6 &$\pm$&    3.1 &$\pm$&    5.1  &    4.9 &$\pm$&    3.6 &$\pm$&    2.1  &    7.1 &$\pm$&    4.3 &$\pm$&    2.8  \\ 
 1.85  &    4.6 &$\pm$&    1.7 &$\pm$&    0.9  &    8.4 &$\pm$&    1.8 &$\pm$&    1.5  &    5.6 &$\pm$&    1.8 &$\pm$&    1.7  &    7.3 &$\pm$&    2.4 &$\pm$&    1.1  &    9.2 &$\pm$&    2.7 &$\pm$&    1.0  &    8.3 &$\pm$&    3.2 &$\pm$&    2.2  \\ 
 1.87  &    4.6 &$\pm$&    1.9 &$\pm$&    1.2  &    4.8 &$\pm$&    2.1 &$\pm$&    1.2  &    5.4 &$\pm$&    2.0 &$\pm$&    1.6  &    6.4 &$\pm$&    1.8 &$\pm$&    0.8  &    7.1 &$\pm$&    2.3 &$\pm$&    2.3  &    6.6 &$\pm$&    2.9 &$\pm$&    1.4  \\ 
 1.89  &      11 $\:\:$&$\pm$&       4 $\:\:$&$\pm$&      11 $\:\:$   &    8.7 &$\pm$&    2.6 &$\pm$&    2.2  &    8.2 &$\pm$&    1.8 &$\pm$&    0.9  &    5.8 &$\pm$&    1.6 &$\pm$&    1.0  &   13.5 &$\pm$&    2.9 &$\pm$&    2.8  &    7.8 &$\pm$&    3.6 &$\pm$&    1.9  \\ 
 1.91  &   10.1 &$\pm$&    2.6 &$\pm$&    1.9  &    4.2 &$\pm$&    2.7 &$\pm$&    1.8  &    5.0 &$\pm$&    1.4 &$\pm$&    1.2  &    7.3 &$\pm$&    1.8 &$\pm$&    1.8  &    5.7 &$\pm$&    3.7 &$\pm$&    1.8  &   10.0 &$\pm$&    5.9 &$\pm$&    3.9  \\ 
 1.93  &    1.4 &$\pm$&    1.7 &$\pm$&    1.9  &    1.5 &$\pm$&    1.4 &$\pm$&    1.8  &    3.2 &$\pm$&    1.4 &$\pm$&    1.2  &    2.5 &$\pm$&    2.2 &$\pm$&    1.8   & \ & \ & \  & \  & \    & \ & \ & \  & \  & \   \\ 
 1.95  &    5.2 &$\pm$&    1.8 &$\pm$&    1.9  &    2.8 &$\pm$&    1.5 &$\pm$&    1.8  &    2.8 &$\pm$&    1.7 &$\pm$&    1.2   & \ & \ & \  & \  & \    & \ & \ & \  & \  & \    & \ & \ & \  & \  & \   \\ 
 1.97  &    3.3 &$\pm$&    2.1 &$\pm$&    1.9  &    6.4 &$\pm$&    2.4 &$\pm$&    1.8   & \ & \ & \  & \  & \    & \ & \ & \  & \  & \    & \ & \ & \  & \  & \    & \ & \ & \  & \  & \   \\ 
\end{tabular}\end{ruledtabular}\end{table*}
\endgroup


\begingroup\squeezetable
\begin{table*}\begin{ruledtabular}
\caption{\label{tablehf:q2-1.0-ctcm2}
H$(e,e'p)\gamma$ cross section  $ d^5\sigma / ( dk'_{lab} \, d[\Omega_e]_{lab} \, d[\Omega_p]_{\mathrm{c.m.}})$ ($\pm$ statistical $\pm$ systematic error) at 
$Q^2$=1.0~GeV$^2$ and $\cos\theta_{\gamma\gamma}^\ast = -0.875$, in  fb/(MeV $\cdot$ sr$^2$).
}
\begin{tabular}{|c|rrrrr|rrrrr|rrrrr|rrrrr|rrrrr|rrrrr|}
$W$ & \multicolumn{5}{c|}{$\phi=15^\circ$}
&\multicolumn{5}{c|}{$\phi=45^\circ$}
&\multicolumn{5}{c|}{$\phi=75^\circ$}
&\multicolumn{5}{c|}{$\phi=105^\circ$}
&\multicolumn{5}{c|}{$\phi=135^\circ$}
&\multicolumn{5}{c|}{$\phi=165^\circ$} \\ 
 (GeV) & &  & &  & &  & &  & &  & &  & & & & & & & & & & & & & & & & & & \\ 
\hline
 0.99   & \ & \ & \  & \  & \   &     548 $\:\:$&$\pm$&     306 $\:\:$&$\pm$&     112 $\:\:$   &     362 $\:\:$&$\pm$&     159 $\:\:$&$\pm$&      75 $\:\:$   &     459 $\:\:$&$\pm$&     107 $\:\:$&$\pm$&      63 $\:\:$   &     453 $\:\:$&$\pm$&      78 $\:\:$&$\pm$&      46 $\:\:$   &     478 $\:\:$&$\pm$&      72 $\:\:$&$\pm$&      39 $\:\:$   \\ 
 1.01   & \ & \ & \  & \  & \   &     230 $\:\:$&$\pm$&      94 $\:\:$&$\pm$&      94 $\:\:$   &     320 $\:\:$&$\pm$&      75 $\:\:$&$\pm$&      54 $\:\:$   &     342 $\:\:$&$\pm$&      54 $\:\:$&$\pm$&      30 $\:\:$   &     361 $\:\:$&$\pm$&      47 $\:\:$&$\pm$&      42 $\:\:$   &     291 $\:\:$&$\pm$&      37 $\:\:$&$\pm$&      28 $\:\:$   \\ 
 1.03  &      65 $\:\:$&$\pm$&      27 $\:\:$&$\pm$&      14 $\:\:$   &     186 $\:\:$&$\pm$&      60 $\:\:$&$\pm$&      26 $\:\:$   &     263 $\:\:$&$\pm$&      48 $\:\:$&$\pm$&      33 $\:\:$   &     224 $\:\:$&$\pm$&      34 $\:\:$&$\pm$&      17 $\:\:$   &     280 $\:\:$&$\pm$&      34 $\:\:$&$\pm$&      20 $\:\:$   &     283 $\:\:$&$\pm$&      32 $\:\:$&$\pm$&      22 $\:\:$   \\ 
 1.05   & \ & \ & \  & \  & \   &     144 $\:\:$&$\pm$&      43 $\:\:$&$\pm$&      13 $\:\:$   &     204 $\:\:$&$\pm$&      35 $\:\:$&$\pm$&      23 $\:\:$   &     197 $\:\:$&$\pm$&      28 $\:\:$&$\pm$&      14 $\:\:$   &     144 $\:\:$&$\pm$&      22 $\:\:$&$\pm$&      18 $\:\:$   &     141 $\:\:$&$\pm$&      21 $\:\:$&$\pm$&      11 $\:\:$   \\ 
 1.07   & \ & \ & \  & \  & \   &     124 $\:\:$&$\pm$&      34 $\:\:$&$\pm$&      18 $\:\:$   &      87 $\:\:$&$\pm$&      19 $\:\:$&$\pm$&       9 $\:\:$   &     154 $\:\:$&$\pm$&      22 $\:\:$&$\pm$&      12 $\:\:$   &     134 $\:\:$&$\pm$&      19 $\:\:$&$\pm$&       6 $\:\:$   &     157 $\:\:$&$\pm$&      20 $\:\:$&$\pm$&      14 $\:\:$   \\ 
 1.09   & \ & \ & \  & \  & \   &     115 $\:\:$&$\pm$&      29 $\:\:$&$\pm$&      10 $\:\:$   &     143 $\:\:$&$\pm$&      23 $\:\:$&$\pm$&       9 $\:\:$   &      96 $\:\:$&$\pm$&      16 $\:\:$&$\pm$&      40 $\:\:$   &     124 $\:\:$&$\pm$&      17 $\:\:$&$\pm$&      14 $\:\:$   &     121 $\:\:$&$\pm$&      19 $\:\:$&$\pm$&      11 $\:\:$   \\ 
 1.11  &      30 $\:\:$&$\pm$&      15 $\:\:$&$\pm$&       7 $\:\:$   &      69 $\:\:$&$\pm$&      20 $\:\:$&$\pm$&       4 $\:\:$   &     106 $\:\:$&$\pm$&      18 $\:\:$&$\pm$&       8 $\:\:$   &     134 $\:\:$&$\pm$&      18 $\:\:$&$\pm$&      10 $\:\:$   &     129 $\:\:$&$\pm$&      21 $\:\:$&$\pm$&      13 $\:\:$   &      81 $\:\:$&$\pm$&      20 $\:\:$&$\pm$&      45 $\:\:$   \\ 
 1.13  &      30 $\:\:$&$\pm$&      14 $\:\:$&$\pm$&       5 $\:\:$   &      85 $\:\:$&$\pm$&      20 $\:\:$&$\pm$&       8 $\:\:$   &      82 $\:\:$&$\pm$&      14 $\:\:$&$\pm$&       6 $\:\:$   &     114 $\:\:$&$\pm$&      17 $\:\:$&$\pm$&       4 $\:\:$   &      78 $\:\:$&$\pm$&      19 $\:\:$&$\pm$&      16 $\:\:$   &     105 $\:\:$&$\pm$&      30 $\:\:$&$\pm$&      11 $\:\:$   \\ 
 1.15  &      59 $\:\:$&$\pm$&      20 $\:\:$&$\pm$&       3 $\:\:$   &      45 $\:\:$&$\pm$&      13 $\:\:$&$\pm$&      12 $\:\:$   &     133 $\:\:$&$\pm$&      19 $\:\:$&$\pm$&       9 $\:\:$   &      99 $\:\:$&$\pm$&      18 $\:\:$&$\pm$&      18 $\:\:$   &      98 $\:\:$&$\pm$&      22 $\:\:$&$\pm$&      13 $\:\:$   &     116 $\:\:$&$\pm$&      25 $\:\:$&$\pm$&      10 $\:\:$   \\ 
 1.17  &      31 $\:\:$&$\pm$&      14 $\:\:$&$\pm$&       8 $\:\:$   &      43 $\:\:$&$\pm$&      14 $\:\:$&$\pm$&      15 $\:\:$   &     110 $\:\:$&$\pm$&      17 $\:\:$&$\pm$&       5 $\:\:$   &     107 $\:\:$&$\pm$&      18 $\:\:$&$\pm$&      11 $\:\:$   &     118 $\:\:$&$\pm$&      22 $\:\:$&$\pm$&      14 $\:\:$   &     106 $\:\:$&$\pm$&      30 $\:\:$&$\pm$&      14 $\:\:$   \\ 
 1.19  &     130 $\:\:$&$\pm$&      28 $\:\:$&$\pm$&      26 $\:\:$   &     118 $\:\:$&$\pm$&      18 $\:\:$&$\pm$&       8 $\:\:$   &     123 $\:\:$&$\pm$&      17 $\:\:$&$\pm$&      11 $\:\:$   &     126 $\:\:$&$\pm$&      18 $\:\:$&$\pm$&      12 $\:\:$   &     145 $\:\:$&$\pm$&      29 $\:\:$&$\pm$&      18 $\:\:$   &     272 $\:\:$&$\pm$&     122 $\:\:$&$\pm$&      90 $\:\:$   \\ 
 1.21  &     103 $\:\:$&$\pm$&      21 $\:\:$&$\pm$&      11 $\:\:$   &     124 $\:\:$&$\pm$&      19 $\:\:$&$\pm$&       7 $\:\:$   &     135 $\:\:$&$\pm$&      17 $\:\:$&$\pm$&       5 $\:\:$   &     183 $\:\:$&$\pm$&      21 $\:\:$&$\pm$&      11 $\:\:$   &     177 $\:\:$&$\pm$&      51 $\:\:$&$\pm$&      18 $\:\:$   &     217 $\:\:$&$\pm$&      70 $\:\:$&$\pm$&      46 $\:\:$   \\ 
 1.23  &     129 $\:\:$&$\pm$&      21 $\:\:$&$\pm$&       8 $\:\:$   &     117 $\:\:$&$\pm$&      18 $\:\:$&$\pm$&       6 $\:\:$   &     135 $\:\:$&$\pm$&      16 $\:\:$&$\pm$&       9 $\:\:$   &     133 $\:\:$&$\pm$&      20 $\:\:$&$\pm$&      12 $\:\:$   &     130 $\:\:$&$\pm$&      24 $\:\:$&$\pm$&      42 $\:\:$   &      68 $\:\:$&$\pm$&      19 $\:\:$&$\pm$&      28 $\:\:$   \\ 
 1.25  &      99 $\:\:$&$\pm$&      17 $\:\:$&$\pm$&       6 $\:\:$   &      95 $\:\:$&$\pm$&      17 $\:\:$&$\pm$&       6 $\:\:$   &      99 $\:\:$&$\pm$&      13 $\:\:$&$\pm$&       7 $\:\:$   &      68 $\:\:$&$\pm$&      14 $\:\:$&$\pm$&       8 $\:\:$   &      88 $\:\:$&$\pm$&      14 $\:\:$&$\pm$&      14 $\:\:$   &      81 $\:\:$&$\pm$&      21 $\:\:$&$\pm$&       8 $\:\:$   \\ 
 1.27  &      54 $\:\:$&$\pm$&      13 $\:\:$&$\pm$&       7 $\:\:$   &      61 $\:\:$&$\pm$&      13 $\:\:$&$\pm$&      14 $\:\:$   &      87 $\:\:$&$\pm$&      13 $\:\:$&$\pm$&       5 $\:\:$   &      71 $\:\:$&$\pm$&      11 $\:\:$&$\pm$&       6 $\:\:$   &      58 $\:\:$&$\pm$&      12 $\:\:$&$\pm$&       7 $\:\:$    & \ & \ & \  & \  & \   \\ 
 1.29  &      35 $\:\:$&$\pm$&      11 $\:\:$&$\pm$&       5 $\:\:$   &      41 $\:\:$&$\pm$&      10 $\:\:$&$\pm$&       9 $\:\:$   &      47 $\:\:$&$\pm$&      10 $\:\:$&$\pm$&       4 $\:\:$   &   70.0 &$\pm$&    9.4 &$\pm$&    4.7  &      58 $\:\:$&$\pm$&      19 $\:\:$&$\pm$&       7 $\:\:$    & \ & \ & \  & \  & \   \\ 
 1.31  &      29 $\:\:$&$\pm$&      10 $\:\:$&$\pm$&       5 $\:\:$   &      51 $\:\:$&$\pm$&      10 $\:\:$&$\pm$&       5 $\:\:$   &   49.2 &$\pm$&    9.0 &$\pm$&    3.1  &   58.3 &$\pm$&    8.8 &$\pm$&    3.7   & \ & \ & \  & \  & \    & \ & \ & \  & \  & \   \\ 
 1.33  &   29.3 &$\pm$&    8.5 &$\pm$&    2.5  &      44 $\:\:$&$\pm$&      10 $\:\:$&$\pm$&       4 $\:\:$   &   35.0 &$\pm$&    6.6 &$\pm$&    4.3  &   39.8 &$\pm$&    8.8 &$\pm$&    2.7   & \ & \ & \  & \  & \    & \ & \ & \  & \  & \   \\ 
 1.35  &   10.4 &$\pm$&    5.0 &$\pm$&    7.7  &   28.9 &$\pm$&    9.7 &$\pm$&    4.6  &   31.2 &$\pm$&    5.4 &$\pm$&    3.3  &      43 $\:\:$&$\pm$&      11 $\:\:$&$\pm$&      10 $\:\:$   &      52 $\:\:$&$\pm$&      19 $\:\:$&$\pm$&      22 $\:\:$    & \ & \ & \  & \  & \   \\ 
 1.37  &   24.0 &$\pm$&    9.9 &$\pm$&    5.9  &      57 $\:\:$&$\pm$&      12 $\:\:$&$\pm$&       4 $\:\:$   &   45.0 &$\pm$&    6.3 &$\pm$&    4.1  &      38 $\:\:$&$\pm$&      12 $\:\:$&$\pm$&       4 $\:\:$   &      24 $\:\:$&$\pm$&      18 $\:\:$&$\pm$&       7 $\:\:$    & \ & \ & \  & \  & \   \\ 
 1.39   & \ & \ & \  & \  & \   &   42.1 &$\pm$&    7.5 &$\pm$&    7.4  &   35.9 &$\pm$&    6.2 &$\pm$&    2.3  &      31 $\:\:$&$\pm$&      10 $\:\:$&$\pm$&       6 $\:\:$    & \ & \ & \  & \  & \    & \ & \ & \  & \  & \   \\ 
 1.41  &      12 $\:\:$&$\pm$&      10 $\:\:$&$\pm$&      13 $\:\:$   &   21.3 &$\pm$&    4.8 &$\pm$&    7.8  &   30.8 &$\pm$&    6.9 &$\pm$&    1.8  &   18.0 &$\pm$&    9.1 &$\pm$&    4.1  &      69 $\:\:$&$\pm$&      37 $\:\:$&$\pm$&       8 $\:\:$    & \ & \ & \  & \  & \   \\ 
 1.43  &   28.6 &$\pm$&    8.0 &$\pm$&    6.1  &   32.1 &$\pm$&    5.7 &$\pm$&    4.2  &   27.6 &$\pm$&    8.0 &$\pm$&    6.6  &      25 $\:\:$&$\pm$&      11 $\:\:$&$\pm$&       6 $\:\:$   &      48 $\:\:$&$\pm$&      21 $\:\:$&$\pm$&      12 $\:\:$    & \ & \ & \  & \  & \   \\ 
 1.45  &   27.8 &$\pm$&    6.5 &$\pm$&    4.0  &   36.9 &$\pm$&    7.3 &$\pm$&    8.8  &   31.3 &$\pm$&    8.6 &$\pm$&    5.5  &      11 $\:\:$&$\pm$&       8 $\:\:$&$\pm$&      10 $\:\:$    & \ & \ & \  & \  & \    & \ & \ & \  & \  & \   \\ 
 1.47   & \ & \ & \  & \  & \   &      44 $\:\:$&$\pm$&      11 $\:\:$&$\pm$&       8 $\:\:$   &      25 $\:\:$&$\pm$&       7 $\:\:$&$\pm$&      10 $\:\:$   &      49 $\:\:$&$\pm$&      12 $\:\:$&$\pm$&       9 $\:\:$    & \ & \ & \  & \  & \    & \ & \ & \  & \  & \   \\ 
 1.49   & \ & \ & \  & \  & \   &      47 $\:\:$&$\pm$&      14 $\:\:$&$\pm$&      14 $\:\:$   &      46 $\:\:$&$\pm$&      10 $\:\:$&$\pm$&       6 $\:\:$   &      70 $\:\:$&$\pm$&      14 $\:\:$&$\pm$&       8 $\:\:$    & \ & \ & \  & \  & \    & \ & \ & \  & \  & \   \\ 
 1.51   & \ & \ & \  & \  & \   &      38 $\:\:$&$\pm$&      10 $\:\:$&$\pm$&       6 $\:\:$   &      57 $\:\:$&$\pm$&      10 $\:\:$&$\pm$&       5 $\:\:$   &      50 $\:\:$&$\pm$&      12 $\:\:$&$\pm$&      16 $\:\:$    & \ & \ & \  & \  & \    & \ & \ & \  & \  & \   \\ 
 1.53   & \ & \ & \  & \  & \   &      40 $\:\:$&$\pm$&      11 $\:\:$&$\pm$&       6 $\:\:$   &      53 $\:\:$&$\pm$&      10 $\:\:$&$\pm$&       3 $\:\:$   &      37 $\:\:$&$\pm$&      10 $\:\:$&$\pm$&      12 $\:\:$    & \ & \ & \  & \  & \    & \ & \ & \  & \  & \   \\ 
 1.55   & \ & \ & \  & \  & \   &      30 $\:\:$&$\pm$&      13 $\:\:$&$\pm$&      16 $\:\:$   &   45.2 &$\pm$&    8.4 &$\pm$&    2.6  &      42 $\:\:$&$\pm$&      10 $\:\:$&$\pm$&       6 $\:\:$    & \ & \ & \  & \  & \    & \ & \ & \  & \  & \   \\ 
 1.57   & \ & \ & \  & \  & \    & \ & \ & \  & \  & \   &   36.7 &$\pm$&    7.8 &$\pm$&    6.8  &      59 $\:\:$&$\pm$&      12 $\:\:$&$\pm$&       5 $\:\:$    & \ & \ & \  & \  & \    & \ & \ & \  & \  & \   \\ 
 1.59   & \ & \ & \  & \  & \   &   15.5 &$\pm$&    7.3 &$\pm$&    9.9  &   22.1 &$\pm$&    6.7 &$\pm$&    3.6  &      37 $\:\:$&$\pm$&      12 $\:\:$&$\pm$&       4 $\:\:$    & \ & \ & \  & \  & \    & \ & \ & \  & \  & \   \\ 
 1.61   & \ & \ & \  & \  & \   &   12.8 &$\pm$&    6.7 &$\pm$&    6.5  &   28.4 &$\pm$&    7.4 &$\pm$&    3.3  &      30 $\:\:$&$\pm$&      11 $\:\:$&$\pm$&       3 $\:\:$    & \ & \ & \  & \  & \    & \ & \ & \  & \  & \   \\ 
 1.63   & \ & \ & \  & \  & \    & \ & \ & \  & \  & \   &   47.3 &$\pm$&    9.1 &$\pm$&    4.3  &      32 $\:\:$&$\pm$&      12 $\:\:$&$\pm$&       6 $\:\:$    & \ & \ & \  & \  & \    & \ & \ & \  & \  & \   \\ 
 1.65   & \ & \ & \  & \  & \    & \ & \ & \  & \  & \   &   48.9 &$\pm$&    9.7 &$\pm$&    6.0  &      27 $\:\:$&$\pm$&      14 $\:\:$&$\pm$&       5 $\:\:$    & \ & \ & \  & \  & \    & \ & \ & \  & \  & \   \\ 
 1.67   & \ & \ & \  & \  & \    & \ & \ & \  & \  & \   &   18.1 &$\pm$&    7.4 &$\pm$&    7.1  &      47 $\:\:$&$\pm$&      13 $\:\:$&$\pm$&      15 $\:\:$    & \ & \ & \  & \  & \    & \ & \ & \  & \  & \   \\ 
 1.69   & \ & \ & \  & \  & \    & \ & \ & \  & \  & \   &   12.6 &$\pm$&    9.1 &$\pm$&    9.4   & \ & \ & \  & \  & \    & \ & \ & \  & \  & \    & \ & \ & \  & \  & \   \\ 
\end{tabular}\end{ruledtabular}\end{table*}
\endgroup


\begingroup\squeezetable
\begin{table*}\begin{ruledtabular}
\caption{\label{tablehf:q2-1.0-ctcm3}
H$(e,e'p)\gamma$ cross section $ d^5\sigma / ( dk'_{lab} \, d[\Omega_e]_{lab} \, d[\Omega_p]_{\mathrm{c.m.}})$ ($\pm$ statistical $\pm$ systematic error) at 
$Q^2$=1.0~GeV$^2$ and $\cos\theta_{\gamma\gamma}^\ast = -0.650$, in  fb/(MeV $\cdot$ sr$^2$).
}
\begin{tabular}{|c|rrrrr|rrrrr|rrrrr|rrrrr|rrrrr|rrrrr|}
$W$ & \multicolumn{5}{c|}{$\phi=15^\circ$}
&\multicolumn{5}{c|}{$\phi=45^\circ$}
&\multicolumn{5}{c|}{$\phi=75^\circ$}
&\multicolumn{5}{c|}{$\phi=105^\circ$}
&\multicolumn{5}{c|}{$\phi=135^\circ$}
&\multicolumn{5}{c|}{$\phi=165^\circ$} \\ 
 (GeV) & &  & &  & &  & &  & &  & &  & & & & & & & & & & & & & & & & & & \\ \hline
 0.99   & \ & \ & \  & \  & \    & \ & \ & \  & \  & \   &     554 &$\pm$&     216 &$\pm$&      88    &     452 &$\pm$&      90 &$\pm$&      38    &     454 &$\pm$&      64 &$\pm$&      23    &     340 &$\pm$&      49 &$\pm$&      26    \\ 
 1.01  &    4710 &$\pm$&    1720 &$\pm$&     565    &    1400 &$\pm$&     552 &$\pm$&     135    &     380 &$\pm$&      87 &$\pm$&      71    &     286 &$\pm$&      46 &$\pm$&      17    &     213 &$\pm$&      30 &$\pm$&      16    &     274 &$\pm$&      31 &$\pm$&      17    \\ 
 1.03  &    2760 &$\pm$&    1140 &$\pm$&     430    &     237 &$\pm$&     109 &$\pm$&      55    &     270 &$\pm$&      50 &$\pm$&      18    &     195 &$\pm$&      28 &$\pm$&      16    &     234 &$\pm$&      27 &$\pm$&      12    &     205 &$\pm$&      25 &$\pm$&      11    \\ 
 1.05   & \ & \ & \  & \  & \   &     277 &$\pm$&      94 &$\pm$&      63    &     234 &$\pm$&      41 &$\pm$&      12    &     211 &$\pm$&      27 &$\pm$&      16    &     167 &$\pm$&      22 &$\pm$&      14    &     173 &$\pm$&      22 &$\pm$&       9    \\ 
 1.07   & \ & \ & \  & \  & \   &      65 &$\pm$&      36 &$\pm$&      19    &     183 &$\pm$&      29 &$\pm$&       8    &     161 &$\pm$&      21 &$\pm$&      10    &     140 &$\pm$&      18 &$\pm$&      11    &     142 &$\pm$&      19 &$\pm$&      20    \\ 
 1.09   & \ & \ & \  & \  & \   &     266 &$\pm$&      59 &$\pm$&      12    &     153 &$\pm$&      24 &$\pm$&       9    &     169 &$\pm$&      20 &$\pm$&       9    &     139 &$\pm$&      19 &$\pm$&      10    &      97 &$\pm$&      21 &$\pm$&     101    \\ 
 1.11   & \ & \ & \  & \  & \   &     189 &$\pm$&      44 &$\pm$&      18    &     139 &$\pm$&      21 &$\pm$&       5    &     108 &$\pm$&      15 &$\pm$&       7    &     152 &$\pm$&      26 &$\pm$&      41     & \ & \ & \  & \  & \   \\ 
 1.13  &     462 &$\pm$&     181 &$\pm$&     270    &     195 &$\pm$&      41 &$\pm$&       9    &     113 &$\pm$&      17 &$\pm$&       5    &     123 &$\pm$&      18 &$\pm$&       7     & \ & \ & \  & \  & \    & \ & \ & \  & \  & \   \\ 
 1.15  &     131 &$\pm$&      69 &$\pm$&      97    &      92 &$\pm$&      26 &$\pm$&       9    &     102 &$\pm$&      15 &$\pm$&       5    &     125 &$\pm$&      20 &$\pm$&      11    &     128 &$\pm$&      27 &$\pm$&     169     & \ & \ & \  & \  & \   \\ 
 1.17  &     134 &$\pm$&      57 &$\pm$&      65    &     132 &$\pm$&      27 &$\pm$&      13    &      95 &$\pm$&      15 &$\pm$&      10    &     135 &$\pm$&      21 &$\pm$&      15     & \ & \ & \  & \  & \    & \ & \ & \  & \  & \   \\ 
 1.19  &     196 &$\pm$&      52 &$\pm$&       3    &     107 &$\pm$&      22 &$\pm$&      24    &     168 &$\pm$&      20 &$\pm$&       5    &     152 &$\pm$&      22 &$\pm$&      27     & \ & \ & \  & \  & \    & \ & \ & \  & \  & \   \\ 
 1.21   & \ & \ & \  & \  & \   &     154 &$\pm$&      25 &$\pm$&      22    &     153 &$\pm$&      18 &$\pm$&      14    &     125 &$\pm$&      21 &$\pm$&      17     & \ & \ & \  & \  & \    & \ & \ & \  & \  & \   \\ 
 1.23   & \ & \ & \  & \  & \   &     104 &$\pm$&      19 &$\pm$&      16    &     132 &$\pm$&      17 &$\pm$&      10    &      87 &$\pm$&      21 &$\pm$&      10     & \ & \ & \  & \  & \    & \ & \ & \  & \  & \   \\ 
 1.25   & \ & \ & \  & \  & \   &     108 &$\pm$&      19 &$\pm$&      12    &      76 &$\pm$&      13 &$\pm$&      12    &      50 &$\pm$&      15 &$\pm$&      59     & \ & \ & \  & \  & \    & \ & \ & \  & \  & \   \\ 
 1.27   & \ & \ & \  & \  & \   &     107 &$\pm$&      20 &$\pm$&      11    &     103 &$\pm$&      15 &$\pm$&      15    &      85 &$\pm$&      16 &$\pm$&      73     & \ & \ & \  & \  & \    & \ & \ & \  & \  & \   \\ 
 1.29   & \ & \ & \  & \  & \   &      51 &$\pm$&      17 &$\pm$&       6    &      79 &$\pm$&      15 &$\pm$&      15    &      43 &$\pm$&      10 &$\pm$&      12     & \ & \ & \  & \  & \    & \ & \ & \  & \  & \   \\ 
 1.31   & \ & \ & \  & \  & \   &      77 &$\pm$&      19 &$\pm$&      23    &      69 &$\pm$&      13 &$\pm$&      14    &      68 &$\pm$&      14 &$\pm$&      32     & \ & \ & \  & \  & \    & \ & \ & \  & \  & \   \\ 
 1.33   & \ & \ & \  & \  & \   &      60 &$\pm$&      16 &$\pm$&      16    &      42 &$\pm$&      10 &$\pm$&      15    &      54 &$\pm$&      15 &$\pm$&      42     & \ & \ & \  & \  & \    & \ & \ & \  & \  & \   \\ 
 1.35   & \ & \ & \  & \  & \   &      62 &$\pm$&      16 &$\pm$&       4    &      65 &$\pm$&      11 &$\pm$&       3     & \ & \ & \  & \  & \    & \ & \ & \  & \  & \    & \ & \ & \  & \  & \   \\ 
 1.37   & \ & \ & \  & \  & \   &      59 &$\pm$&      18 &$\pm$&      11    &      29 &$\pm$&       7 &$\pm$&      16     & \ & \ & \  & \  & \    & \ & \ & \  & \  & \    & \ & \ & \  & \  & \   \\ 
 1.39   & \ & \ & \  & \  & \    & \ & \ & \  & \  & \   &      43 &$\pm$&       8 &$\pm$&      25     & \ & \ & \  & \  & \    & \ & \ & \  & \  & \    & \ & \ & \  & \  & \   \\ 
 1.41   & \ & \ & \  & \  & \    & \ & \ & \  & \  & \   &      45 &$\pm$&      10 &$\pm$&      23     & \ & \ & \  & \  & \    & \ & \ & \  & \  & \    & \ & \ & \  & \  & \   \\ 
 1.43   & \ & \ & \  & \  & \    & \ & \ & \  & \  & \   &      36 &$\pm$&      11 &$\pm$&      18     & \ & \ & \  & \  & \    & \ & \ & \  & \  & \    & \ & \ & \  & \  & \   \\ 
\end{tabular}\end{ruledtabular}\end{table*}
\endgroup



\begingroup\squeezetable
\begin{table*}\begin{ruledtabular}
\caption{\label{tab-gam-q2dep-1}
H$(e,e'p)\gamma$ cross section $ d^5\sigma / ( dk'_{lab} \, d[\Omega_e]_{lab} \, d[\Omega_p]_{\mathrm{c.m.}})$ ($\pm$ statistical error) at $\cos\theta_{\gamma\gamma}^\ast = -0.975$ and $W=$ 1.45, 1.47, 1.49, 1.51 GeV, in  fb/(MeV $\cdot$ sr$^2$). The systematic error is globally $\pm$ 12\% on each point. 
}
\begin{tabular}{|c|c|rrr|rrr|rrr|rrr|rrr|rrr|}
$W$ & $Q^2$
&\multicolumn{3}{c|}{$\phi=15^\circ$}
&\multicolumn{3}{c|}{$\phi=45^\circ$}
&\multicolumn{3}{c|}{$\phi=75^\circ$}
&\multicolumn{3}{c|}{$\phi=105^\circ$}
&\multicolumn{3}{c|}{$\phi=135^\circ$}
&\multicolumn{3}{c|}{$\phi=165^\circ$} \\ 
(GeV) & (GeV$^2$) &    & & &     & & &     & & &     & & &     & & &     & & \\
\hline
  1.45 &   0.700   & \ & \ & \     & \ & \ & \     & \ & \ & \     & \ & \ & \     & \ & \ & \    &     153 $\:\:$&$\pm$&      61 $\:\:$   \\ 
  1.45 &   0.875  &      58 $\:\:$&$\pm$&      47 $\:\:$   &      49 $\:\:$&$\pm$&      24 $\:\:$   &      43 $\:\:$&$\pm$&      17 $\:\:$   &     124 $\:\:$&$\pm$&      46 $\:\:$    & \ & \ & \     & \ & \ & \    \\ 
  1.45 &   0.925   & \ & \ & \    &      17 $\:\:$&$\pm$&      12 $\:\:$   &      26 $\:\:$&$\pm$&      12 $\:\:$   &      98 $\:\:$&$\pm$&      43 $\:\:$    & \ & \ & \     & \ & \ & \    \\ 
  1.45 &   0.975  &      34 $\:\:$&$\pm$&      17 $\:\:$   &   16.2 &$\pm$&    9.8  &      30 $\:\:$&$\pm$&      11 $\:\:$   &     137 $\:\:$&$\pm$&     102 $\:\:$    & \ & \ & \     & \ & \ & \    \\ 
  1.45 &   1.025   & \ & \ & \    &      33 $\:\:$&$\pm$&      16 $\:\:$   &      49 $\:\:$&$\pm$&      17 $\:\:$   &   12.6 &$\pm$&    9.3  &      35 $\:\:$&$\pm$&      12 $\:\:$   &      62 $\:\:$&$\pm$&      19 $\:\:$   \\ 
  1.45 &   1.075   & \ & \ & \     & \ & \ & \     & \ & \ & \    &   28.4 &$\pm$&    9.7  &   31.1 &$\pm$&    9.3  &      18 $\:\:$&$\pm$&      10 $\:\:$   \\ 
  1.45 &   1.125   & \ & \ & \     & \ & \ & \     & \ & \ & \    &   19.8 &$\pm$&    8.4  &   15.1 &$\pm$&    7.7  &      24 $\:\:$&$\pm$&      13 $\:\:$   \\ 
  1.45 &   2.150   & \ & \ & \     & \ & \ & \    &    9.5 &$\pm$&    5.6  &    2.3 &$\pm$&    1.2  &    2.9 &$\pm$&    1.2  &    2.0 &$\pm$&    1.4  \\ 
 \hline       
  1.47 &   0.633   & \ & \ & \     & \ & \ & \     & \ & \ & \    &    1040 $\:\:$&$\pm$&     909 $\:\:$   &     119 $\:\:$&$\pm$&     117 $\:\:$    & \ & \ & \    \\ 
  1.47 &   0.700   & \ & \ & \     & \ & \ & \     & \ & \ & \    &      84 $\:\:$&$\pm$&      79 $\:\:$   &     145 $\:\:$&$\pm$&      38 $\:\:$   &      72 $\:\:$&$\pm$&      36 $\:\:$   \\ 
  1.47 &   0.767   & \ & \ & \     & \ & \ & \     & \ & \ & \     & \ & \ & \     & \ & \ & \    &     156 $\:\:$&$\pm$&      95 $\:\:$   \\ 
  1.47 &   0.875  &      31 $\:\:$&$\pm$&      24 $\:\:$   &      60 $\:\:$&$\pm$&      18 $\:\:$   &      49 $\:\:$&$\pm$&      17 $\:\:$    & \ & \ & \     & \ & \ & \     & \ & \ & \    \\ 
  1.47 &   0.925  &      73 $\:\:$&$\pm$&      21 $\:\:$   &      46 $\:\:$&$\pm$&      14 $\:\:$   &      48 $\:\:$&$\pm$&      18 $\:\:$    & \ & \ & \     & \ & \ & \     & \ & \ & \    \\ 
  1.47 &   0.975  &      40 $\:\:$&$\pm$&      19 $\:\:$   &      19 $\:\:$&$\pm$&      10 $\:\:$   &      44 $\:\:$&$\pm$&      22 $\:\:$   &      61 $\:\:$&$\pm$&      17 $\:\:$   &      48 $\:\:$&$\pm$&      17 $\:\:$   &      29 $\:\:$&$\pm$&      19 $\:\:$   \\ 
  1.47 &   1.025   & \ & \ & \     & \ & \ & \    &      40 $\:\:$&$\pm$&      17 $\:\:$   &   28.6 &$\pm$&    9.4  &      37 $\:\:$&$\pm$&      11 $\:\:$   &      46 $\:\:$&$\pm$&      19 $\:\:$   \\ 
  1.47 &   1.075  &     154 $\:\:$&$\pm$&     111 $\:\:$    & \ & \ & \    &      63 $\:\:$&$\pm$&      21 $\:\:$   &   27.5 &$\pm$&    8.7  &      22 $\:\:$&$\pm$&      10 $\:\:$   &      34 $\:\:$&$\pm$&      18 $\:\:$   \\ 
  1.47 &   1.125   & \ & \ & \     & \ & \ & \    &      22 $\:\:$&$\pm$&      13 $\:\:$   &   14.1 &$\pm$&    7.0  &      46 $\:\:$&$\pm$&      16 $\:\:$   &      25 $\:\:$&$\pm$&      19 $\:\:$   \\ 
  1.47 &   2.050   & \ & \ & \     & \ & \ & \     & \ & \ & \    &    3.5 &$\pm$&    1.6  &    1.7 &$\pm$&    1.1  &    4.1 &$\pm$&    2.4  \\ 
  1.47 &   2.150   & \ & \ & \     & \ & \ & \    &    2.0 &$\pm$&    1.5  &    2.2 &$\pm$&    1.1  &    2.2 &$\pm$&    1.6  &    2.4 &$\pm$&    1.5  \\ 
 \hline       
  1.49 &   0.567   & \ & \ & \     & \ & \ & \     & \ & \ & \     & \ & \ & \    &    3280 $\:\:$&$\pm$&    2850 $\:\:$   &    2500 $\:\:$&$\pm$&    2110 $\:\:$   \\ 
  1.49 &   0.633   & \ & \ & \     & \ & \ & \     & \ & \ & \    &     335 $\:\:$&$\pm$&      75 $\:\:$   &     124 $\:\:$&$\pm$&      46 $\:\:$   &     169 $\:\:$&$\pm$&      61 $\:\:$   \\ 
  1.49 &   0.700   & \ & \ & \     & \ & \ & \     & \ & \ & \    &     140 $\:\:$&$\pm$&      46 $\:\:$   &      82 $\:\:$&$\pm$&      31 $\:\:$   &      98 $\:\:$&$\pm$&      45 $\:\:$   \\ 
  1.49 &   0.767   & \ & \ & \     & \ & \ & \     & \ & \ & \    &     124 $\:\:$&$\pm$&      92 $\:\:$   &     144 $\:\:$&$\pm$&      75 $\:\:$   &     108 $\:\:$&$\pm$&     112 $\:\:$   \\ 
  1.49 &   0.875  &      39 $\:\:$&$\pm$&      15 $\:\:$   &      83 $\:\:$&$\pm$&      18 $\:\:$   &      51 $\:\:$&$\pm$&      26 $\:\:$    & \ & \ & \     & \ & \ & \     & \ & \ & \    \\ 
  1.49 &   0.925  &      89 $\:\:$&$\pm$&      23 $\:\:$   &      56 $\:\:$&$\pm$&      17 $\:\:$   &      76 $\:\:$&$\pm$&      23 $\:\:$   &      44 $\:\:$&$\pm$&      17 $\:\:$   &      63 $\:\:$&$\pm$&      28 $\:\:$   &      86 $\:\:$&$\pm$&      43 $\:\:$   \\ 
  1.49 &   0.975  &      66 $\:\:$&$\pm$&      37 $\:\:$   &      77 $\:\:$&$\pm$&      36 $\:\:$   &      44 $\:\:$&$\pm$&      14 $\:\:$   &      40 $\:\:$&$\pm$&      11 $\:\:$   &      25 $\:\:$&$\pm$&      14 $\:\:$   &      29 $\:\:$&$\pm$&      23 $\:\:$   \\ 
  1.49 &   1.025  &      95 $\:\:$&$\pm$&      51 $\:\:$   &      26 $\:\:$&$\pm$&      21 $\:\:$   &      68 $\:\:$&$\pm$&      15 $\:\:$   &      41 $\:\:$&$\pm$&      10 $\:\:$   &      63 $\:\:$&$\pm$&      21 $\:\:$   &      30 $\:\:$&$\pm$&      24 $\:\:$   \\ 
  1.49 &   1.075   & \ & \ & \    &      42 $\:\:$&$\pm$&      24 $\:\:$   &      46 $\:\:$&$\pm$&      12 $\:\:$   &      40 $\:\:$&$\pm$&      10 $\:\:$   &      21 $\:\:$&$\pm$&      19 $\:\:$   &      57 $\:\:$&$\pm$&      27 $\:\:$   \\ 
  1.49 &   1.125   & \ & \ & \     & \ & \ & \    &      24 $\:\:$&$\pm$&      11 $\:\:$   &    8.0 &$\pm$&    8.1  &      42 $\:\:$&$\pm$&      11 $\:\:$   &      51 $\:\:$&$\pm$&      10 $\:\:$   \\ 
  1.49 &   1.950   & \ & \ & \     & \ & \ & \    &      24 $\:\:$&$\pm$&      14 $\:\:$   &    6.2 &$\pm$&    4.4   & \ & \ & \    &    5.8 &$\pm$&    3.5  \\ 
  1.49 &   2.050   & \ & \ & \    &    8.2 &$\pm$&    5.9  &    2.2 &$\pm$&    1.4  &    2.2 &$\pm$&    1.1  &    5.1 &$\pm$&    1.6  &    3.1 &$\pm$&    1.5  \\ 
  1.49 &   2.150   & \ & \ & \    &    9.1 &$\pm$&    6.5  &    2.8 &$\pm$&    1.4  &    2.5 &$\pm$&    1.8   & \ & \ & \     & \ & \ & \    \\ 
 \hline       
  1.51 &   0.567   & \ & \ & \    &     488 $\:\:$&$\pm$&     318 $\:\:$   &     370 $\:\:$&$\pm$&     137 $\:\:$   &     151 $\:\:$&$\pm$&      77 $\:\:$   &     332 $\:\:$&$\pm$&     126 $\:\:$   &     465 $\:\:$&$\pm$&     267 $\:\:$   \\ 
  1.51 &   0.633   & \ & \ & \    &     483 $\:\:$&$\pm$&     379 $\:\:$   &     129 $\:\:$&$\pm$&      82 $\:\:$   &     207 $\:\:$&$\pm$&      39 $\:\:$   &     204 $\:\:$&$\pm$&      53 $\:\:$   &     171 $\:\:$&$\pm$&      74 $\:\:$   \\ 
  1.51 &   0.700   & \ & \ & \     & \ & \ & \    &     159 $\:\:$&$\pm$&     145 $\:\:$   &      88 $\:\:$&$\pm$&      29 $\:\:$   &     150 $\:\:$&$\pm$&      46 $\:\:$   &     112 $\:\:$&$\pm$&      65 $\:\:$   \\ 
  1.51 &   0.767   & \ & \ & \     & \ & \ & \     & \ & \ & \    &      98 $\:\:$&$\pm$&      80 $\:\:$   &     227 $\:\:$&$\pm$&     157 $\:\:$   &     808 $\:\:$&$\pm$&     674 $\:\:$   \\ 
  1.51 &   0.875  &      55 $\:\:$&$\pm$&      15 $\:\:$   &      26 $\:\:$&$\pm$&      11 $\:\:$   &      42 $\:\:$&$\pm$&      24 $\:\:$   &      99 $\:\:$&$\pm$&      31 $\:\:$    & \ & \ & \     & \ & \ & \    \\ 
  1.51 &   0.925  &      33 $\:\:$&$\pm$&      19 $\:\:$   &      56 $\:\:$&$\pm$&      19 $\:\:$   &      35 $\:\:$&$\pm$&      12 $\:\:$   &      49 $\:\:$&$\pm$&      14 $\:\:$   &      55 $\:\:$&$\pm$&      33 $\:\:$   &     277 $\:\:$&$\pm$&     104 $\:\:$   \\ 
  1.51 &   0.975  &      54 $\:\:$&$\pm$&      25 $\:\:$   &      59 $\:\:$&$\pm$&      20 $\:\:$   &      76 $\:\:$&$\pm$&      14 $\:\:$   &      64 $\:\:$&$\pm$&      16 $\:\:$   &      50 $\:\:$&$\pm$&      35 $\:\:$   &     106 $\:\:$&$\pm$&      73 $\:\:$   \\ 
  1.51 &   1.025   & \ & \ & \    &      25 $\:\:$&$\pm$&      13 $\:\:$   &   36.9 &$\pm$&    9.8  &      34 $\:\:$&$\pm$&      13 $\:\:$    & \ & \ & \     & \ & \ & \    \\ 
  1.51 &   1.075  &     128 $\:\:$&$\pm$&      66 $\:\:$   &      54 $\:\:$&$\pm$&      22 $\:\:$   &      30 $\:\:$&$\pm$&      10 $\:\:$   &      55 $\:\:$&$\pm$&      13 $\:\:$   &   27.3 &$\pm$&    8.5  &      68 $\:\:$&$\pm$&      13 $\:\:$   \\ 
  1.51 &   1.125   & \ & \ & \     & \ & \ & \    &      62 $\:\:$&$\pm$&      36 $\:\:$   &   13.4 &$\pm$&    7.6  &   42.6 &$\pm$&    8.6  &      52 $\:\:$&$\pm$&      10 $\:\:$   \\ 
  1.51 &   1.950   & \ & \ & \     & \ & \ & \     & \ & \ & \    &    4.2 &$\pm$&    1.8  &    8.4 &$\pm$&    2.6  &    2.5 &$\pm$&    1.6  \\ 
  1.51 &   2.050   & \ & \ & \     & \ & \ & \    &    5.3 &$\pm$&    1.7  &    4.1 &$\pm$&    1.4  &    2.0 &$\pm$&    1.2  &    3.3 &$\pm$&    2.2  \\ 
  1.51 &   2.150  &    9.1 &$\pm$&    6.6   & \ & \ & \     & \ & \ & \    &    5.2 &$\pm$&    3.1   & \ & \ & \     & \ & \ & \    \\ 
\end{tabular}\end{ruledtabular}\end{table*}
\endgroup


\begingroup\squeezetable
\begin{table*}\begin{ruledtabular}
\caption{\label{tab-gam-q2dep-2}
H$(e,e'p)\gamma$ cross section $ d^5\sigma / ( dk'_{lab} \, d[\Omega_e]_{lab} \, d[\Omega_p]_{\mathrm{c.m.}})$ ($\pm$ statistical error) at 
$\cos\theta_{\gamma\gamma}^\ast = -0.975$ and $W=$ 1.53, 1.55, 1.57, 1.59 GeV, in  fb/(MeV $\cdot$ sr$^2$). The systematic error is globally $\pm$ 12\% on each 
point. 
}
\begin{tabular}{|c|c|rrr|rrr|rrr|rrr|rrr|rrr|}
$W$ & $Q^2$
&\multicolumn{3}{c|}{$\phi=15^\circ$}
&\multicolumn{3}{c|}{$\phi=45^\circ$}
&\multicolumn{3}{c|}{$\phi=75^\circ$}
&\multicolumn{3}{c|}{$\phi=105^\circ$}
&\multicolumn{3}{c|}{$\phi=135^\circ$}
&\multicolumn{3}{c|}{$\phi=165^\circ$} \\ 
(GeV) & (GeV$^2$) &    & & &     & & &     & & &     & & &     & & &     & & \\
\hline
  1.53 &   0.500  &     405 $\:\:$&$\pm$&     383 $\:\:$   &     449 $\:\:$&$\pm$&     312 $\:\:$   &     346 $\:\:$&$\pm$&     199 $\:\:$   &     907 $\:\:$&$\pm$&     472 $\:\:$    & \ & \ & \     & \ & \ & \    \\ 
  1.53 &   0.567  &      50 $\:\:$&$\pm$&      95 $\:\:$   &     145 $\:\:$&$\pm$&      91 $\:\:$   &      68 $\:\:$&$\pm$&      43 $\:\:$   &     146 $\:\:$&$\pm$&      45 $\:\:$   &     382 $\:\:$&$\pm$&     134 $\:\:$   &     258 $\:\:$&$\pm$&     173 $\:\:$   \\ 
  1.53 &   0.633   & \ & \ & \    &     192 $\:\:$&$\pm$&     135 $\:\:$   &     165 $\:\:$&$\pm$&      45 $\:\:$   &     117 $\:\:$&$\pm$&      33 $\:\:$   &     103 $\:\:$&$\pm$&      66 $\:\:$   &     461 $\:\:$&$\pm$&     165 $\:\:$   \\ 
  1.53 &   0.700   & \ & \ & \    &     361 $\:\:$&$\pm$&     304 $\:\:$   &      62 $\:\:$&$\pm$&      56 $\:\:$   &      87 $\:\:$&$\pm$&      33 $\:\:$   &     170 $\:\:$&$\pm$&      94 $\:\:$   &     260 $\:\:$&$\pm$&     164 $\:\:$   \\ 
  1.53 &   0.875  &      70 $\:\:$&$\pm$&      17 $\:\:$   &      63 $\:\:$&$\pm$&      17 $\:\:$   &      55 $\:\:$&$\pm$&      16 $\:\:$   &      64 $\:\:$&$\pm$&      25 $\:\:$    & \ & \ & \     & \ & \ & \    \\ 
  1.53 &   0.925  &      56 $\:\:$&$\pm$&      20 $\:\:$   &   10.1 &$\pm$&    9.1  &      38 $\:\:$&$\pm$&      11 $\:\:$   &      31 $\:\:$&$\pm$&      18 $\:\:$    & \ & \ & \     & \ & \ & \    \\ 
  1.53 &   0.975  &      33 $\:\:$&$\pm$&      14 $\:\:$   &      81 $\:\:$&$\pm$&      17 $\:\:$   &   30.9 &$\pm$&    9.3  &      67 $\:\:$&$\pm$&      27 $\:\:$    & \ & \ & \    &      95 $\:\:$&$\pm$&      70 $\:\:$   \\ 
  1.53 &   1.025  &      26 $\:\:$&$\pm$&      14 $\:\:$   &      31 $\:\:$&$\pm$&      12 $\:\:$   &      49 $\:\:$&$\pm$&      12 $\:\:$   &      42 $\:\:$&$\pm$&      12 $\:\:$   &      52 $\:\:$&$\pm$&      12 $\:\:$   &      46 $\:\:$&$\pm$&      14 $\:\:$   \\ 
  1.53 &   1.075   & \ & \ & \    &      81 $\:\:$&$\pm$&      58 $\:\:$   &      47 $\:\:$&$\pm$&      24 $\:\:$   &   40.0 &$\pm$&    9.2  &   17.0 &$\pm$&    6.2  &   22.4 &$\pm$&    8.8  \\ 
  1.53 &   1.125   & \ & \ & \     & \ & \ & \     & \ & \ & \    &   15.0 &$\pm$&    6.2  &   17.4 &$\pm$&    6.3  &      36 $\:\:$&$\pm$&      11 $\:\:$   \\ 
  1.53 &   1.850   & \ & \ & \     & \ & \ & \     & \ & \ & \    &   11.2 &$\pm$&    7.2   & \ & \ & \     & \ & \ & \    \\ 
  1.53 &   1.950   & \ & \ & \    &    4.3 &$\pm$&    2.7  &    2.9 &$\pm$&    1.4  &    4.5 &$\pm$&    1.6  &    3.0 &$\pm$&    1.6  &    2.6 &$\pm$&    1.9  \\ 
  1.53 &   2.050   & \ & \ & \     & \ & \ & \    &    3.3 &$\pm$&    1.3  &    1.5 &$\pm$&    0.9   & \ & \ & \     & \ & \ & \    \\ 
  1.53 &   2.150  &    7.2 &$\pm$&    5.1  &    3.8 &$\pm$&    2.3   & \ & \ & \    &    4.6 &$\pm$&    3.3   & \ & \ & \     & \ & \ & \    \\ 
 \hline       
  1.55 &   0.500  &     315 $\:\:$&$\pm$&     181 $\:\:$   &     290 $\:\:$&$\pm$&     144 $\:\:$   &     235 $\:\:$&$\pm$&      69 $\:\:$   &     236 $\:\:$&$\pm$&     126 $\:\:$    & \ & \ & \     & \ & \ & \    \\ 
  1.55 &   0.567  &     221 $\:\:$&$\pm$&      75 $\:\:$   &     268 $\:\:$&$\pm$&      69 $\:\:$   &     170 $\:\:$&$\pm$&      39 $\:\:$   &     261 $\:\:$&$\pm$&      63 $\:\:$   &     621 $\:\:$&$\pm$&     511 $\:\:$   &    2290 $\:\:$&$\pm$&    1550 $\:\:$   \\ 
  1.55 &   0.633  &     137 $\:\:$&$\pm$&     129 $\:\:$   &     171 $\:\:$&$\pm$&      69 $\:\:$   &      75 $\:\:$&$\pm$&      30 $\:\:$   &     100 $\:\:$&$\pm$&      43 $\:\:$   &     588 $\:\:$&$\pm$&     401 $\:\:$    & \ & \ & \    \\ 
  1.55 &   0.700   & \ & \ & \     & \ & \ & \    &     123 $\:\:$&$\pm$&     116 $\:\:$   &      77 $\:\:$&$\pm$&     104 $\:\:$    & \ & \ & \     & \ & \ & \    \\ 
  1.55 &   0.875  &      48 $\:\:$&$\pm$&      16 $\:\:$   &      50 $\:\:$&$\pm$&      13 $\:\:$   &      28 $\:\:$&$\pm$&      10 $\:\:$   &      99 $\:\:$&$\pm$&      75 $\:\:$    & \ & \ & \     & \ & \ & \    \\ 
  1.55 &   0.925  &      54 $\:\:$&$\pm$&      15 $\:\:$   &      36 $\:\:$&$\pm$&      10 $\:\:$   &      45 $\:\:$&$\pm$&      12 $\:\:$    & \ & \ & \     & \ & \ & \     & \ & \ & \    \\ 
  1.55 &   0.975  &      39 $\:\:$&$\pm$&      14 $\:\:$   &      30 $\:\:$&$\pm$&      10 $\:\:$   &   20.8 &$\pm$&    8.8  &      37 $\:\:$&$\pm$&      11 $\:\:$   &      57 $\:\:$&$\pm$&      14 $\:\:$   &      54 $\:\:$&$\pm$&      18 $\:\:$   \\ 
  1.55 &   1.025  &     147 $\:\:$&$\pm$&      71 $\:\:$   &      48 $\:\:$&$\pm$&      23 $\:\:$   &   10.9 &$\pm$&    7.2  &   41.3 &$\pm$&    8.8  &      48 $\:\:$&$\pm$&      10 $\:\:$   &      80 $\:\:$&$\pm$&      18 $\:\:$   \\ 
  1.55 &   1.075   & \ & \ & \    &      63 $\:\:$&$\pm$&      44 $\:\:$   &      46 $\:\:$&$\pm$&      15 $\:\:$   &   31.6 &$\pm$&    7.6  &      61 $\:\:$&$\pm$&      12 $\:\:$   &      27 $\:\:$&$\pm$&      13 $\:\:$   \\ 
  1.55 &   1.125   & \ & \ & \     & \ & \ & \    &      19 $\:\:$&$\pm$&      10 $\:\:$   &   20.1 &$\pm$&    6.3  &   22.5 &$\pm$&    9.1  &      29 $\:\:$&$\pm$&      16 $\:\:$   \\ 
  1.55 &   1.850   & \ & \ & \     & \ & \ & \    &    9.3 &$\pm$&    6.6  &    6.6 &$\pm$&    3.1   & \ & \ & \     & \ & \ & \    \\ 
  1.55 &   1.950   & \ & \ & \    &    2.3 &$\pm$&    1.4  &    1.2 &$\pm$&    0.7  &    2.0 &$\pm$&    1.0  &    6.0 &$\pm$&    4.3   & \ & \ & \    \\ 
  1.55 &   2.050   & \ & \ & \    &    2.0 &$\pm$&    1.1  &    4.5 &$\pm$&    1.5  &    2.6 &$\pm$&    1.6   & \ & \ & \     & \ & \ & \    \\ 
 \hline       
  1.57 &   0.500  &     257 $\:\:$&$\pm$&      78 $\:\:$   &     146 $\:\:$&$\pm$&      57 $\:\:$   &     400 $\:\:$&$\pm$&      69 $\:\:$    & \ & \ & \     & \ & \ & \     & \ & \ & \    \\ 
  1.57 &   0.567  &      97 $\:\:$&$\pm$&      79 $\:\:$   &     155 $\:\:$&$\pm$&      47 $\:\:$   &     191 $\:\:$&$\pm$&      38 $\:\:$    & \ & \ & \     & \ & \ & \     & \ & \ & \    \\ 
  1.57 &   0.633  &     146 $\:\:$&$\pm$&     142 $\:\:$   &     101 $\:\:$&$\pm$&      85 $\:\:$   &      37 $\:\:$&$\pm$&      35 $\:\:$   &     476 $\:\:$&$\pm$&     247 $\:\:$    & \ & \ & \     & \ & \ & \    \\ 
  1.57 &   0.875  &      30 $\:\:$&$\pm$&      10 $\:\:$   &   30.6 &$\pm$&    9.4  &      49 $\:\:$&$\pm$&      15 $\:\:$    & \ & \ & \     & \ & \ & \     & \ & \ & \    \\ 
  1.57 &   0.925  &   25.3 &$\pm$&    9.9  &      40 $\:\:$&$\pm$&      10 $\:\:$   &      36 $\:\:$&$\pm$&      12 $\:\:$   &      50 $\:\:$&$\pm$&      14 $\:\:$   &      52 $\:\:$&$\pm$&      18 $\:\:$   &      37 $\:\:$&$\pm$&      26 $\:\:$   \\ 
  1.57 &   0.975  &      46 $\:\:$&$\pm$&      28 $\:\:$   &      36 $\:\:$&$\pm$&      19 $\:\:$   &      36 $\:\:$&$\pm$&      10 $\:\:$   &   39.1 &$\pm$&    8.8  &      44 $\:\:$&$\pm$&      13 $\:\:$   &      45 $\:\:$&$\pm$&      21 $\:\:$   \\ 
  1.57 &   1.025  &      65 $\:\:$&$\pm$&      46 $\:\:$    & \ & \ & \    &   26.4 &$\pm$&    8.5  &   25.5 &$\pm$&    7.1  &      17 $\:\:$&$\pm$&      10 $\:\:$   &      27 $\:\:$&$\pm$&      18 $\:\:$   \\ 
  1.57 &   1.075   & \ & \ & \    &      34 $\:\:$&$\pm$&      17 $\:\:$   &   21.8 &$\pm$&    7.5  &   39.9 &$\pm$&    8.8  &      29 $\:\:$&$\pm$&      13 $\:\:$   &      22 $\:\:$&$\pm$&      21 $\:\:$   \\ 
  1.57 &   1.125   & \ & \ & \     & \ & \ & \    &   20.5 &$\pm$&    7.9  &   18.8 &$\pm$&    7.2  &      37 $\:\:$&$\pm$&      16 $\:\:$   &      21 $\:\:$&$\pm$&      14 $\:\:$   \\ 
  1.57 &   1.850   & \ & \ & \    &    3.1 &$\pm$&    2.3  &    1.8 &$\pm$&    1.2  &    5.4 &$\pm$&    3.3  &   13.5 &$\pm$&    9.7   & \ & \ & \    \\ 
  1.57 &   1.950  &    2.4 &$\pm$&    1.3  &    3.0 &$\pm$&    1.2  &    4.7 &$\pm$&    1.5  &    3.4 &$\pm$&    1.8   & \ & \ & \     & \ & \ & \    \\ 
  1.57 &   2.050   & \ & \ & \     & \ & \ & \    &    2.3 &$\pm$&    1.4   & \ & \ & \     & \ & \ & \     & \ & \ & \    \\ 
 \hline       
  1.59 &   0.433   & \ & \ & \    &     121 $\:\:$&$\pm$&     177 $\:\:$   &     440 $\:\:$&$\pm$&     410 $\:\:$    & \ & \ & \     & \ & \ & \     & \ & \ & \    \\ 
  1.59 &   0.500  &     118 $\:\:$&$\pm$&      58 $\:\:$   &     141 $\:\:$&$\pm$&      48 $\:\:$   &     187 $\:\:$&$\pm$&      73 $\:\:$    & \ & \ & \     & \ & \ & \     & \ & \ & \    \\ 
  1.59 &   0.567  &     117 $\:\:$&$\pm$&      63 $\:\:$   &      64 $\:\:$&$\pm$&      38 $\:\:$   &     173 $\:\:$&$\pm$&      56 $\:\:$    & \ & \ & \     & \ & \ & \     & \ & \ & \    \\ 
  1.59 &   0.633   & \ & \ & \    &    6070 $\:\:$&$\pm$&    6330 $\:\:$    & \ & \ & \     & \ & \ & \     & \ & \ & \     & \ & \ & \    \\ 
  1.59 &   0.875  &   26.6 &$\pm$&    8.2  &   37.4 &$\pm$&    9.0  &      45 $\:\:$&$\pm$&      16 $\:\:$   &      49 $\:\:$&$\pm$&      18 $\:\:$   &      93 $\:\:$&$\pm$&      45 $\:\:$   &     156 $\:\:$&$\pm$&     129 $\:\:$   \\ 
  1.59 &   0.925  &      74 $\:\:$&$\pm$&      23 $\:\:$   &      46 $\:\:$&$\pm$&      13 $\:\:$   &   26.1 &$\pm$&    9.3  &      53 $\:\:$&$\pm$&      12 $\:\:$   &      14 $\:\:$&$\pm$&      14 $\:\:$   &     108 $\:\:$&$\pm$&      78 $\:\:$   \\ 
  1.59 &   0.975  &      33 $\:\:$&$\pm$&      22 $\:\:$   &      30 $\:\:$&$\pm$&      12 $\:\:$   &   27.9 &$\pm$&    7.8  &   35.1 &$\pm$&    9.3  &      32 $\:\:$&$\pm$&      30 $\:\:$   &      54 $\:\:$&$\pm$&      42 $\:\:$   \\ 
  1.59 &   1.025  &      21 $\:\:$&$\pm$&      18 $\:\:$   &      32 $\:\:$&$\pm$&      11 $\:\:$   &   29.1 &$\pm$&    7.5  &      44 $\:\:$&$\pm$&      10 $\:\:$    & \ & \ & \     & \ & \ & \    \\ 
  1.59 &   1.075  &      23 $\:\:$&$\pm$&      14 $\:\:$   &      34 $\:\:$&$\pm$&      14 $\:\:$   &   30.0 &$\pm$&    7.7  &      38 $\:\:$&$\pm$&      11 $\:\:$   &      60 $\:\:$&$\pm$&      24 $\:\:$   &      29 $\:\:$&$\pm$&      14 $\:\:$   \\ 
  1.59 &   1.125   & \ & \ & \     & \ & \ & \    &      56 $\:\:$&$\pm$&      22 $\:\:$   &      16 $\:\:$&$\pm$&      11 $\:\:$   &   18.8 &$\pm$&    6.3  &   12.2 &$\pm$&    5.5  \\ 
  1.59 &   1.850   & \ & \ & \    &    2.0 &$\pm$&    1.3  &    1.7 &$\pm$&    1.0  &    5.8 &$\pm$&    2.6   & \ & \ & \     & \ & \ & \    \\ 
  1.59 &   1.950   & \ & \ & \    &    2.5 &$\pm$&    1.5  &    2.0 &$\pm$&    1.5   & \ & \ & \     & \ & \ & \     & \ & \ & \    \\ 
\end{tabular}\end{ruledtabular}\end{table*}
\endgroup


\begingroup\squeezetable
\begin{table*}\begin{ruledtabular}
\caption{\label{tab-piz-q2dep-1}
H$(e,e'p)\pi^0$ two-fold cross section \ $ d^2 \sigma / d [ \Omega_p ] _{\mathrm{c.m.}} $ \  ($\pm$ statistical error) at 
$\cthcm = -0.975$ and $W=$ 1.45, 1.47, 1.49, 1.51 GeV, in nb/sr. The systematic error is globally $\pm$ 4\% on each point.
}
\begin{tabular}{|c|c|rrr|rrr|rrr|rrr|rrr|rrr|}
$W$ & $Q^2$
&\multicolumn{3}{c|}{$\phi=15^\circ$}
&\multicolumn{3}{c|}{$\phi=45^\circ$}
&\multicolumn{3}{c|}{$\phi=75^\circ$}
&\multicolumn{3}{c|}{$\phi=105^\circ$}
&\multicolumn{3}{c|}{$\phi=135^\circ$}
&\multicolumn{3}{c|}{$\phi=165^\circ$} \\ 
(GeV) & (GeV$^2$) &    & & &     & & &     & & &     & & &     & & &     & & \\
\hline 
  1.45 &   0.700   & \ & \ & \     & \ & \ & \     & \ & \ & \     & \ & \ & \    &     788 $\:\:$&$\pm$&     498 $\:\:$   &     289 $\:\:$&$\pm$&     158 $\:\:$   \\ 
  1.45 &   0.767   & \ & \ & \     & \ & \ & \     & \ & \ & \     & \ & \ & \    &     407 $\:\:$&$\pm$&     120 $\:\:$   &     590 $\:\:$&$\pm$&     148 $\:\:$   \\ 
  1.45 &   0.875  &     120 $\:\:$&$\pm$&      28 $\:\:$   &     105 $\:\:$&$\pm$&      23 $\:\:$   &     268 $\:\:$&$\pm$&      34 $\:\:$   &     167 $\:\:$&$\pm$&      51 $\:\:$    & \ & \ & \     & \ & \ & \    \\ 
  1.45 &   0.925  &     105 $\:\:$&$\pm$&      16 $\:\:$   &     134 $\:\:$&$\pm$&      16 $\:\:$   &     201 $\:\:$&$\pm$&      18 $\:\:$   &     355 $\:\:$&$\pm$&      51 $\:\:$   &     771 $\:\:$&$\pm$&     497 $\:\:$   &     561 $\:\:$&$\pm$&     445 $\:\:$   \\ 
  1.45 &   0.975  &     128 $\:\:$&$\pm$&      17 $\:\:$   &     154 $\:\:$&$\pm$&      15 $\:\:$   &     239 $\:\:$&$\pm$&      19 $\:\:$   &     230 $\:\:$&$\pm$&      56 $\:\:$   &    1360 $\:\:$&$\pm$&    1090 $\:\:$    & \ & \ & \    \\ 
  1.45 &   1.025  &     184 $\:\:$&$\pm$&      22 $\:\:$   &     144 $\:\:$&$\pm$&      16 $\:\:$   &     252 $\:\:$&$\pm$&      22 $\:\:$   &     167 $\:\:$&$\pm$&      48 $\:\:$   &     327 $\:\:$&$\pm$&      77 $\:\:$   &     532 $\:\:$&$\pm$&     113 $\:\:$   \\ 
  1.45 &   1.075  &     177 $\:\:$&$\pm$&      33 $\:\:$   &     195 $\:\:$&$\pm$&      26 $\:\:$   &     210 $\:\:$&$\pm$&      25 $\:\:$   &     266 $\:\:$&$\pm$&      24 $\:\:$   &     395 $\:\:$&$\pm$&      30 $\:\:$   &     423 $\:\:$&$\pm$&      38 $\:\:$   \\ 
  1.45 &   1.125  &     150 $\:\:$&$\pm$&     102 $\:\:$   &     157 $\:\:$&$\pm$&      82 $\:\:$   &     246 $\:\:$&$\pm$&      48 $\:\:$   &     315 $\:\:$&$\pm$&      24 $\:\:$   &     382 $\:\:$&$\pm$&      28 $\:\:$   &     461 $\:\:$&$\pm$&      42 $\:\:$   \\ 
  1.45 &   2.050   & \ & \ & \     & \ & \ & \     & \ & \ & \     & \ & \ & \     & \ & \ & \    &     267 $\:\:$&$\pm$&     186 $\:\:$   \\ 
  1.45 &   2.150   & \ & \ & \    &     314 $\:\:$&$\pm$&     146 $\:\:$   &     103 $\:\:$&$\pm$&      26 $\:\:$   &      91 $\:\:$&$\pm$&      16 $\:\:$   &     130 $\:\:$&$\pm$&      18 $\:\:$   &     164 $\:\:$&$\pm$&      21 $\:\:$   \\ 
  1.45 &   2.250   & \ & \ & \     & \ & \ & \     & \ & \ & \    &     124 $\:\:$&$\pm$&      49 $\:\:$   &      71 $\:\:$&$\pm$&      35 $\:\:$   &      96 $\:\:$&$\pm$&      44 $\:\:$   \\ 
 \hline       
  1.47 &   0.700   & \ & \ & \     & \ & \ & \     & \ & \ & \    &     497 $\:\:$&$\pm$&      87 $\:\:$   &     525 $\:\:$&$\pm$&      43 $\:\:$   &     663 $\:\:$&$\pm$&      51 $\:\:$   \\ 
  1.47 &   0.767   & \ & \ & \     & \ & \ & \     & \ & \ & \    &     272 $\:\:$&$\pm$&      95 $\:\:$   &     475 $\:\:$&$\pm$&      63 $\:\:$   &     653 $\:\:$&$\pm$&      91 $\:\:$   \\ 
  1.47 &   0.875  &     209 $\:\:$&$\pm$&      23 $\:\:$   &     166 $\:\:$&$\pm$&      17 $\:\:$   &     209 $\:\:$&$\pm$&      21 $\:\:$   &     483 $\:\:$&$\pm$&     308 $\:\:$    & \ & \ & \     & \ & \ & \    \\ 
  1.47 &   0.925  &     120 $\:\:$&$\pm$&      15 $\:\:$   &     167 $\:\:$&$\pm$&      15 $\:\:$   &     236 $\:\:$&$\pm$&      23 $\:\:$   &     918 $\:\:$&$\pm$&     770 $\:\:$    & \ & \ & \     & \ & \ & \    \\ 
  1.47 &   0.975  &     174 $\:\:$&$\pm$&      19 $\:\:$   &     220 $\:\:$&$\pm$&      19 $\:\:$   &     257 $\:\:$&$\pm$&      28 $\:\:$   &     194 $\:\:$&$\pm$&      84 $\:\:$   &     402 $\:\:$&$\pm$&     104 $\:\:$   &     299 $\:\:$&$\pm$&     104 $\:\:$   \\ 
  1.47 &   1.025  &     154 $\:\:$&$\pm$&      22 $\:\:$   &     235 $\:\:$&$\pm$&      23 $\:\:$   &     225 $\:\:$&$\pm$&      23 $\:\:$   &     334 $\:\:$&$\pm$&      26 $\:\:$   &     411 $\:\:$&$\pm$&      34 $\:\:$   &     523 $\:\:$&$\pm$&      53 $\:\:$   \\ 
  1.47 &   1.075  &     138 $\:\:$&$\pm$&      51 $\:\:$   &     179 $\:\:$&$\pm$&      41 $\:\:$   &     229 $\:\:$&$\pm$&      25 $\:\:$   &     295 $\:\:$&$\pm$&      20 $\:\:$   &     439 $\:\:$&$\pm$&      31 $\:\:$   &     642 $\:\:$&$\pm$&      59 $\:\:$   \\ 
  1.47 &   1.125  &      84 $\:\:$&$\pm$&      66 $\:\:$   &     191 $\:\:$&$\pm$&      50 $\:\:$   &     223 $\:\:$&$\pm$&      26 $\:\:$   &     320 $\:\:$&$\pm$&      23 $\:\:$   &     502 $\:\:$&$\pm$&      44 $\:\:$   &     523 $\:\:$&$\pm$&      66 $\:\:$   \\ 
  1.47 &   2.050   & \ & \ & \     & \ & \ & \    &      33 $\:\:$&$\pm$&      19 $\:\:$   &      94 $\:\:$&$\pm$&      24 $\:\:$   &     206 $\:\:$&$\pm$&      35 $\:\:$   &     178 $\:\:$&$\pm$&      32 $\:\:$   \\ 
  1.47 &   2.150  &      35 $\:\:$&$\pm$&      21 $\:\:$   &      78 $\:\:$&$\pm$&      24 $\:\:$   &      65 $\:\:$&$\pm$&      12 $\:\:$   &     134 $\:\:$&$\pm$&      15 $\:\:$   &     165 $\:\:$&$\pm$&      18 $\:\:$   &     189 $\:\:$&$\pm$&      23 $\:\:$   \\ 
  1.47 &   2.250   & \ & \ & \     & \ & \ & \    &     108 $\:\:$&$\pm$&      68 $\:\:$   &     447 $\:\:$&$\pm$&     209 $\:\:$   &     102 $\:\:$&$\pm$&      73 $\:\:$    & \ & \ & \    \\ 
 \hline       
  1.49 &   0.633   & \ & \ & \     & \ & \ & \     & \ & \ & \    &     530 $\:\:$&$\pm$&      69 $\:\:$   &     667 $\:\:$&$\pm$&      66 $\:\:$   &     802 $\:\:$&$\pm$&      88 $\:\:$   \\ 
  1.49 &   0.700   & \ & \ & \     & \ & \ & \     & \ & \ & \    &     528 $\:\:$&$\pm$&      35 $\:\:$   &     605 $\:\:$&$\pm$&      30 $\:\:$   &     695 $\:\:$&$\pm$&      43 $\:\:$   \\ 
  1.49 &   0.767   & \ & \ & \     & \ & \ & \     & \ & \ & \    &     451 $\:\:$&$\pm$&      74 $\:\:$   &     488 $\:\:$&$\pm$&      65 $\:\:$   &     440 $\:\:$&$\pm$&      95 $\:\:$   \\ 
  1.49 &   0.875  &     164 $\:\:$&$\pm$&      15 $\:\:$   &     198 $\:\:$&$\pm$&      16 $\:\:$   &     283 $\:\:$&$\pm$&      33 $\:\:$    & \ & \ & \     & \ & \ & \     & \ & \ & \    \\ 
  1.49 &   0.925  &     145 $\:\:$&$\pm$&      15 $\:\:$   &     209 $\:\:$&$\pm$&      18 $\:\:$   &     260 $\:\:$&$\pm$&      39 $\:\:$   &     344 $\:\:$&$\pm$&     109 $\:\:$   &     969 $\:\:$&$\pm$&     270 $\:\:$   &     436 $\:\:$&$\pm$&     201 $\:\:$   \\ 
  1.49 &   0.975  &     145 $\:\:$&$\pm$&      17 $\:\:$   &     230 $\:\:$&$\pm$&      20 $\:\:$   &     231 $\:\:$&$\pm$&      22 $\:\:$   &     363 $\:\:$&$\pm$&      27 $\:\:$   &     510 $\:\:$&$\pm$&      51 $\:\:$   &     544 $\:\:$&$\pm$&      77 $\:\:$   \\ 
  1.49 &   1.025  &     146 $\:\:$&$\pm$&      31 $\:\:$   &     155 $\:\:$&$\pm$&      24 $\:\:$   &     272 $\:\:$&$\pm$&      20 $\:\:$   &     327 $\:\:$&$\pm$&      21 $\:\:$   &     485 $\:\:$&$\pm$&      44 $\:\:$   &     437 $\:\:$&$\pm$&      64 $\:\:$   \\ 
  1.49 &   1.075  &     132 $\:\:$&$\pm$&      33 $\:\:$   &     151 $\:\:$&$\pm$&      27 $\:\:$   &     245 $\:\:$&$\pm$&      19 $\:\:$   &     354 $\:\:$&$\pm$&      23 $\:\:$   &     351 $\:\:$&$\pm$&      44 $\:\:$   &     460 $\:\:$&$\pm$&      80 $\:\:$   \\ 
  1.49 &   1.125  &     128 $\:\:$&$\pm$&      33 $\:\:$   &     133 $\:\:$&$\pm$&      26 $\:\:$   &     248 $\:\:$&$\pm$&      22 $\:\:$   &     375 $\:\:$&$\pm$&      29 $\:\:$   &     427 $\:\:$&$\pm$&      55 $\:\:$   &     304 $\:\:$&$\pm$&      50 $\:\:$   \\ 
  1.49 &   1.950   & \ & \ & \     & \ & \ & \     & \ & \ & \     & \ & \ & \    &     543 $\:\:$&$\pm$&     324 $\:\:$    & \ & \ & \    \\ 
  1.49 &   2.050   & \ & \ & \    &      52 $\:\:$&$\pm$&      19 $\:\:$   &      70 $\:\:$&$\pm$&      13 $\:\:$   &      98 $\:\:$&$\pm$&      13 $\:\:$   &     145 $\:\:$&$\pm$&      17 $\:\:$   &     218 $\:\:$&$\pm$&      25 $\:\:$   \\ 
  1.49 &   2.150  &      34 $\:\:$&$\pm$&      15 $\:\:$   &      68 $\:\:$&$\pm$&      16 $\:\:$   &      88 $\:\:$&$\pm$&      12 $\:\:$   &      97 $\:\:$&$\pm$&      13 $\:\:$   &     154 $\:\:$&$\pm$&      22 $\:\:$   &     135 $\:\:$&$\pm$&      28 $\:\:$   \\ 
 \hline       
  1.51 &   0.567  &     706 $\:\:$&$\pm$&     555 $\:\:$   &     426 $\:\:$&$\pm$&     351 $\:\:$   &     679 $\:\:$&$\pm$&     187 $\:\:$   &     664 $\:\:$&$\pm$&     136 $\:\:$   &     878 $\:\:$&$\pm$&     179 $\:\:$   &    1020 $\:\:$&$\pm$&     276 $\:\:$   \\ 
  1.51 &   0.633  &     410 $\:\:$&$\pm$&     173 $\:\:$   &     495 $\:\:$&$\pm$&     146 $\:\:$   &     490 $\:\:$&$\pm$&      52 $\:\:$   &     617 $\:\:$&$\pm$&      29 $\:\:$   &     717 $\:\:$&$\pm$&      39 $\:\:$   &     721 $\:\:$&$\pm$&      56 $\:\:$   \\ 
  1.51 &   0.700  &     463 $\:\:$&$\pm$&     353 $\:\:$   &     552 $\:\:$&$\pm$&     325 $\:\:$   &     481 $\:\:$&$\pm$&      84 $\:\:$   &     584 $\:\:$&$\pm$&      27 $\:\:$   &     702 $\:\:$&$\pm$&      38 $\:\:$   &     680 $\:\:$&$\pm$&      57 $\:\:$   \\ 
  1.51 &   0.767   & \ & \ & \     & \ & \ & \    &    1200 $\:\:$&$\pm$&     496 $\:\:$   &     496 $\:\:$&$\pm$&      61 $\:\:$   &     470 $\:\:$&$\pm$&      84 $\:\:$   &     821 $\:\:$&$\pm$&     196 $\:\:$   \\ 
  1.51 &   0.875  &     248 $\:\:$&$\pm$&      17 $\:\:$   &     283 $\:\:$&$\pm$&      19 $\:\:$   &     371 $\:\:$&$\pm$&      67 $\:\:$   &     249 $\:\:$&$\pm$&     111 $\:\:$   &    1270 $\:\:$&$\pm$&     552 $\:\:$   &     876 $\:\:$&$\pm$&     514 $\:\:$   \\ 
  1.51 &   0.925  &     248 $\:\:$&$\pm$&      19 $\:\:$   &     285 $\:\:$&$\pm$&      22 $\:\:$   &     347 $\:\:$&$\pm$&      28 $\:\:$   &     441 $\:\:$&$\pm$&      35 $\:\:$   &     522 $\:\:$&$\pm$&      81 $\:\:$   &     495 $\:\:$&$\pm$&     124 $\:\:$   \\ 
  1.51 &   0.975  &     234 $\:\:$&$\pm$&      29 $\:\:$   &     301 $\:\:$&$\pm$&      27 $\:\:$   &     345 $\:\:$&$\pm$&      20 $\:\:$   &     441 $\:\:$&$\pm$&      27 $\:\:$   &     527 $\:\:$&$\pm$&      73 $\:\:$   &     478 $\:\:$&$\pm$&     102 $\:\:$   \\ 
  1.51 &   1.025  &     240 $\:\:$&$\pm$&      33 $\:\:$   &     252 $\:\:$&$\pm$&      25 $\:\:$   &     330 $\:\:$&$\pm$&      19 $\:\:$   &     428 $\:\:$&$\pm$&      28 $\:\:$   &     468 $\:\:$&$\pm$&      81 $\:\:$   &     574 $\:\:$&$\pm$&     139 $\:\:$   \\ 
  1.51 &   1.075  &     170 $\:\:$&$\pm$&      27 $\:\:$   &     266 $\:\:$&$\pm$&      27 $\:\:$   &     350 $\:\:$&$\pm$&      22 $\:\:$   &     320 $\:\:$&$\pm$&      28 $\:\:$   &     372 $\:\:$&$\pm$&      55 $\:\:$   &     431 $\:\:$&$\pm$&      66 $\:\:$   \\ 
  1.51 &   1.125  &     244 $\:\:$&$\pm$&      53 $\:\:$   &     185 $\:\:$&$\pm$&      32 $\:\:$   &     316 $\:\:$&$\pm$&      28 $\:\:$   &     391 $\:\:$&$\pm$&      29 $\:\:$   &     424 $\:\:$&$\pm$&      25 $\:\:$   &     529 $\:\:$&$\pm$&      31 $\:\:$   \\ 
  1.51 &   1.950  &      76 $\:\:$&$\pm$&      38 $\:\:$   &     115 $\:\:$&$\pm$&      41 $\:\:$   &     132 $\:\:$&$\pm$&      30 $\:\:$   &      84 $\:\:$&$\pm$&      20 $\:\:$   &     144 $\:\:$&$\pm$&      28 $\:\:$   &     199 $\:\:$&$\pm$&      39 $\:\:$   \\ 
  1.51 &   2.050  &      51 $\:\:$&$\pm$&      15 $\:\:$   &      96 $\:\:$&$\pm$&      17 $\:\:$   &     120 $\:\:$&$\pm$&      13 $\:\:$   &     146 $\:\:$&$\pm$&      13 $\:\:$   &     173 $\:\:$&$\pm$&      18 $\:\:$   &     169 $\:\:$&$\pm$&      24 $\:\:$   \\ 
  1.51 &   2.150  &      30 $\:\:$&$\pm$&      12 $\:\:$   &      79 $\:\:$&$\pm$&      15 $\:\:$   &      84 $\:\:$&$\pm$&      13 $\:\:$   &     116 $\:\:$&$\pm$&      17 $\:\:$   &     115 $\:\:$&$\pm$&      30 $\:\:$   &     197 $\:\:$&$\pm$&      58 $\:\:$   \\ 
\end{tabular}\end{ruledtabular}\end{table*}
\endgroup


\begingroup\squeezetable
\begin{table*}\begin{ruledtabular}
\caption{\label{tab-piz-q2dep-2}
H$(e,e'p)\pi^0$ two-fold cross section \ $ d^2 \sigma / d [ \Omega_p ] _{\mathrm{c.m.}} $ \  ($\pm$ statistical error) at 
$\cthcm = -0.975$ and $W=$ 1.53, 1.55, 1.57, 1.59 GeV, in nb/sr. The systematic error is globally $\pm$ 4\% on each point. 
}
\begin{tabular}{|c|c|rrr|rrr|rrr|rrr|rrr|rrr|}
$W$ & $Q^2$
&\multicolumn{3}{c|}{$\phi=15^\circ$}
&\multicolumn{3}{c|}{$\phi=45^\circ$}
&\multicolumn{3}{c|}{$\phi=75^\circ$}
&\multicolumn{3}{c|}{$\phi=105^\circ$}
&\multicolumn{3}{c|}{$\phi=135^\circ$}
&\multicolumn{3}{c|}{$\phi=165^\circ$} \\ 
(GeV) & (GeV$^2$) &    & & &     & & &     & & &     & & &     & & &     & & \\
 \hline 
  1.53 &   0.567  &     248 $\:\:$&$\pm$&      53 $\:\:$   &     409 $\:\:$&$\pm$&      54 $\:\:$   &     503 $\:\:$&$\pm$&      34 $\:\:$   &     580 $\:\:$&$\pm$&      34 $\:\:$   &     600 $\:\:$&$\pm$&      70 $\:\:$   &     406 $\:\:$&$\pm$&      81 $\:\:$   \\ 
  1.53 &   0.633  &     385 $\:\:$&$\pm$&      79 $\:\:$   &     499 $\:\:$&$\pm$&      69 $\:\:$   &     510 $\:\:$&$\pm$&      30 $\:\:$   &     578 $\:\:$&$\pm$&      24 $\:\:$   &     522 $\:\:$&$\pm$&      45 $\:\:$   &     557 $\:\:$&$\pm$&      70 $\:\:$   \\ 
  1.53 &   0.700  &     296 $\:\:$&$\pm$&     120 $\:\:$   &     336 $\:\:$&$\pm$&      91 $\:\:$   &     448 $\:\:$&$\pm$&      38 $\:\:$   &     561 $\:\:$&$\pm$&      28 $\:\:$   &     519 $\:\:$&$\pm$&      55 $\:\:$   &     519 $\:\:$&$\pm$&      85 $\:\:$   \\ 
  1.53 &   0.767   & \ & \ & \     & \ & \ & \    &     384 $\:\:$&$\pm$&     146 $\:\:$   &     427 $\:\:$&$\pm$&      76 $\:\:$   &     441 $\:\:$&$\pm$&     195 $\:\:$   &     890 $\:\:$&$\pm$&     506 $\:\:$   \\ 
  1.53 &   0.875  &     311 $\:\:$&$\pm$&      19 $\:\:$   &     299 $\:\:$&$\pm$&      22 $\:\:$   &     409 $\:\:$&$\pm$&      34 $\:\:$   &     403 $\:\:$&$\pm$&      52 $\:\:$   &     291 $\:\:$&$\pm$&     149 $\:\:$   &    2850 $\:\:$&$\pm$&    1830 $\:\:$   \\ 
  1.53 &   0.925  &     269 $\:\:$&$\pm$&      26 $\:\:$   &     313 $\:\:$&$\pm$&      23 $\:\:$   &     360 $\:\:$&$\pm$&      20 $\:\:$   &     510 $\:\:$&$\pm$&      42 $\:\:$   &     501 $\:\:$&$\pm$&     159 $\:\:$   &     630 $\:\:$&$\pm$&     278 $\:\:$   \\ 
  1.53 &   0.975  &     262 $\:\:$&$\pm$&      27 $\:\:$   &     340 $\:\:$&$\pm$&      23 $\:\:$   &     430 $\:\:$&$\pm$&      22 $\:\:$   &     368 $\:\:$&$\pm$&      37 $\:\:$   &     148 $\:\:$&$\pm$&     103 $\:\:$   &    1320 $\:\:$&$\pm$&     806 $\:\:$   \\ 
  1.53 &   1.025  &     285 $\:\:$&$\pm$&      29 $\:\:$   &     344 $\:\:$&$\pm$&      24 $\:\:$   &     365 $\:\:$&$\pm$&      21 $\:\:$   &     419 $\:\:$&$\pm$&      44 $\:\:$   &     229 $\:\:$&$\pm$&      46 $\:\:$   &     282 $\:\:$&$\pm$&      56 $\:\:$   \\ 
  1.53 &   1.075  &     357 $\:\:$&$\pm$&      46 $\:\:$   &     294 $\:\:$&$\pm$&      28 $\:\:$   &     358 $\:\:$&$\pm$&      25 $\:\:$   &     366 $\:\:$&$\pm$&      23 $\:\:$   &     407 $\:\:$&$\pm$&      24 $\:\:$   &     407 $\:\:$&$\pm$&      29 $\:\:$   \\ 
  1.53 &   1.125  &     175 $\:\:$&$\pm$&     130 $\:\:$   &     294 $\:\:$&$\pm$&      91 $\:\:$   &     339 $\:\:$&$\pm$&      48 $\:\:$   &     400 $\:\:$&$\pm$&      22 $\:\:$   &     392 $\:\:$&$\pm$&      22 $\:\:$   &     396 $\:\:$&$\pm$&      29 $\:\:$   \\ 
  1.53 &   1.850  &    1030 $\:\:$&$\pm$&     804 $\:\:$    & \ & \ & \    &     514 $\:\:$&$\pm$&     380 $\:\:$    & \ & \ & \     & \ & \ & \     & \ & \ & \    \\ 
  1.53 &   1.950  &      83 $\:\:$&$\pm$&      23 $\:\:$   &     164 $\:\:$&$\pm$&      27 $\:\:$   &     156 $\:\:$&$\pm$&      18 $\:\:$   &     126 $\:\:$&$\pm$&      14 $\:\:$   &     179 $\:\:$&$\pm$&      21 $\:\:$   &     126 $\:\:$&$\pm$&      23 $\:\:$   \\ 
  1.53 &   2.050  &     103 $\:\:$&$\pm$&      17 $\:\:$   &     135 $\:\:$&$\pm$&      16 $\:\:$   &     174 $\:\:$&$\pm$&      14 $\:\:$   &     163 $\:\:$&$\pm$&      15 $\:\:$   &     143 $\:\:$&$\pm$&      22 $\:\:$   &     117 $\:\:$&$\pm$&      28 $\:\:$   \\ 
  1.53 &   2.150  &     116 $\:\:$&$\pm$&      25 $\:\:$   &     122 $\:\:$&$\pm$&      21 $\:\:$   &      94 $\:\:$&$\pm$&      17 $\:\:$   &     176 $\:\:$&$\pm$&      33 $\:\:$   &     263 $\:\:$&$\pm$&      97 $\:\:$   &     166 $\:\:$&$\pm$&     106 $\:\:$   \\ 
 \hline       
  1.55 &   0.500  &     286 $\:\:$&$\pm$&      49 $\:\:$   &     447 $\:\:$&$\pm$&      53 $\:\:$   &     511 $\:\:$&$\pm$&      46 $\:\:$   &     622 $\:\:$&$\pm$&      97 $\:\:$   &     879 $\:\:$&$\pm$&     504 $\:\:$   &    3240 $\:\:$&$\pm$&    2290 $\:\:$   \\ 
  1.55 &   0.567  &     328 $\:\:$&$\pm$&      37 $\:\:$   &     436 $\:\:$&$\pm$&      32 $\:\:$   &     503 $\:\:$&$\pm$&      21 $\:\:$   &     474 $\:\:$&$\pm$&      30 $\:\:$   &     273 $\:\:$&$\pm$&      88 $\:\:$   &     272 $\:\:$&$\pm$&     124 $\:\:$   \\ 
  1.55 &   0.633  &     278 $\:\:$&$\pm$&      46 $\:\:$   &     345 $\:\:$&$\pm$&      38 $\:\:$   &     480 $\:\:$&$\pm$&      23 $\:\:$   &     487 $\:\:$&$\pm$&      28 $\:\:$   &     306 $\:\:$&$\pm$&      79 $\:\:$   &     365 $\:\:$&$\pm$&     117 $\:\:$   \\ 
  1.55 &   0.700  &      62 $\:\:$&$\pm$&      70 $\:\:$   &     258 $\:\:$&$\pm$&      83 $\:\:$   &     427 $\:\:$&$\pm$&      37 $\:\:$   &     412 $\:\:$&$\pm$&      39 $\:\:$   &     193 $\:\:$&$\pm$&     114 $\:\:$   &     274 $\:\:$&$\pm$&     169 $\:\:$   \\ 
  1.55 &   0.767   & \ & \ & \     & \ & \ & \     & \ & \ & \    &     440 $\:\:$&$\pm$&     454 $\:\:$    & \ & \ & \     & \ & \ & \    \\ 
  1.55 &   0.875  &     235 $\:\:$&$\pm$&      18 $\:\:$   &     301 $\:\:$&$\pm$&      20 $\:\:$   &     313 $\:\:$&$\pm$&      19 $\:\:$   &     271 $\:\:$&$\pm$&      79 $\:\:$    & \ & \ & \     & \ & \ & \    \\ 
  1.55 &   0.925  &     237 $\:\:$&$\pm$&      20 $\:\:$   &     312 $\:\:$&$\pm$&      19 $\:\:$   &     403 $\:\:$&$\pm$&      23 $\:\:$   &     133 $\:\:$&$\pm$&      57 $\:\:$    & \ & \ & \     & \ & \ & \    \\ 
  1.55 &   0.975  &     272 $\:\:$&$\pm$&      22 $\:\:$   &     334 $\:\:$&$\pm$&      20 $\:\:$   &     387 $\:\:$&$\pm$&      24 $\:\:$   &     339 $\:\:$&$\pm$&      58 $\:\:$   &     286 $\:\:$&$\pm$&      60 $\:\:$   &     302 $\:\:$&$\pm$&      72 $\:\:$   \\ 
  1.55 &   1.025  &     306 $\:\:$&$\pm$&      30 $\:\:$   &     358 $\:\:$&$\pm$&      24 $\:\:$   &     328 $\:\:$&$\pm$&      21 $\:\:$   &     318 $\:\:$&$\pm$&      20 $\:\:$   &     361 $\:\:$&$\pm$&      25 $\:\:$   &     225 $\:\:$&$\pm$&      24 $\:\:$   \\ 
  1.55 &   1.075  &     277 $\:\:$&$\pm$&      71 $\:\:$   &     244 $\:\:$&$\pm$&      42 $\:\:$   &     336 $\:\:$&$\pm$&      27 $\:\:$   &     339 $\:\:$&$\pm$&      18 $\:\:$   &     313 $\:\:$&$\pm$&      21 $\:\:$   &     274 $\:\:$&$\pm$&      27 $\:\:$   \\ 
  1.55 &   1.125  &     278 $\:\:$&$\pm$&     102 $\:\:$   &     189 $\:\:$&$\pm$&      61 $\:\:$   &     366 $\:\:$&$\pm$&      32 $\:\:$   &     298 $\:\:$&$\pm$&      18 $\:\:$   &     270 $\:\:$&$\pm$&      23 $\:\:$   &     243 $\:\:$&$\pm$&      31 $\:\:$   \\ 
  1.55 &   1.850  &     158 $\:\:$&$\pm$&      79 $\:\:$   &     138 $\:\:$&$\pm$&      44 $\:\:$   &     182 $\:\:$&$\pm$&      35 $\:\:$   &     101 $\:\:$&$\pm$&      24 $\:\:$   &     107 $\:\:$&$\pm$&      29 $\:\:$   &     134 $\:\:$&$\pm$&      41 $\:\:$   \\ 
  1.55 &   1.950  &     101 $\:\:$&$\pm$&      17 $\:\:$   &     116 $\:\:$&$\pm$&      15 $\:\:$   &     137 $\:\:$&$\pm$&      13 $\:\:$   &     144 $\:\:$&$\pm$&      13 $\:\:$   &     125 $\:\:$&$\pm$&      20 $\:\:$   &      62 $\:\:$&$\pm$&      20 $\:\:$   \\ 
  1.55 &   2.050  &      77 $\:\:$&$\pm$&      12 $\:\:$   &     121 $\:\:$&$\pm$&      13 $\:\:$   &     121 $\:\:$&$\pm$&      12 $\:\:$   &     145 $\:\:$&$\pm$&      18 $\:\:$   &      26 $\:\:$&$\pm$&      18 $\:\:$    & \ & \ & \    \\ 
  1.55 &   2.150  &      92 $\:\:$&$\pm$&      31 $\:\:$   &     138 $\:\:$&$\pm$&      33 $\:\:$   &     118 $\:\:$&$\pm$&      32 $\:\:$   &      83 $\:\:$&$\pm$&      54 $\:\:$    & \ & \ & \     & \ & \ & \    \\ 
 \hline       
  1.57 &   0.433  &     342 $\:\:$&$\pm$&     195 $\:\:$   &     100 $\:\:$&$\pm$&     147 $\:\:$   &     293 $\:\:$&$\pm$&     332 $\:\:$    & \ & \ & \     & \ & \ & \     & \ & \ & \    \\ 
  1.57 &   0.500  &     421 $\:\:$&$\pm$&      33 $\:\:$   &     464 $\:\:$&$\pm$&      28 $\:\:$   &     548 $\:\:$&$\pm$&      27 $\:\:$   &     488 $\:\:$&$\pm$&     145 $\:\:$    & \ & \ & \     & \ & \ & \    \\ 
  1.57 &   0.567  &     286 $\:\:$&$\pm$&      28 $\:\:$   &     422 $\:\:$&$\pm$&      24 $\:\:$   &     440 $\:\:$&$\pm$&      19 $\:\:$   &     407 $\:\:$&$\pm$&      63 $\:\:$    & \ & \ & \     & \ & \ & \    \\ 
  1.57 &   0.633  &     284 $\:\:$&$\pm$&      43 $\:\:$   &     339 $\:\:$&$\pm$&      33 $\:\:$   &     474 $\:\:$&$\pm$&      24 $\:\:$   &     340 $\:\:$&$\pm$&      57 $\:\:$    & \ & \ & \     & \ & \ & \    \\ 
  1.57 &   0.700  &     324 $\:\:$&$\pm$&     414 $\:\:$   &     299 $\:\:$&$\pm$&     205 $\:\:$   &     431 $\:\:$&$\pm$&      85 $\:\:$   &     327 $\:\:$&$\pm$&     150 $\:\:$    & \ & \ & \     & \ & \ & \    \\ 
  1.57 &   0.875  &     222 $\:\:$&$\pm$&      16 $\:\:$   &     258 $\:\:$&$\pm$&      14 $\:\:$   &     333 $\:\:$&$\pm$&      22 $\:\:$    & \ & \ & \     & \ & \ & \     & \ & \ & \    \\ 
  1.57 &   0.925  &     254 $\:\:$&$\pm$&      18 $\:\:$   &     299 $\:\:$&$\pm$&      17 $\:\:$   &     341 $\:\:$&$\pm$&      27 $\:\:$   &     332 $\:\:$&$\pm$&      67 $\:\:$   &     145 $\:\:$&$\pm$&      47 $\:\:$   &     174 $\:\:$&$\pm$&      73 $\:\:$   \\ 
  1.57 &   0.975  &     275 $\:\:$&$\pm$&      22 $\:\:$   &     328 $\:\:$&$\pm$&      19 $\:\:$   &     316 $\:\:$&$\pm$&      20 $\:\:$   &     322 $\:\:$&$\pm$&      20 $\:\:$   &     286 $\:\:$&$\pm$&      27 $\:\:$   &     178 $\:\:$&$\pm$&      30 $\:\:$   \\ 
  1.57 &   1.025  &     180 $\:\:$&$\pm$&      36 $\:\:$   &     306 $\:\:$&$\pm$&      33 $\:\:$   &     301 $\:\:$&$\pm$&      19 $\:\:$   &     298 $\:\:$&$\pm$&      16 $\:\:$   &     234 $\:\:$&$\pm$&      23 $\:\:$   &     167 $\:\:$&$\pm$&      28 $\:\:$   \\ 
  1.57 &   1.075  &     238 $\:\:$&$\pm$&      51 $\:\:$   &     299 $\:\:$&$\pm$&      42 $\:\:$   &     300 $\:\:$&$\pm$&      20 $\:\:$   &     295 $\:\:$&$\pm$&      16 $\:\:$   &     220 $\:\:$&$\pm$&      24 $\:\:$   &     278 $\:\:$&$\pm$&      40 $\:\:$   \\ 
  1.57 &   1.125  &     194 $\:\:$&$\pm$&      52 $\:\:$   &     188 $\:\:$&$\pm$&      36 $\:\:$   &     282 $\:\:$&$\pm$&      20 $\:\:$   &     284 $\:\:$&$\pm$&      20 $\:\:$   &     222 $\:\:$&$\pm$&      34 $\:\:$   &     138 $\:\:$&$\pm$&      37 $\:\:$   \\ 
  1.57 &   1.850  &     117 $\:\:$&$\pm$&      27 $\:\:$   &     123 $\:\:$&$\pm$&      22 $\:\:$   &     165 $\:\:$&$\pm$&      20 $\:\:$   &     131 $\:\:$&$\pm$&      18 $\:\:$   &      84 $\:\:$&$\pm$&      23 $\:\:$   &      46 $\:\:$&$\pm$&      33 $\:\:$   \\ 
  1.57 &   1.950  &      78 $\:\:$&$\pm$&      12 $\:\:$   &     131 $\:\:$&$\pm$&      13 $\:\:$   &     119 $\:\:$&$\pm$&      11 $\:\:$   &     121 $\:\:$&$\pm$&      14 $\:\:$   &      97 $\:\:$&$\pm$&      30 $\:\:$   &      78 $\:\:$&$\pm$&      36 $\:\:$   \\ 
  1.57 &   2.050  &      87 $\:\:$&$\pm$&      14 $\:\:$   &     132 $\:\:$&$\pm$&      15 $\:\:$   &     123 $\:\:$&$\pm$&      16 $\:\:$   &      73 $\:\:$&$\pm$&      23 $\:\:$    & \ & \ & \    &    1080 $\:\:$&$\pm$&     823 $\:\:$   \\ 
 \hline       
  1.59 &   0.433  &     592 $\:\:$&$\pm$&     101 $\:\:$   &     514 $\:\:$&$\pm$&      84 $\:\:$   &     605 $\:\:$&$\pm$&     160 $\:\:$    & \ & \ & \     & \ & \ & \     & \ & \ & \    \\ 
  1.59 &   0.500  &     457 $\:\:$&$\pm$&      28 $\:\:$   &     471 $\:\:$&$\pm$&      23 $\:\:$   &     571 $\:\:$&$\pm$&      31 $\:\:$    & \ & \ & \     & \ & \ & \     & \ & \ & \    \\ 
  1.59 &   0.567  &     414 $\:\:$&$\pm$&      30 $\:\:$   &     446 $\:\:$&$\pm$&      22 $\:\:$   &     456 $\:\:$&$\pm$&      25 $\:\:$    & \ & \ & \     & \ & \ & \     & \ & \ & \    \\ 
  1.59 &   0.633  &     398 $\:\:$&$\pm$&     113 $\:\:$   &     406 $\:\:$&$\pm$&      62 $\:\:$   &     463 $\:\:$&$\pm$&      48 $\:\:$    & \ & \ & \     & \ & \ & \     & \ & \ & \    \\ 
  1.59 &   0.875  &     282 $\:\:$&$\pm$&      15 $\:\:$   &     281 $\:\:$&$\pm$&      13 $\:\:$   &     244 $\:\:$&$\pm$&      26 $\:\:$   &     315 $\:\:$&$\pm$&      83 $\:\:$   &      99 $\:\:$&$\pm$&      72 $\:\:$   &     132 $\:\:$&$\pm$&      95 $\:\:$   \\ 
  1.59 &   0.925  &     315 $\:\:$&$\pm$&      21 $\:\:$   &     337 $\:\:$&$\pm$&      20 $\:\:$   &     372 $\:\:$&$\pm$&      23 $\:\:$   &     326 $\:\:$&$\pm$&      24 $\:\:$   &     231 $\:\:$&$\pm$&      38 $\:\:$   &     229 $\:\:$&$\pm$&      48 $\:\:$   \\ 
  1.59 &   0.975  &     330 $\:\:$&$\pm$&      38 $\:\:$   &     358 $\:\:$&$\pm$&      29 $\:\:$   &     354 $\:\:$&$\pm$&      18 $\:\:$   &     367 $\:\:$&$\pm$&      20 $\:\:$   &     210 $\:\:$&$\pm$&      34 $\:\:$   &     209 $\:\:$&$\pm$&      46 $\:\:$   \\ 
  1.59 &   1.025  &     298 $\:\:$&$\pm$&      42 $\:\:$   &     327 $\:\:$&$\pm$&      30 $\:\:$   &     314 $\:\:$&$\pm$&      17 $\:\:$   &     302 $\:\:$&$\pm$&      19 $\:\:$   &     277 $\:\:$&$\pm$&      47 $\:\:$   &     131 $\:\:$&$\pm$&      42 $\:\:$   \\ 
  1.59 &   1.075  &     311 $\:\:$&$\pm$&      43 $\:\:$   &     366 $\:\:$&$\pm$&      35 $\:\:$   &     350 $\:\:$&$\pm$&      19 $\:\:$   &     295 $\:\:$&$\pm$&      21 $\:\:$   &     240 $\:\:$&$\pm$&      55 $\:\:$   &     151 $\:\:$&$\pm$&      59 $\:\:$   \\ 
  1.59 &   1.125  &     195 $\:\:$&$\pm$&      50 $\:\:$   &     300 $\:\:$&$\pm$&      42 $\:\:$   &     310 $\:\:$&$\pm$&      23 $\:\:$   &     245 $\:\:$&$\pm$&      26 $\:\:$   &     212 $\:\:$&$\pm$&      26 $\:\:$   &     145 $\:\:$&$\pm$&      20 $\:\:$   \\ 
  1.59 &   1.750   & \ & \ & \    &     257 $\:\:$&$\pm$&     124 $\:\:$   &     115 $\:\:$&$\pm$&      61 $\:\:$   &     200 $\:\:$&$\pm$&      96 $\:\:$    & \ & \ & \     & \ & \ & \    \\ 
  1.59 &   1.850  &     129 $\:\:$&$\pm$&      19 $\:\:$   &     124 $\:\:$&$\pm$&      15 $\:\:$   &     166 $\:\:$&$\pm$&      15 $\:\:$   &     133 $\:\:$&$\pm$&      17 $\:\:$   &     120 $\:\:$&$\pm$&      44 $\:\:$   &     285 $\:\:$&$\pm$&     133 $\:\:$   \\ 
  1.59 &   1.950  &     121 $\:\:$&$\pm$&      13 $\:\:$   &     143 $\:\:$&$\pm$&      12 $\:\:$   &     132 $\:\:$&$\pm$&      12 $\:\:$   &     139 $\:\:$&$\pm$&      27 $\:\:$    & \ & \ & \     & \ & \ & \    \\ 
  1.59 &   2.050  &      49 $\:\:$&$\pm$&      28 $\:\:$   &     208 $\:\:$&$\pm$&      52 $\:\:$   &     174 $\:\:$&$\pm$&      67 $\:\:$    & \ & \ & \     & \ & \ & \     & \ & \ & \    \\ 
\end{tabular}\end{ruledtabular}\end{table*}
\endgroup


\begin{table}\begin{ruledtabular}
\caption{\label{tab-q2dep-figs}
The reduced cross section $ \langle d^2 \sigma \rangle $ ($\pm$ statistical error) as a function of $Q^2$, at $W=1.53$ GeV and $\cthcm =-0.975$, for the two processes H$(e,e'p)\gamma$ and H$(e,e'p)\pi^0$. The systematic error on each cross-section point is globally $\pm$ 15\% for the H$(e,e'p)\gamma$ process and $\pm$ 4\% for the H$(e,e'p)\pi^0$ process.
}
\begin{tabular}{|c|rrr|rrr|}
\multicolumn{1}{|c|}{ $Q^2$ }
& \multicolumn{3}{c|}{ $ \langle  d^2 \sigma_{\gamma}  \rangle $ \ }
& \multicolumn{3}{c|}{ $ \langle  d^2 \sigma_{\pi^0}   \rangle $ \ } \\
(GeV$^2$) & \multicolumn{3}{c|}{(nb/sr)}
& \multicolumn{3}{c|}{(nb/sr)} \\
\hline
 0.567   \ &      35.4   \ & $\pm$ \ &       7.0   \ &     480.9   \ & $\pm$ \ &      18.6 \ \\
 0.633   \ &      30.5   \ & $\pm$ \ &       5.4   \ &     499.9   \ & $\pm$ \ &      15.0 \ \\
 0.700   \ &      25.9   \ & $\pm$ \ &      13.2   \ &     468.8   \ & $\pm$ \ &      18.2 \ \\
 0.767   \ &      \      \ &       \ &  \          \ &     383.6   \ & $\pm$ \ &      57.1 \ \\
 \hline
0.875   \ &       23.7   \ & $\pm$ \ &       4.1   \ &     394.9   \ & $\pm$ \ &      15.7 \ \\
 0.925   \ &      20.4   \ & $\pm$ \ &       3.8   \ &     382.4   \ & $\pm$ \ &      14.0 \ \\
 0.975   \ &      27.3   \ & $\pm$ \ &       4.0   \ &     393.6   \ & $\pm$ \ &      14.3 \ \\
 1.025   \ &      22.2   \ & $\pm$ \ &       3.4   \ &     352.4   \ & $\pm$ \ &      13.5 \ \\
 1.075   \ &      20.7   \ & $\pm$ \ &       3.0   \ &     347.1   \ & $\pm$ \ &      10.6 \ \\
 1.125   \ &      16.0   \ & $\pm$ \ &       3.0   \ &     338.1   \ & $\pm$ \ &      11.7 \ \\
\hline
 1.950   \ &       7.8   \ & $\pm$ \ &       2.1   \ &     131.9   \ & $\pm$ \ &       7.9 \ \\
 2.050   \ &       8.6   \ & $\pm$ \ &       2.4   \ &     149.4   \ & $\pm$ \ &       7.5 \ \\
 2.150   \ &      \      \ &       \ &  \          \ &     136.0   \ & $\pm$ \ &      12.6 \ \\
\end{tabular}\end{ruledtabular}\end{table}


\begingroup\squeezetable
\begin{table*}\begin{ruledtabular}
\caption{\label{tab-ratio}
The ratio  $r$ of $\gamma$ to $\pi^0$ reduced cross sections ($\pm$ statistical error),  at  $Q^2=1$ GeV$^2$ and $\cthcm=-0.975$. The next columns contain the reduced cross section $\langle d^2 \sigma _{\gamma} \rangle$ ($\pm$ statistical $\pm$ systematic error) at these kinematics, in terms of either $ \langle d^2 \sigma_{\gamma} / d [ \Omega_p ] _{\mathrm{c.m.}} \rangle $ or $ \langle d \sigma_{\gamma} / dt \rangle $. The systematic error on $\langle d \sigma \rangle $ is obtained by  averaging over $\phigg$ the systematic error given in Table~\ref{tablehf:q2-1.0-ctcm1}.
}
\begin{tabular}{|c|rrr|rrrrr|rrrrr|}
$W$ & \multicolumn{3}{c|}{$r= \langle  d^2 \sigma_{\gamma}  \rangle \ / \ \langle  d^2 \sigma_{\pi^0}   \rangle $}
&\multicolumn{5}{c|}{$ \langle d^2 \sigma_{\gamma} / d [ \Omega_p ] _{\mathrm{c.m.}} \rangle $}
&\multicolumn{5}{c|}{$ \langle d \sigma_{\gamma} / dt \rangle $} \\  
 (GeV) & \multicolumn{3}{c|}{ \ } & \multicolumn{5}{c|}{ (nb/sr) }  & \multicolumn{5}{c|}{ (nb/GeV$^2$) } \\
\hline
 0.99  & \ & \ & \   &     497 $\:\:$&$\pm$&      67 $\:\:$&$\pm$&     283 $\:\:$  \   &   28200 $\:\:$&$\pm$&    3810 $\:\:$&$\pm$&   16100 $\:\:$  \   \\ 
 1.01  & \ & \ & \   &     262 $\:\:$&$\pm$&      23 $\:\:$&$\pm$&      81 $\:\:$  \   &   10900 $\:\:$&$\pm$&     995 $\:\:$&$\pm$&    3390 $\:\:$  \   \\ 
 1.03  & \ & \ & \   &     142 $\:\:$&$\pm$&      12 $\:\:$&$\pm$&      33 $\:\:$  \   &    4750 $\:\:$&$\pm$&     411 $\:\:$&$\pm$&    1120 $\:\:$  \   \\ 
 1.05  & \ & \ & \   &     100 $\:\:$&$\pm$&       8 $\:\:$&$\pm$&      32 $\:\:$  \   &    2800 $\:\:$&$\pm$&     242 $\:\:$&$\pm$&     906 $\:\:$  \   \\ 
 1.07  & \ & \ & \   &      88 $\:\:$&$\pm$&       7 $\:\:$&$\pm$&      19 $\:\:$  \   &    2120 $\:\:$&$\pm$&     168 $\:\:$&$\pm$&     465 $\:\:$  \   \\ 
 1.09  & \ & \ & \   &      55 $\:\:$&$\pm$&       4 $\:\:$&$\pm$&      11 $\:\:$  \   &    1180 $\:\:$&$\pm$&     103 $\:\:$&$\pm$&     247 $\:\:$  \   \\ 
 1.11 &   0.3136 & $\pm$ &   0.0315 \   &      60 $\:\:$&$\pm$&       4 $\:\:$&$\pm$&      10 $\:\:$  \   &    1140 $\:\:$&$\pm$&      92 $\:\:$&$\pm$&     199 $\:\:$  \   \\ 
 1.13 &   0.1083 & $\pm$ &   0.0099 \   &   54.3 &$\pm$&    4.7 &$\pm$&    9.2  \   &     939 $\:\:$&$\pm$&      80 $\:\:$&$\pm$&     159 $\:\:$  \   \\ 
 1.15 &   0.0644 & $\pm$ &   0.0051 \   &      67 $\:\:$&$\pm$&       5 $\:\:$&$\pm$&      18 $\:\:$  \   &    1070 $\:\:$&$\pm$&      82 $\:\:$&$\pm$&     288 $\:\:$  \   \\ 
 1.17 &   0.0462 & $\pm$ &   0.0030 \   &      91 $\:\:$&$\pm$&       5 $\:\:$&$\pm$&      10 $\:\:$  \   &    1350 $\:\:$&$\pm$&      83 $\:\:$&$\pm$&     147 $\:\:$  \   \\ 
 1.19 &   0.0309 & $\pm$ &   0.0018 \   &     100 $\:\:$&$\pm$&       5 $\:\:$&$\pm$&      10 $\:\:$  \   &    1370 $\:\:$&$\pm$&      76 $\:\:$&$\pm$&     143 $\:\:$  \   \\ 
 1.21 &   0.0234 & $\pm$ &   0.0013 \   &      94 $\:\:$&$\pm$&       5 $\:\:$&$\pm$&      10 $\:\:$  \   &    1210 $\:\:$&$\pm$&      68 $\:\:$&$\pm$&     131 $\:\:$  \   \\ 
 1.23 &   0.0212 & $\pm$ &   0.0013 \   &   72.4 &$\pm$&    4.3 &$\pm$&    9.5  \   &     874 $\:\:$&$\pm$&      51 $\:\:$&$\pm$&     115 $\:\:$  \   \\ 
 1.25 &   0.0212 & $\pm$ &   0.0014 \   &   51.3 &$\pm$&    3.2 &$\pm$&    7.4  \   &     586 $\:\:$&$\pm$&      36 $\:\:$&$\pm$&      84 $\:\:$  \   \\ 
 1.27 &   0.0192 & $\pm$ &   0.0014 \   &   32.8 &$\pm$&    2.3 &$\pm$&    3.9  \   &     355 $\:\:$&$\pm$&      24 $\:\:$&$\pm$&      41 $\:\:$  \   \\ 
 1.29 &   0.0188 & $\pm$ &   0.0014 \   &   22.7 &$\pm$&    1.6 &$\pm$&    2.6  \   &     233 $\:\:$&$\pm$&      16 $\:\:$&$\pm$&      26 $\:\:$  \   \\ 
 1.31 &   0.0239 & $\pm$ &   0.0019 \   &   20.9 &$\pm$&    1.6 &$\pm$&    4.2  \   &     205 $\:\:$&$\pm$&      15 $\:\:$&$\pm$&      41 $\:\:$  \   \\ 
 1.33 &   0.0264 & $\pm$ &   0.0022 \   &   18.7 &$\pm$&    1.5 &$\pm$&    2.6  \   &     175 $\:\:$&$\pm$&      13 $\:\:$&$\pm$&      24 $\:\:$  \   \\ 
 1.35 &   0.0292 & $\pm$ &   0.0025 \   &   16.4 &$\pm$&    1.4 &$\pm$&    3.3  \   &     147 $\:\:$&$\pm$&      12 $\:\:$&$\pm$&      29 $\:\:$  \   \\ 
 1.37 &   0.0277 & $\pm$ &   0.0028 \   &   12.9 &$\pm$&    1.3 &$\pm$&    3.6  \   &     111 $\:\:$&$\pm$&      10 $\:\:$&$\pm$&      31 $\:\:$  \   \\ 
 1.39 &   0.0354 & $\pm$ &   0.0035 \   &   14.1 &$\pm$&    1.3 &$\pm$&    2.6  \   &     116 $\:\:$&$\pm$&      11 $\:\:$&$\pm$&      21 $\:\:$  \   \\ 
 1.41 &   0.0444 & $\pm$ &   0.0044 \   &   14.7 &$\pm$&    1.4 &$\pm$&    2.0  \   &     117 $\:\:$&$\pm$&      11 $\:\:$&$\pm$&      16 $\:\:$  \   \\ 
 1.43 &   0.0441 & $\pm$ &   0.0047 \   &   13.1 &$\pm$&    1.4 &$\pm$&    2.2  \   &     101 $\:\:$&$\pm$&      10 $\:\:$&$\pm$&      16 $\:\:$  \   \\ 
 1.45 &   0.0594 & $\pm$ &   0.0055 \   &   17.2 &$\pm$&    1.6 &$\pm$&    2.6  \   &     128 $\:\:$&$\pm$&      11 $\:\:$&$\pm$&      18 $\:\:$  \   \\ 
 1.47 &   0.0677 & $\pm$ &   0.0054 \   &   21.3 &$\pm$&    1.7 &$\pm$&    2.3  \   &     152 $\:\:$&$\pm$&      11 $\:\:$&$\pm$&      16 $\:\:$  \   \\ 
 1.49 &   0.0874 & $\pm$ &   0.0057 \   &   27.6 &$\pm$&    1.7 &$\pm$&    3.9  \   &     191 $\:\:$&$\pm$&      11 $\:\:$&$\pm$&      27 $\:\:$  \   \\ 
 1.51 &   0.0722 & $\pm$ &   0.0043 \   &   27.8 &$\pm$&    1.6 &$\pm$&    3.3  \   &     186 $\:\:$&$\pm$&      10 $\:\:$&$\pm$&      22 $\:\:$  \   \\ 
 1.53 &   0.0616 & $\pm$ &   0.0039 \   &   23.2 &$\pm$&    1.4 &$\pm$&    2.7  \   &     151 $\:\:$&$\pm$&       9 $\:\:$&$\pm$&      17 $\:\:$  \   \\ 
 1.55 &   0.0704 & $\pm$ &   0.0044 \   &   22.0 &$\pm$&    1.3 &$\pm$&    2.4  \   &     139 $\:\:$&$\pm$&       8 $\:\:$&$\pm$&      15 $\:\:$  \   \\ 
 1.57 &   0.0659 & $\pm$ &   0.0045 \   &   17.5 &$\pm$&    1.2 &$\pm$&    3.2  \   &     107 $\:\:$&$\pm$&       7 $\:\:$&$\pm$&      19 $\:\:$  \   \\ 
 1.59 &   0.0700 & $\pm$ &   0.0046 \   &   18.3 &$\pm$&    1.2 &$\pm$&    2.8  \   &     108 $\:\:$&$\pm$&       7 $\:\:$&$\pm$&      16 $\:\:$  \   \\ 
 1.61 &   0.0503 & $\pm$ &   0.0037 \   &   15.6 &$\pm$&    1.1 &$\pm$&    2.2  \   &      89 $\:\:$&$\pm$&       6 $\:\:$&$\pm$&      12 $\:\:$  \   \\ 
 1.63 &   0.0427 & $\pm$ &   0.0028 \   &   17.5 &$\pm$&    1.1 &$\pm$&    1.8  \   &      98 $\:\:$&$\pm$&       6 $\:\:$&$\pm$&      10 $\:\:$  \   \\ 
 1.65 &   0.0306 & $\pm$ &   0.0020 \   &   17.4 &$\pm$&    1.1 &$\pm$&    2.4  \   &      94 $\:\:$&$\pm$&       6 $\:\:$&$\pm$&      12 $\:\:$  \   \\ 
 1.67 &   0.0180 & $\pm$ &   0.0015 \   &   12.7 &$\pm$&    1.0 &$\pm$&    2.0  \   &      67 $\:\:$&$\pm$&       5 $\:\:$&$\pm$&      10 $\:\:$  \   \\ 
 1.69 &   0.0165 & $\pm$ &   0.0016 \   &   11.8 &$\pm$&    1.1 &$\pm$&    4.2  \   &      60 $\:\:$&$\pm$&       5 $\:\:$&$\pm$&      21 $\:\:$  \   \\ 
 1.71 &   0.0141 & $\pm$ &   0.0017 \   &    8.8 &$\pm$&    1.0 &$\pm$&    1.9  \   &   44.2 &$\pm$&    5.2 &$\pm$&    9.3  \   \\ 
 1.73 &   0.0149 & $\pm$ &   0.0019 \   &    6.7 &$\pm$&    0.8 &$\pm$&    2.6  \   &      32 $\:\:$&$\pm$&       4 $\:\:$&$\pm$&      12 $\:\:$  \   \\ 
 1.75 &   0.0167 & $\pm$ &   0.0021 \   &    6.2 &$\pm$&    0.8 &$\pm$&    1.2  \   &   29.5 &$\pm$&    3.6 &$\pm$&    5.6  \   \\ 
 1.77 &   0.0153 & $\pm$ &   0.0020 \   &    5.2 &$\pm$&    0.7 &$\pm$&    2.0  \   &   24.4 &$\pm$&    3.2 &$\pm$&    9.1  \   \\ 
 1.79 &   0.0160 & $\pm$ &   0.0021 \   &    5.3 &$\pm$&    0.7 &$\pm$&    2.5  \   &      24 $\:\:$&$\pm$&       3 $\:\:$&$\pm$&      11 $\:\:$  \   \\ 
 1.81 &   0.0199 & $\pm$ &   0.0021 \   &    7.0 &$\pm$&    0.7 &$\pm$&    2.3  \   &      31 $\:\:$&$\pm$&       3 $\:\:$&$\pm$&      10 $\:\:$  \   \\ 
 1.83 &   0.0156 & $\pm$ &   0.0020 \   &    5.6 &$\pm$&    0.7 &$\pm$&    2.0  \   &   24.0 &$\pm$&    3.1 &$\pm$&    8.7  \   \\ 
 1.85 &   0.0159 & $\pm$ &   0.0020 \   &    5.7 &$\pm$&    0.7 &$\pm$&    1.1  \   &   23.9 &$\pm$&    3.0 &$\pm$&    4.9  \   \\ 
 1.87 &   0.0139 & $\pm$ &   0.0021 \   &    4.8 &$\pm$&    0.7 &$\pm$&    1.2  \   &   19.7 &$\pm$&    2.9 &$\pm$&    4.9  \   \\ 
 1.89 &   0.0210 & $\pm$ &   0.0025 \   &    6.7 &$\pm$&    0.8 &$\pm$&    2.6  \   &      27 $\:\:$&$\pm$&       3 $\:\:$&$\pm$&      10 $\:\:$  \   \\ 
 1.91 &   0.0188 & $\pm$ &   0.0028 \   &    5.5 &$\pm$&    0.8 &$\pm$&    1.7  \   &   21.7 &$\pm$&    3.2 &$\pm$&    6.5  \   \\ 
 1.93 &   0.0080 & $\pm$ &   0.0029 \   &    2.1 &$\pm$&    0.8 &$\pm$&    0.6  \   &    8.1 &$\pm$&    3.0 &$\pm$&    2.4  \   \\ 
 1.95 &   0.0153 & $\pm$ &   0.0043 \   &    3.4 &$\pm$&    1.0 &$\pm$&    1.0  \   &   12.9 &$\pm$&    3.6 &$\pm$&    3.9  \   \\ 
 1.97 &   0.0278 & $\pm$ &   0.0093 \   &    5.2 &$\pm$&    1.7 &$\pm$&    1.6  \   &   19.1 &$\pm$&    6.4 &$\pm$&    5.7  \   \\  
\end{tabular}\end{ruledtabular}\end{table*}
\endgroup


\end{document}